%% file: main.tex
\documentclass[twocolumn]{aastex631}
\usepackage{natbib}
\usepackage{enumitem}
\defcitealias{peroux2021}{PH20}
\defcitealias{madau2014}{MD14}
\defcitealias{lehner2022}{L22}
\defcitealias{bellstedt2020}{B20}
\defcitealias{lehner2016}{L16}
\defcitealias{omeara2013}{O13}
\defcitealias{ribaudo2011}{R11}
\defcitealias{deepak2025}{D25}
\defcitealias{wotta2019}{W19}
\defcitealias{hm05}{HM05}

\usepackage{amsmath}
\renewcommand{\H}{\ensuremath{\mathrm{H}}}
\newcommand{\HI}{\ensuremath{\mathrm{H\,\mathrm{I}}}}
\newcommand{\HII}{\ensuremath{\mathrm{H\,\mathrm{II}}}}
\newcommand{\OI}{\ensuremath{\mbox{\ion{O}{1}}}}
\newcommand{\OIV}{\ensuremath{\mbox{\ion{O}{4}}}}
\newcommand{\OVI}{\ensuremath{\mbox{\ion{O}{6}}}}
\newcommand{\SiII}{\ensuremath{\mbox{\ion{Si}{2}}}}
\newcommand{\CIII}{\ensuremath{\mbox{\ion{C}{3}}}}
\newcommand{\FeII}{\ensuremath{\mbox{\ion{Fe}{2}}}}
\newcommand{\FeIII}{\ensuremath{\mbox{\ion{Fe}{3}}}}
\newcommand{\SiIII}{\ensuremath{\mbox{\ion{Si}{3}}}}
\newcommand{\CII}{\ensuremath{\mbox{\ion{C}{2}}}}
\newcommand{\CIV}{\ensuremath{\mbox{\ion{C}{4}}}}
\newcommand{\SiIV}{\ensuremath{\mbox{\ion{Si}{4}}}}
\newcommand{\MgI}{\ensuremath{\mbox{\ion{Mg}{1}}}}
\newcommand{\MgII}{\ensuremath{\mbox{\ion{Mg}{2}}}}
\newcommand{\NI}{\ensuremath{\mbox{\ion{N}{1}}}}
\newcommand{\NII}{\ensuremath{\mbox{\ion{N}{2}}}}
\newcommand{\SII}{\ensuremath{\mbox{\ion{S}{2}}}}
\newcommand{\SIII}{\ensuremath{\mbox{\ion{S}{3}}}}

\newcommand{\NV}{\ensuremath{\mbox{\ion{N}{5}}}}
\newcommand{\AlII}{\ensuremath{\mbox{\ion{Al}{2}}}}
\newcommand{\ZnII}{\ensuremath{\mbox{\ion{Zn}{2}}}}
\newcommand{\AlIII}{\ensuremath{\mbox{\ion{Al}{3}}}}

\newcommand{\numberabs}{105}

\newcommand{\kms}{\ensuremath{{\rm km\,s}^{-1}}}

\shorttitle{CGM metals at cosmic noon}
\shortauthors{Deepak et al.}

\begin{document}

\title{The BRIDGE Survey. I. Metallicities and Physical Properties of Gas Probing the Circumgalactic Medium at Cosmic Noon}

\author[0000-0003-4203-6223]{Saloni Deepak}
\affiliation{Department of Physics and Astronomy, University of Notre Dame, Notre Dame, IN 46556}

\author[0000-0001-9158-0829]{Nicolas Lehner}
\affiliation{Department of Physics and Astronomy, University
of Notre Dame, Notre Dame, IN 46556}

\author[0000-0002-2591-3792]{J. Christopher Howk}
\affiliation{Department of Physics and Astronomy, University of Notre Dame, Notre Dame, IN 46556}

\author[0000-0002-7893-1054]{John O'Meara}
\affiliation{W.M. Keck Observatory 65-1120 Mamalahoa Highway, Kamuela, HI 96743, USA}

\author[0000-0002-4288-599X]{C\'eline P\'eroux}
\affiliation{European Southern Observatory, Karl-Schwarzschild-Str. 2, 85748 Garching-bei-M\"unchen, Germany; email: cperoux@eso.org }
\affiliation{Aix Marseille Universit\'e, CNRS, LAM (Laboratoire d'Astrophysique de Marseille) UMR 7326, 13388, Marseille, France}

\author[0000-0002-7738-6875]{J. Xavier Prochaska}
\affiliation{Department of Astronomy \& Astrophysics, UCO/Lick Observatory, University of California, 1156 High Street, Santa Cruz, CA 95064, USA}
\affiliation{Kavli Institute for the Physics and Mathematics of the Universe (Kavli IPMU), 5-1-5 Kashiwanoha, Kashiwa, 277-8583, Japan}
\affiliation{Division of Science, National Astronomical Observatory of Japan, 2-21-1 Osawa, Mitaka, Tokyo 181-8588, Japan}

\begin{abstract}
We present BRIDGE-I, a metallicity survey of \HI-selected absorbers probing ``CGM-like'' gas ($16.4\le\log N_{\HI}<20.3$) at Cosmic Noon ($1\le z\le2.2$), the epoch of peak star formation and AGN activity. Combining archival HST UV spectra with Keck/HIRES and VLT/UVES optical spectra, we estimate metallicities and physical conditions for \numberabs\ absorbers. The Cosmic Noon metallicity distribution (median $[\rm{X/H}]=-1.19$) is statistically indistinguishable from that at $z<1$ ($-1.18$) but differs significantly from $z>2.2$ ($-2.02$). This plateau is driven by the most diffuse absorbers ($16.4\le\log N_{\HI}<17.2$), which reach their $z<1$ enrichment level by Cosmic Noon, whereas denser absorbers remain more metal-poor than their $z<1$ counterparts and continue to be enriched at later times. Unlike at $z>2.2$, no absorber falls below $[\rm{X/H}]=-3.5$, indicating no pristine CGM-like gas at Cosmic Noon. Absorbers in BRIDGE-I trace overdensities of $400\lesssim \delta_b\lesssim2200$, higher than at $z>2.2$. Mean $N_{\HI}$ differs by a factor of 400 across absorber types, but mean total hydrogen column by only 2.5, as higher $N_{\HI}$ is offset by a higher neutral fraction. The absorber classes thus differ in ionization state and metallicity rather than the total gas content. A metallicity floor of $[\rm{X/H}]\sim-2.0$ emerges for $\log N_{\HI}>19.0$ absorbers, indicating that denser, more ISM-like gas is well mixed as a population, while more diffuse gas retains a metal-poor population that is not yet fully mixed.
\end{abstract}

\section{Introduction}
\label{sec:intro}
The chemical history of the gas in and around galaxies offers key insights into the physical processes---accretion, star formation, large-scale outflows, AGN feedback, mergers, tidal stripping, and so on---that shape its large-scale properties across cosmic time. The atomic, cool, and warm-hot ($T<10^5$ K), diffuse gas in the circumgalactic medium (CGM) and intergalactic medium (IGM) fuels sustained star formation in galaxies through accretion of metal-poor gas along large-scale filaments and eventual condensation into cold molecular clouds \citep{leroy2008,schruba2011,bouche2013}. These multi-phase gaseous media also transport enriched material from the interstellar medium (ISM) to the outer regions of the galaxy via feedback-driven outflows and galactic fountains, powered by supernovae and AGN \citep{heckman2017,puglisi2021,rubin2022,liou2026}. This interplay has motivated comprehensive metallicity studies across all three media: the ISM \citep{rao2006,kulkarni2007,rafelski2012,som2013,decia2018,decia2021,peroux2021}, the CGM \citep{werk2014,lehner2016,lehner2018,lehner2022,wotta2019,wilde2021,zahedy2021}, and the IGM \citep{songaila1998,ellison2000,schaye2003,aguirre2004,simcoe2004,simcoe2011,shull2014}. Together with surveys of stellar-phase metallicities \citep{kirby2013,gallazzi2014,cullen2019,chartab2024} and the hot gas in the intra-group (IGrM) and intra-cluster media (ICM; \citealt{molendi2016,mantz2017,yates2017,liu2020,ghizzardi2021}), these studies trace the evolution of the cosmic baryon and metal budgets \citep{peeples2014, shull2014,peroux2021,molendi2024,deepak2025}.  Baryons condense from the diffuse IGM into large-scale structures, whereas metals cycle through multiple reservoirs. They remain locked in long-lived stars, enrich star-forming clouds, and are redistributed outward through supernova and AGN feedback. As a result, the metal census of the Universe, which is dominated by cool, dense, neutral gas at high redshift, diversifies with cosmic time, with stars and the IGrM+ICM emerging as the dominant reservoirs by $z<1$ \citep{peroux2021,deepak2025}. 

Early studies of the cosmic metal budget identified a ``missing metals problem" at $z\sim2.5$ \citep{ferrara2005,bouche2005,bouche2007}, where the observed total global metal mass density of the Universe fell an order of magnitude short of the expected value. This problem was revisited several times and attributed to metals residing in the diffuse, ionized gas phases in the galaxy halos \citep{lehner2009,fox2013,shull2014,peeples2014}. We also addressed this problem in \cite{deepak2025} (hereafter \citetalias{deepak2025}), finding that the cool, diffuse, ionized gas indeed contributes 20\% of all metals at $z\gtrsim2.2$--3.6, while at even higher redshift ($z\gtrsim4$), almost all metals reside in the cool, neutral gas probed by damped Lyman-$\alpha$ absorbers (DLAs). We concluded that there is no strong evidence for a missing metals problem at these redshifts. This census identified a gap in our understanding of the metallicity distribution of cool, diffuse, ionized gas tracing the CGM and IGM at $z=1$--2.2, or Cosmic Noon, a period of peak cosmic star formation and AGN activity \citep{madau2014,mcleoud2021}. While extensive surveys of the CGM and IGM exist at low ($z<1$; \citealt{prochaska2017,shull2017,lehner2018,wotta2019,lehner2019,berg2019,berg2023,burchett2019,zahedy2021}) and high redshift ($z>2.2$; \citealt{ellison2000,aguirre2004,lehner2014,lehner2022,prochaska2014,prochaska2015,fumagalli2016,dodorico2022}), a gap remains at these intermediate redshifts ($1\le z\le2.2$). There is virtually no information on the metal distribution of ionized absorbers with \HI\, column densities $\log N_{\HI} < 19.0$ at Cosmic Noon. These gaps leave several key questions unresolved: whether any pristine gas survives during the peak of cosmic star formation, how the enrichment timescales of the CGM and IGM compare to those of the ISM, and how completely metals are mixed into the diffuse gas at these redshifts. Because the extragalactic UV background peaks near Cosmic Noon, the physical conditions inferred for cool, diffuse gas may differ systematically from those at lower and higher redshift. Constraining these properties, such as the ionization parameter and neutral fraction, at Cosmic Noon is therefore also essential.

The lack of a comprehensive metal survey in the diffuse, ionized gas surrounding galaxies at Cosmic Noon stems from observational constraints. Estimating the metallicities in this gas requires measuring the column densities of hydrogen and several metal ions in absorbers detected along quasar spectra. At low redshift ($z<1$), both hydrogen and most metal lines lie in the UV band (with some in the NUV, see \citealt{lehner2019}) and are accessible with high-resolution, space-based UV spectra. At high redshift ($z>2.2$), these same lines are redshifted into the optical band, where high-resolution ground-based optical spectra are used to estimate both hydrogen and metal ion column densities. At intermediate redshifts ($1\leq z\leq2.2$), the \HI\, lines fall in the near-UV while most of the metal lines remain in the optical. Estimating metallicities of diffuse gas at Cosmic Noon therefore requires both near-UV and optical spectroscopy. 

In this paper, we combine archival Hubble Space Telescope (HST) UV spectra with high-resolution, high-quality optical spectra from Keck High Resolution Echelle Spectrometer (HIRES) and Very Large Telescope (VLT) Ultraviolet and Visual Echelle Spectrograph (UVES) for 56 quasars, constructing a sample of \numberabs\, absorbers at Cosmic Noon. Here, we present the first in a series of papers on the metallicity distribution of CGM and IGM gas at Cosmic Noon: the BRIDGE survey. This paper (BRIDGE-I) focuses on metals in ``CGM-like" gas, which we define as absorbers with \HI\, column densities: $16.2\le\log N_{\HI}<20.3$, which often probe overdensities $\delta_b\gtrsim200$ \citep{mcquinn2016,tumlinson2017}. Our next paper in this series will cover metals in ``IGM-like" gas ($\delta_b<200$) traced by absorbers with $\log N_{\rm{HI}}<16.2$. Ultimately, our goal is to update the global metal budget presented in \citetalias{deepak2025} to bridge the gap at Cosmic Noon. 

A statistical study of global metallicity distribution and global metal mass densities requires an unbiased survey of metals.  Our survey therefore uses only \HI-selected absorbers: we first detect an absorber via its \HI\, absorption in UV spectra, then search for corresponding metal absorption in the optical spectra. This \HI-selection has the advantage of being agnostic to the environments being probed, allowing us to study gas metallicities as functions of physical properties rather than location. By contrast, metal-line selection and a galaxy-centered search of absorbers bias metallicity distributions to higher values. For example, absorbers in galaxy-absorber association studies, which target gas close to galaxies, show systematically higher metallicities \citep{battisti2012,rao2017,nielsen2022,berg2023}. 

The method used to estimate $N_{\HI}$ at Cosmic Noon depends on both $N_{\HI}$ itself and the absorber's redshift $z_{\rm abs}$, and the absorber classes in common use are named for the characteristic spectral feature that each regime makes available. Two diagnostics are relevant here. The first is the flux decrement at the Lyman limit ($\sim912$ \AA), which sets the classification. Absorbers with $\tau_{912}<1$ corresponding to $16.2\le\log N_{\HI}<17.2$, are partial Lyman limit systems (pLLSs), for which the partial decrement measures $N_{\HI}$ directly. Absorbers with $\tau_{912}>1$ and $17.2\le\log N_{\HI}<19.0$ are called Lyman limit systems (LLSs), the upper bound being conventional rather than physical. At $\tau_{912}\gg1$ and $\log N_{\HI}\ge19.0$, the limit is fully saturated and no flux remains blueward of it, so the decrement alone yields only a lower bound on $N_{\HI}$; these are super-LLSs (SLLS, also known as sub-DLAs). The second is the Ly$\alpha$ profile, which is accessible in the high-resolution optical spectra at $z_{\rm abs}\gtrsim1.7$. Its damping wings are prominent for $\log N_{\HI}\ge20.3$, the defining threshold for DLAs, and weaken toward lower columns. For SLLSs, the weakness or absence of damping wings therefore places an upper bound on $N_{\HI}$ which, combined with the lower bound from the saturated Lyman limit, brackets the column density. Such absorbers are also referred to as sub-DLAs ($19.0\le\log N_{\HI}<20.3$) in the literature. Our sample comprises pLLSs, LLSs, and SLLSs.

In \S\ref{sec:data}, we describe our survey design and database of CGM absorbers, along with data reduction and systematic uncertainties for each dataset adopted from the literature. In \S\ref{sec:HI} and \S\ref{sec:metal_Cols}, we discuss our treatment of \HI\, column densities and metal ion column densities, respectively. In \S\ref{sec:metallicity}, we estimate metallicities via photoionization modeling and define our assumptions and priors. Our results for the metallicity distribution of CGM-like absorbers, and their estimated physical properties are presented in \S\ref{sec:results}. In \S\ref{sec:discussion}, we discuss the timescale of metallicity evolution for CGM-like absorbers, overdensities probed by these absorbers, metal mixing in the ISM- and CGM-like gas, and the ionization properties of gas in the context of an evolving UV background. We summarize our conclusions in \S\ref{sec:summary}.

We adopt a concordance $\Lambda\rm{CDM}$ cosmology, notably $H_0 = 70.0$ \kms$\, {\rm Mpc^{-1}}$, $\Omega_m = 0.3$ and $\Omega_{\Lambda} = 0.7$. We use the solar abundance from \cite{asplund2009}, particularly the proto-solar mass fraction of heavy elements, $Z_\odot=0.0142$. 

\section{Survey Design and Database}\label{sec:data}

Estimating the metallicity\footnote{We define metallicity as the logarithmic abundance relative to solar: $[\rm{X}/\rm{H}]\equiv\log (N_{\rm{X}}/ N_{\rm{H}}) - \log (N_{\rm{X}}/ N_{\rm{H}})_{\odot}$, where X denotes a metal element. For gas-phase metallicities, X is typically an $\alpha$-element such as O or Si. At Cosmic Noon, we may also use carbon to constrain the metallicity.} of an ionized absorber requires measurements of both metal ion column densities and the \HI\, column density, which together can be used to determine elemental column densities via ionization modeling. We aim to estimate the metallicities for an \HI-selected sample; that is, we first detect the absorbers in \HI\, absorption and then search for corresponding metal lines at the redshift of the absorber $z_{\rm{abs}}$. At the redshifts spanning Cosmic Noon ($1\leq z\leq2.2$), several Lyman series lines and the Lyman limit lie in the UV, below Earth's atmospheric cutoff, so space-based UV spectra from HST are required to estimate $N_{\HI}$. Most metal lines, by contrast, are accessible from the ground, as is Ly$\alpha$ for absorbers at $z\gtrsim1.7$ (and Ly$\beta$ only near the upper end of the redshift range), together with most metal lines are accessible from the ground. We therefore estimate $N_{\HI}$ from HST UV spectra, and metal ion column densities using high-resolution, ground-based, optical spectra obtained with the Keck/HIRES and the VLT/UVES.

\subsection{Space-based UV Data: HST}\label{subsec:uv_spec}

Our database of \HI\, absorbers is primarily drawn from two surveys of \HI-selected pLLSs and LLSs---\cite{ribaudo2011} (hereafter \citetalias{ribaudo2011}) and \cite{omeara2013} (hereafter \citetalias{omeara2013}). Both use the flux decrement at the Lyman limit to detect an absorber and estimate its \HI\, column density down to $\log N_{\HI}\gtrsim16.2$. We also include a sample of \HI-selected SLLSs from the literature \citep{dessauges2003,som2013,prochaska2015}. Below we discuss the data used by \citetalias{ribaudo2011} and \citetalias{omeara2013} as well as the treatment of systematic uncertainties specific to each instrument.

\subsubsection{R11: FOS and STIS}\label{subsub:r11}

\citetalias{ribaudo2011} searched for pLLSs and LLSs using archival low-resolution UV/near-UV spectra from HST STIS and FOS. For the STIS spectra, they combined multiple exposures (weighted by exposure times) taken with G140L (1150--1700 \AA) and G230L (1600--3100 \AA), which yielded a spectral resolution of $R\sim1000$. Lower resolution ($R\sim250$, ``FOS-L") FOS spectra were processed similarly, and supplemented by \cite{bechtold2002} reductions of higher-resolution FOS observations (``FOS-H"; $R\sim1300$) from G130H (1140--1600 \AA), G190H (1575--2330 \AA), and G270H (2220--3300 \AA). For sightlines with both FOS-L and FOS-H data, \citetalias{ribaudo2011} found a mean wavelength shift of 4 \AA\ between the two. For 20 such sightlines, they shifted the G160L spectra to align with the higher resolution FOS-H spectra.\footnote{Wavelength shifts can introduce substantial redshift uncertainty, especially when no Lyman series lines are available and $z_{\rm abs}$ relies solely on the Lyman limit. Since we use \HI\, to identify corresponding metal lines in the higher-resolution optical data, this raises the risk of misidentifying associated metal lines. When Ly$\alpha$ is available, we pay special attention to identifying the \HI\, component dominating the Lyman limit flux decrement.} \citetalias{ribaudo2011} note that background subtraction uncertainty in the reduced FOS spectra can be as high as $\sim30\%$. This dominates the flux error budget in strongly absorbed regions blueward of the Lyman limit, contributing to the uncertainty in the estimated $N_{\HI}$. To account for this, they estimate the background flux as the product of the inverse sensitivity function and the count rate for each grating, adopting $30\%$ of this value as its uncertainty, which propagates into their $N_{\HI}$ error budget.

To estimate $N_{\HI}$, \citetalias{ribaudo2011} model the QSO continuum using the \cite{zheng1997} composite spectrum scaled to match the observed spectra over relatively absorption-free wavelength ranges and applying a running $\chi^2$ goodness-of-fit test over $\sim$30 \AA\ windows to identify continuum deviations (excluding false positives from strong absorption lines). Spectra poorly fit by this routine were individually examined for Lyman limits. The final $N_{\HI}$ uncertainties combine statistical flux error with the uncertainty due to background subtraction. The systematic uncertainties in $z_{\rm{abs}}$ were assessed by comparing low- and high-resolution spectra. 

\citetalias{ribaudo2011} constructed their sample from the $\sim700$ quasars of \cite{bechtold2002}, removing poor quality data or quasars for which the Lyman limit fell outside the wavelength coverage. This yielded a final sample of 249 quasars, along which \citetalias{ribaudo2011} detected 206 absorbers. For our survey of \HI-selected ``CGM-like" absorbers, we exclude absorbers that were detected toward quasars that were specially targeted because of previously known absorption features such as the presence of DLAs (9) or strong \MgII\, absorbers (49). Including these may bias our survey to higher metallicities. This leaves 148 \HI-selected absorbers along 113 quasars, for which we adopt the $N_{\HI}$ and uncertainties derived by \citetalias{ribaudo2011}. 

\subsubsection{O13: WFC3 and ACS}\label{subsub:o13}

We also include \HI-absorbers from \citetalias{omeara2013} detected in WFC3 UVIS-G280 grism (53 sightlines) and ACS PR200L prism spectra (18 sightlines) of 71 quasars. WFC3 and ACS cover 2000--6000 \AA\ and  1500--5000 \AA, respectively, with relatively good signal-to-noise ($\gtrsim10$ per pixel) down to 2000 \AA\ and 1800 \AA, respectively. Both instruments have non-uniform wavelength dispersions ($\Delta\lambda\sim14$ \AA\ for WFC3 and $\sim22$ \AA\ for ACS at 2500 \AA), and \citet{omeara2011} estimate a wavelength calibration error of 2 pixels; together these introduce non-uniform redshift uncertainties. \citetalias{omeara2013} detected 145 absorbers along these 71 spectra via the flux decrement at the Lyman limit, of which 5 were also detected by \citetalias{ribaudo2011} (along J080620+504124 and J162548+264658). For these absorbers, we adopt $z_{\rm{abs}}$ and $N_{\HI}$ determined by \citetalias{ribaudo2011} using higher-resolution STIS and FOS spectra. 

\citetalias{omeara2013} used a custom tool to identify absorbers and fit models of the Lyman limit opacity at $z_{\rm{abs}}$ to estimate $N_{\HI}$. Based on inter-comparisons of models from different authors, they report rms uncertainties of $\sim0.05$ and $\sim0.02$ in optical depth and $z_{\rm{abs}}$, respectively. They note that continuum placement and IGM line blending dominate their error budget, contributing up to 15\% uncertainty on $N_{\HI}$. Since the \citetalias{omeara2013} analysis did not provide specific uncertainties for the individual measurements of $N_{\HI}$, we developed our own fitting process to obtain robust uncertainties on $N_{\HI}$. In particular, we simultaneously model the quasar continuum and absorption from all detected pLLSs and LLSs along each sightline using a Markov Chain Monte Carlo (MCMC) approach. This allows us to account for covariance between the continuum (including uncertainties in the adopted quasar template), redshift uncertainty (from the large wavelength dispersion in the WFC3 and ACS spectra, and wavelength calibration error), and $N_{\HI}$ (see \S\ref{sec:HI}). 

\subsection{Literature Sample}\label{subsec:lit_sample}

We also draw a sample of 17 \HI-selected SLLSs ($19.0\le\log N_{\HI}<20.3$) at Cosmic Noon from the HD-LLS survey \citep{prochaska2015,fumagalli2016}. These include 12 absorbers from  \cite{prochaska2015}, 3 from \cite{dessauges2003} and 2 from \cite{som2013}. Column densities for these absorbers were estimated for \HI\, (Ly$\alpha$ profile fitting) and available metal ions (including \AlII, \AlIII, \SiII, \ZnII, \MgI, \MgII, \FeII) in high-resolution ground-based optical spectra obtained with VLT/UVES \citep{dessauges2003}, the Magellan Inamori Kyocera Echelle (MIKE) spectrograph \citep{som2013}, and MIKE + Keck (HIRES and ESI) \citep{prochaska2015}. \cite{fumagalli2016} estimated metallicities for these absorbers, however, they used the \cite{hm12} UV background. We adopt the literature column densities and uncertainties for these absorbers, and re-estimate metallicities using \cite{hm05} UV background for consistency (see \S\ref{sec:metallicity} and Appendix~\ref{app:euvb}). Our results and comparison with \cite{fumagalli2016} metallicities are presented in Appendix~\ref{app:lit_sample}.

The velocity ranges for integration adopted by \cite{som2013} ($|v|\lesssim250$ \kms), \cite{dessauges2003} ($|v|\lesssim500$ \kms), and \cite{prochaska2015} ($|v|<500$ \kms, though rarely exceeding $200$ \kms) are comparable to our adopted range of $\pm250$ \kms\ (see \S\ref{sec:metal_Cols}). Of the 17 absorbers included in our sample, only one has an integration range $|\Delta v|\sim550$ \kms\ ($z_{\rm{abs}}=1.762$ along J131119$-$012030; see \citealt{som2013}). We summarize metal-ion column densities for the literature sample in Table~\ref{tab:lit_met_cols}.

\subsection{Ground-based Optical Data: Keck/HIRES and VLT/UVES}\label{subsec:opt_spec}

\input{table1_hiresqsos}

Metallicity determination in ionized gas requires observations of multiple metal ions across different ionization states to constrain photoionization models. At Cosmic Noon ($1\leq z \le 2.2$) many of the metal ions tracing photoionized gas (e.g., \CII, \CIII, \CIV, \SiII, \SiIII, \SiIV, \FeII, and so on) are redshifted into the optical band. We therefore searched for high-resolution, high signal-to-noise optical spectra of quasars in our sample in two archival databases: the Keck Observatory Database of Ionized Absorption toward Quasars (KODIAQ) DR2 and the UVES Spectral Quasar Absorption Database (SQUAD) DR1. Both KODIAQ DR2 \citep{omeara2015,omeara2017} and SQUAD DR1 \citep{murphy2019} provide fully reduced, continuum-normalized high-resolution ($R\gtrsim45000$), high-quality ($\rm{S/N}\sim30$--100) quasar spectra. KODIAQ spectra were observed with HIRES on Keck and reduced using the HIRedux\footnote{\url{http://www.ucolick.org/~xavier/HIRedux}} software package \citep{omeara2015,omeara2017}. The database comprises 593 spectra for 300 quasars contributed by the KODIAQ science team. The SQUAD spectra were observed with UVES on the VLT and reduced with UVES POPLER \citep{murphy2019}, a modified version of the UVES Common Pipeline Library (CPL), specially designed to combine multiple echelle exposures into a single spectrum. The SQUAD DR1 comprises uniformly reduced, coadded spectra for 467 quasars ($z\sim1$--5) observed before November 2016. 

The two databases differ in their treatment of continuum normalization during the reduction process. In KODIAQ, individual echelle orders are continuum-normalized (fitted separately using Legendre polynomials) before coaddition \citep{omeara2015}. This makes KODIAQ less suitable for studies requiring accurate flux measurements over wide wavelength ranges (as in the measurements of the Lyman limit or very strong Ly$\alpha$) since the normalization-before-coaddition approach erases large-scale flux calibration. SQUAD fits continua to individual orders during data extraction, but then propagates them through the coaddition procedure, preserving the spectral energy distribution across the full wavelength range \citep{murphy2019}. The flux-calibrated SQUAD spectra are therefore more reliable for measurement of very strong Ly$\alpha$ absorbers. For metal lines, we can locally fit the continuum and measure precise equivalent widths and column densities.

We cross-match our sample of 113 \citetalias{ribaudo2011} and 71 \citetalias{omeara2013} quasars with KODIAQ and SQUAD. For \citetalias{ribaudo2011}, 18 quasars matched KODIAQ and another 21 matched SQUAD (3 quasars matched both). We also cross-matched our list with the Keck Observatory Archive (KOA) and ESO Archive Science Portal and found 23 quasars with unreduced HIRES spectra and 3 quasar sightlines with Phase 3 UVES spectra. Two of the latter correspond to a gravitationally lensed pair (HE1104--1805A and HE1104--1805B), observed together in multiple exposures. We obtained extracted, coadded spectra for both sightlines, kindly provided by Sebastián López and Hugo Cortés (priv. comm. 2025).\footnote{Both the sightlines show two absorbers at $z_{\rm{abs}}=1.66$ and 2.20. Along A we detect a DLA and an LLS; along B an LLS and a pLLS. Since BRIDGE focuses on pLLSs, LLSs, and SLLSs, we use sightline B in our sample; as there was no information about these sightlines prior to observation, it does not introduce a bias.} We coadded 22 UVES exposures (Phase 3) for J012017+213346. With this, we have high-resolution, high S/N optical spectra for 30 quasars with 57 absorbers from \citetalias{ribaudo2011}. 

For \citetalias{omeara2013}, 3 of 71 quasars matched KODIAQ DR2, and none matched SQUAD. Twenty-six additional fully-reduced HIRES spectra were acquired and reduced by JMO.\footnote{We will release this data and facilitate its addition to the existing KODIAQ database.} We list these data in Table \ref{tab:hires}. There are 55 absorbers along these 26 sightlines. 

Of the full sample of 112 (57 + 55) absorbers, 44 have only lower limits on $N_{\HI}$; 20 of these have at least Ly$\alpha$ coverage in the optical spectra. Where possible, we use available Lyman series lines to constrain $z_{\rm{abs}}$ and $N_{\HI}$, or, in the absence of damping wings, set an upper limit on $N_{\HI}$ (see Appendix \ref{app:vpfit}). This sample of 88 absorbers combined with 17 SLLSs from the literature gives a final sample of \numberabs\, absorbers. All sightlines and their absorbers are listed in Table~\ref{tab:full_sample}, along with \HI\, column densities (either adopted from the literature or reprocessed in this work; \S\ref{sec:HI}), and the UV/optical data sources for each sightline.

\subsection{Comparison sample}\label{subsec:comp_sample}

The BRIDGE survey aims to characterize the evolution of the global metallicity distribution in cool, ionized CGM-like gas with redshift, as well as the overdensity traced by $N_{\HI}$. The BRIDGE-I survey covers $16.4\le \log N_{\HI}<20.3$; we supplement this with dust-corrected DLA metallicities ($\log N_{\HI}>20.3$) at Cosmic Noon ($1\le z\le2.2$) from \cite{peroux2021} (hereafter, \citetalias{peroux2021}), which are largely drawn from \cite{decia2018}. For comparison at other redshifts, we use the KODIAQ-Z survey at $2.2<z\le3.6$ \citep{lehner2016,lehner2022}, the COS CGM Compendium (CCC) at $z<1$ \citep{lehner2018,wotta2019}, and the DLA sample of \citetalias{peroux2021} and references therein.\footnote{The KODIAQ-Z sample of 321 absorbers includes  157 from the HD-LLS survey \citep{prochaska2015,fumagalli2016} and 77 from the literature (see \citealt{fumagalli2016}). The metallicities for these absorbers were re-estimated in \cite{wotta2019} and \cite{lehner2022} with the HM05 \citep{hm05} UV background for uniform comparison across all redshifts.}

\section{\HI\, Column Densities}\label{sec:HI}

\begin{figure*}
\begin{centering}
\includegraphics[trim= 10 9 10 11,clip,scale=0.45]{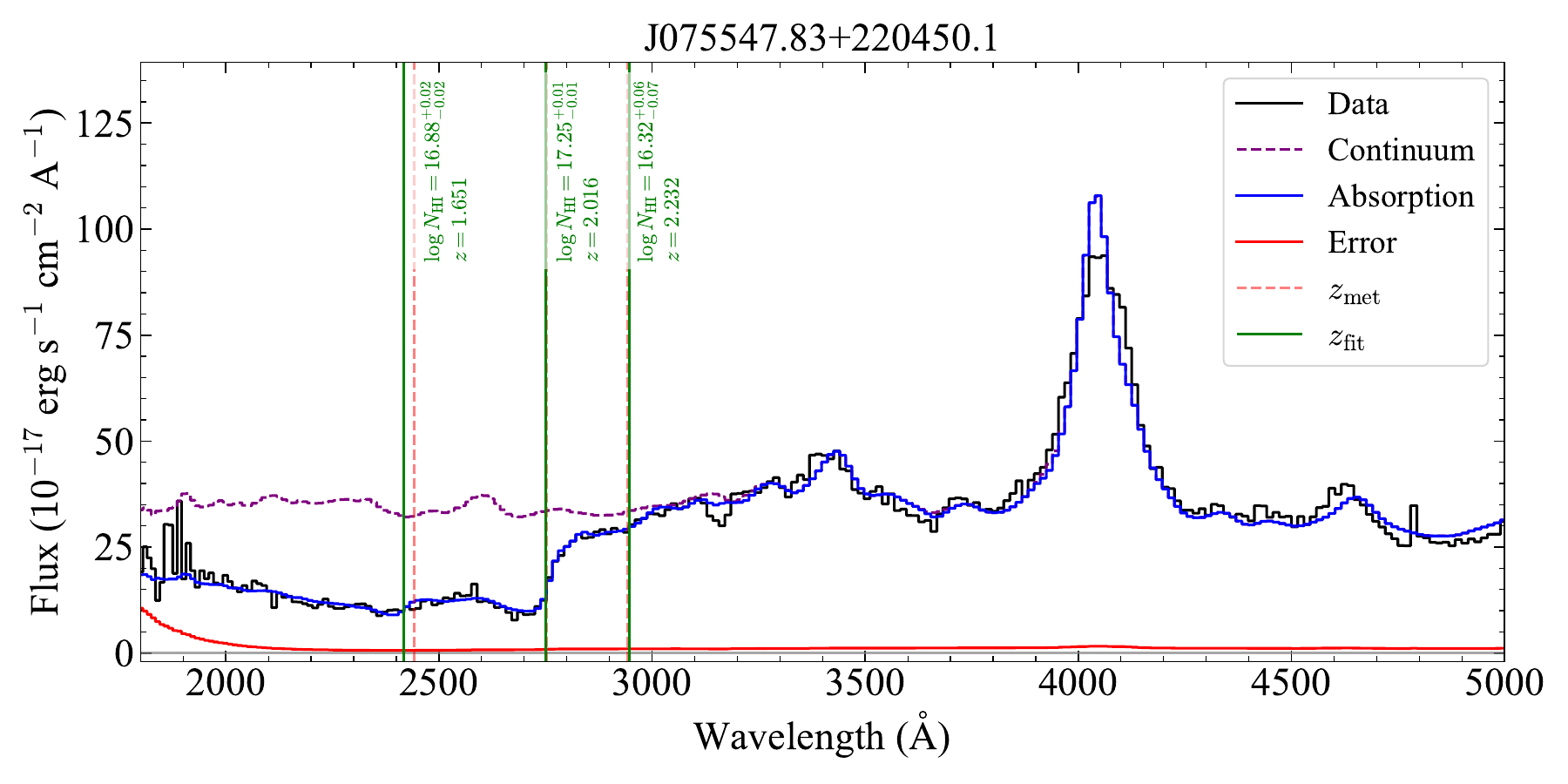}
\par\end{centering}
\begin{centering}
\caption{
{WFC3/UVIS-G280 spectrum (solid black) ($R=200$ at 2500 \AA\ and $\sim320$ at 5000 \AA) and uncertainties (solid red) for J075547+220450. The dashed purple line is the modeled quasar continuum, and the solid blue line is the modeled Lyman limit absorption due to 3 \HI\, absorbers along the sightline. The vertical green lines mark the best-fit redshifts from our fitting code, and the vertical dotted red lines represent the redshifts determined from metal lines in the optical spectra.
}}\label{fig:wfc3_fig}
\par\end{centering}
\end{figure*}

To estimate the metallicity of an absorber, we need the total (neutral+ionized) hydrogen column density. The absorbers in our sample ($16.4\le \log N_{\HI} <20.3$) probe cool (a few $\times 10^4$ K), largely ionized gas, so converting $N_{\HI}$ to $N_{\rm{H}}$ requires large ionization corrections. The total hydrogen column density is: $N_{\rm{H}}=N_{\HI}+N_{\HII}$, and $f_{\HI}=N_{\HI}/N_{\rm{H}}$ is the neutral fraction of the gas cloud, derived via photoionization modeling. Uncertainties in $N_{\HI}$ propagate to $N_{\rm{H}}$, and subsequently to the derived metallicity. 

Both \citetalias{ribaudo2011} and \citetalias{omeara2013} estimated $N_{\HI}$ by modeling the flux decrement at Lyman limit and its recovery at lower wavelengths. They primarily differ in the spectral resolution. Given the relatively higher resolution of FOS and STIS ($R\sim1000$--1300), \citetalias{ribaudo2011} performed detailed uncertainty analysis for their sample, and we adopt their $N_{\HI}$ and uncertainties. The higher-resolution FOS and STIS spectra also allowed for the accurate determination of $z_{\rm{abs}}$ for $\sim90\%$ of the absorbers, for which resolved Lyman series lines were available. For the remaining, $z_{\rm{abs}}$ was based on the Lyman limit only.

For \citetalias{omeara2013}, the S/N ($\gtrsim10$) of the WFC3/ACS data is sufficient to reliably constrain $N_{\HI}$, up to $\log N_{\HI}\lesssim18.2$. However, this S/N is high enough that propagating error from the observed flux alone yields unrealistically small uncertainties on the \HI\, column density. \citetalias{omeara2013} also note that the systematic uncertainty of continuum placement by hand (and to a lesser extent, line blending in the LAF), is the most significant uncertainty in the estimated $N_{\HI}$. Thus, determining reliable uncertainties requires accounting for statistical uncertainties and reducing systematic uncertainties during model-fitting. 

The lower-resolution WFC3 and ACS spectra ($R=200$ at 2500 \AA) also result in a large uncertainty in $z_{\rm{abs}}$. For example, at $2500$ \AA, WFC3 has wavelength dispersion $\delta\lambda\sim14$ \AA. For a 2 pixel uncertainty in the wavelength calibration of WFC3 G280, this translates to an uncertainty of $\pm0.03$ in redshift or $\sim3300$ \kms (for the Lyman limit at $z\sim1.8$). This redshift uncertainty makes it difficult to match metal lines in the optical spectra and kinematically resolve individual components of the \HI-selected absorber. Additionally, both WFC3 grism and ACS prism spectra have varying $\delta\lambda$ that increases with wavelength. This means longer wavelengths are condensed into fewer pixels, making the determination of $z_{\rm{abs}}$ unreliable with increasing redshift. This is worse for ACS data where $\delta\lambda$ is as high as $100$ \AA\ at $\lambda_{\rm obs}=3500$ \AA, which also limits pixels available for continuum fitting. In such cases, if the $z_{\rm{abs}}$ is high enough, Lyman series lines may shift to the optical band and can be used for redshift determination and accurately ascertaining the \HI-metal line pairs. 

For lower-redshift absorbers with uncontaminated metal lines in the optical data, we determine $z_{\rm{abs}}$ by searching near the redshift estimated from the Lyman-limit. When metal lines from multiple absorbers fall within a single HST resolution element, it becomes difficult to uniquely identify the associated \HI-metal line pairs. If the velocity structures for several metal ions are similar, and their mean velocities agree within the HST redshift uncertainty, we treat the complex of metal lines as one absorber associated with the dominant \HI\, component causing the flux decrement at the Lyman limit. This is the case for $\sim10\%$ (10/105) of our sample.

We refit all the WFC3 and ACS sightlines with available high-resolution, optical spectra. \citetalias{omeara2013}'s model for \HI\, absorption includes the flux decrement at the Lyman limit and its recovery at lower wavelengths, but not the flux attenuation from the Lyman series lines for each absorber. This attenuation is typically small, but because the uncertainties in the continuum model parameters and absorber optical depths are correlated, the slight flux suppression at higher wavelengths from lower-order Lyman series lines (Ly$\alpha$, Ly$\beta$, etc.), along with the gradual fall-off near the Lyman limit from higher-order lines, can help break this degeneracy. We therefore perform MCMC optimization to simultaneously fit the continuum parameters and the Lyman limit optical depth for multiple absorbers, while accounting for the Lyman series attenuation from each. In addition to the statistical uncertainties in the observed flux, the MCMC analysis also accounts for uncertainties in the assumed quasar template from the variation in the shape of the composite quasar spectrum, as well as from the galactic extinction and IGM transmission corrections applied to the stacked spectra and reported by \cite{telfer2002}. 

In almost all cases, we determine precise absorber redshifts from metal lines (and/or Lyman series lines) in the high-resolution optical spectra. We adopt these as flat priors, allowing $z_{\rm{abs}}$ to vary within the redshift equivalent of two pixels on either side, set by the WFC3 wavelength calibration uncertainty.\footnote{There are only 3 ACS sightlines with optical data in our sample, all treated identically.} Comparing redshifts determined from the Lyman limit alone with those from resolved Lyman series or metal lines, we find $|\Delta z|\leq0.03$, within the 2 pixel uncertainty. 

Figure~\ref{fig:wfc3_fig} shows an example of the fitting of three intervening \HI\, absorbers along the sightline J075547+220450, where the fitted redshifts (green) agree closely with those measured from the metal lines (dotted red). We also show the estimated quasar continuum (dashed purple), the modeled Lyman limit absorption (solid blue), and the estimated $N_{\HI}$ with uncertainties for each absorber. Refitted models for all quasar sightlines with optical spectra in the BRIDGE sample are shown in Appendix~\ref{appendix:figs} (Figure~\ref{fig:wfc3_fits_mosaic} and \ref{fig:acs_fits}). 

For sightlines with $\log N_{\HI}\gtrsim17.8$ absorbers, the flux is completely absorbed below the Lyman limit yielding only a lower limit on $N_{\HI}$. When lower-order Lyman series lines are accessible in the high-resolution optical spectra, we use damping wings (when $\log N_{\HI}>19.0$) to estimate $N_{\HI}$. When damping wings are absent, we estimate an upper bound on $N_{\HI}$. We describe our modeling approach and treatment of lower limits on $N_{\HI}$ in Appendix~\ref{app:hicols} and \ref{app:vpfit}, respectively. 

\begin{figure}
\begin{centering}
\includegraphics[trim= 13 9 12 0,clip,scale=0.34]{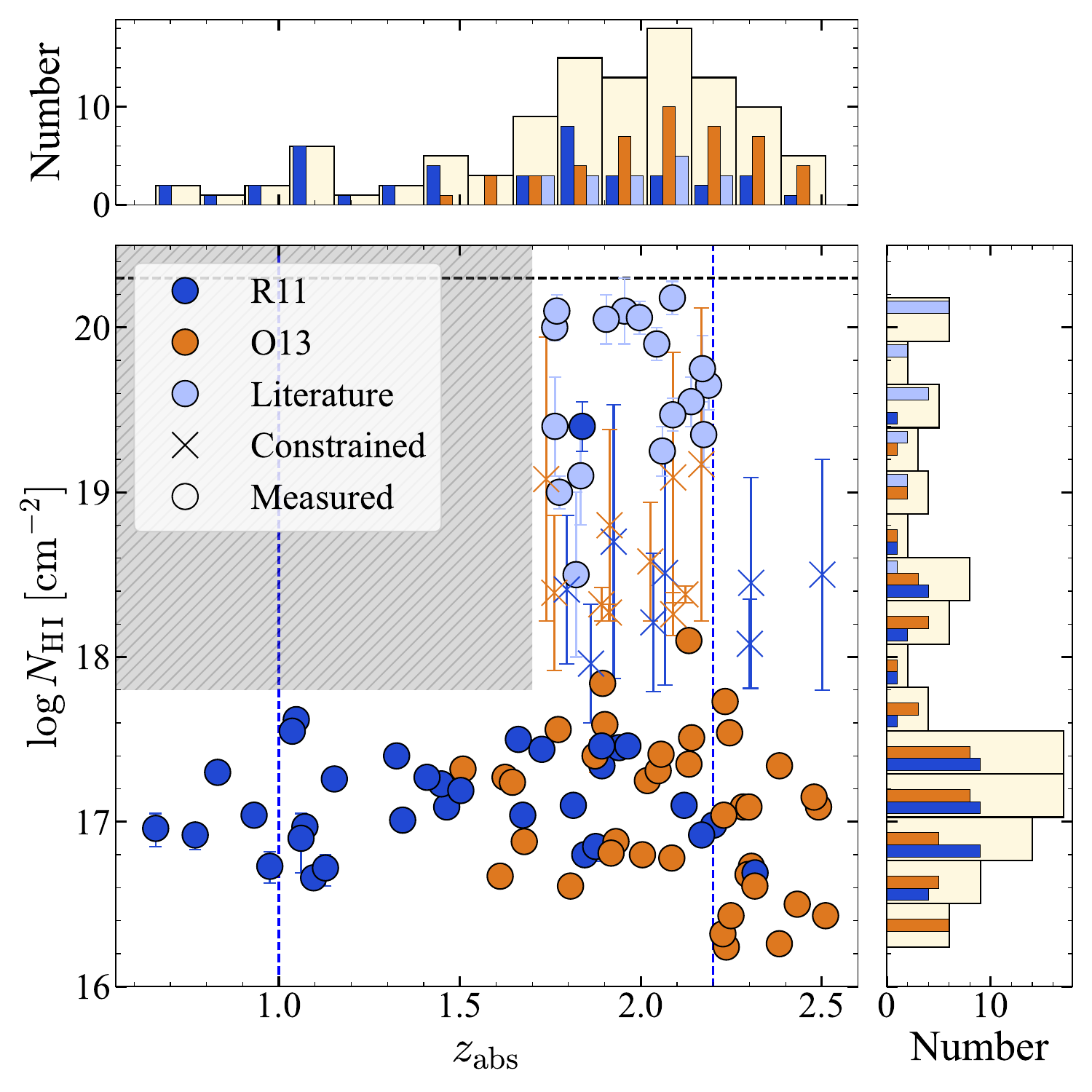}
\par\end{centering}
\begin{centering}
\caption{
{The full sample of absorbers with $N_{\HI}$ for which archival high-resolution optical spectra are available. The sample is largely drawn from \citetalias{ribaudo2011} (dark blue) and \citetalias{omeara2013} (orange), with an additional 17 absorbers from the literature (light blue; see \S\ref{subsec:lit_sample}). Circles represent \HI\ absorbers with $N_{\HI}$ estimated from the Lyman limit (error bars show the 68\% CI); crosses mark central values for strong absorbers whose lower and upper bounds on $N_{\HI}$ (error bars) are constrained using the Lyman limit and Lyman series lines. The dashed blue lines mark the boundaries of the Cosmic Noon sample (80/105 absorbers). The dashed black line marks the DLA threshold, $\log N_{\HI}=20.3$, and the hatched gray region shows the regime that remains observationally inaccessible.}\label{fig:absorbers}
}
\par\end{centering}
\end{figure}

Figure~\ref{fig:absorbers} summarizes our complete sample of \numberabs\, \HI-selected absorbers showing $\log N_{\HI}$ vs $z_{\rm abs}$. Of these 80 absorbers lie in the Cosmic Noon redshift interval $1\le z\le2.2$, with a mean redshift of $\langle z\rangle =1.81\pm0.31$. The saturated absorbers at the Lyman limit for which we derive lower and upper bounds on $N_{\HI}$ are labeled as ``Constrained." For absorbers with $\log N_{\HI}\gtrsim18$ and $z_{\rm{abs}}\lesssim1.7$, Ly$\alpha$ is not accessible in optical spectra, and we only have lower limits on $N_{\HI}$. This explains the lack of absorbers in this regime (hatched gray region). We also note that our sample has fewer low $N_{\HI}$ absorbers ($\log N_{\HI}<16.6$) at $z_{\rm{abs}}<2$. This is due to sensitivity issues. If the Lyman limit of a lower-$z$ pLLS falls within the flux recovery regime of a higher-$z$ LLS, the S/N of the data is too poor to detect it. A pLLS at higher-$z$ is detected as long as its Lyman limit is not blended with that of a LLS. If blended, the pLLS adds to the uncertainty in the $z_{\rm{abs}}$ of the LLS. Together, these effects lead to a lower detection rate of pLLSs at lower redshifts in both surveys. \citetalias{ribaudo2011} stress that the pLLS sample in their survey is not complete due to the poor S/N and resolution of the data, especially at lower wavelengths.

\begin{figure*}
\begin{centering}
\includegraphics[width=\textwidth,trim= 10 9 10 0,clip]{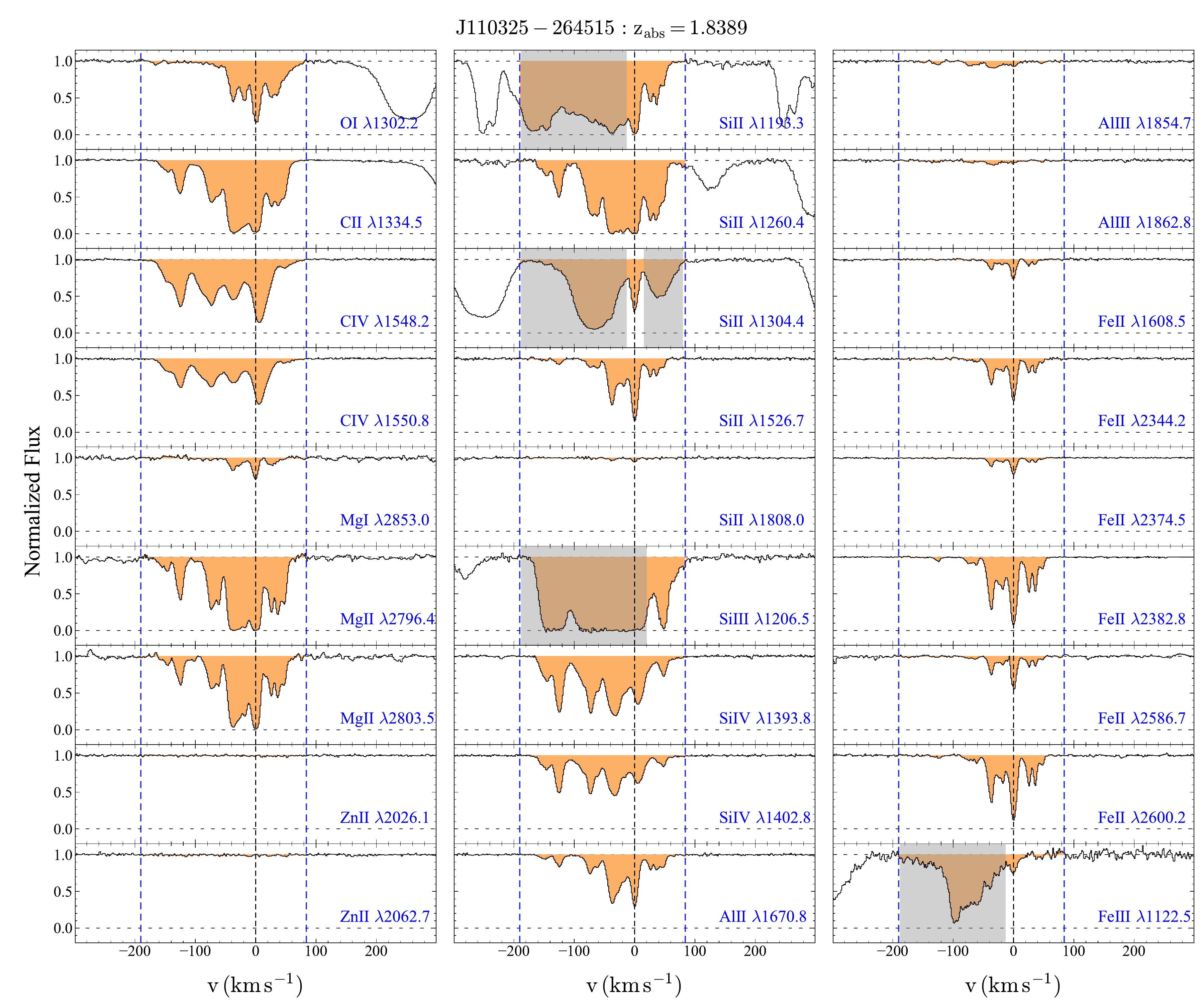}
\par\end{centering}
\begin{centering}
\caption{
{Example of normalized UVES spectra ($R\sim60000$, S/N$\,=97$) against the rest frame velocity of an absorber at $z=1.8389$ toward J110325-264515. Access to multiple transitions of the same ion (e.g., \SiII) allows us to compare the corresponding $N_a(v)$ profiles to check for saturation and contamination. The dashed blue lines represent the integration range in velocity $\left[v_{\rm{min}},v_{\rm{max}}\right]$, and the shaded grey regions represent contamination.}
}\label{fig:absorber}
\par\end{centering}
\end{figure*}

\section{Metal Column Densities}\label{sec:metal_Cols}

We now discuss our treatment of metal ions associated with our \HI-selected sample. For each absorber, we search for metal absorption in the high-resolution optical spectra. For all absorbers, we find associated metal lines within 0.03 ($\sim3000$ \kms at $z\sim1.8$) of the redshift determined using the Lyman limit. We estimate metal ion column densities using the apparent optical depth (AOD) method \citep{savage1991}. We first model the quasar continuum ($F_c$) by fitting the unabsorbed regions of the observed spectra ($F$) to Legendre polynomials, done locally for each transition using a modified version of the python package {\tt rbcodes} \citep{rbcodes}.\footnote{We use our own code for the AOD analysis, implemented as a third panel in the GUI showing the $N_a(v)$ profile for interactively marking contamination.} We then convert the absorption profiles to the apparent optical depth per unit velocity, $\tau_a(v)=\ln F_c(v)/F(v)$, which is related to the apparent column density per unit velocity $N_a(v)$ via $N_a(v)=3.768\times10^{14}\,\tau_a(v)/(f\lambda \, (\rm{\AA}))\,\rm{cm}^{-2}(km\,s^{-1})^{-1}$, where $f$ is the oscillator strength of the transition and $\lambda$ is the wavelength in \AA. The total column density is the integrated $N_a(v)$ profile $N_a=\int^{v_{\rm{max}}}_{v_{\rm{min}}}N_a(v)dv$, where [$v_{\rm{min}}$, $v_{\rm{max}}$] specify the velocity interval or the boundaries of the absorption. 

An absorber is almost always a complex of several components. These components may be resolved in the high-resolution optical spectra for metal ions but cannot be separated for \HI\, in the low-resolution ($R\lesssim300$) WFC3 and ACS data. Even when Ly$\alpha$ is available (e.g., Figure~\ref{fig:voigt}), the combination of high $b$-values, high $f\lambda$, and high cosmic hydrogen abundance produces a severely blended absorption profile. 

Given this mismatch, we must be clear on our approach to defining an absorber. Where possible, we define an absorber via the metal line integration range [$v_{\rm{min}}$, $v_{\rm{max}}$], including all components identified from the higher-resolution metal-ion absorption profiles. \citet{prochaska2015} adopt an interval of $\pm 500$ \kms, while \citet{lehner2022} (hereafter \citetalias{lehner2022}) adopt an interval based on the width of Lyman series lines. Both find that metal-line absorption rarely exceeds $\pm200$ \kms relative to the central redshift. The BRIDGE absorbers show the same behavior, and we adopt $\pm250$ \kms as a conservative maximum interval. All absorption features within this range are treated as a single absorber unless they are obvious contaminants and are cleanly separable. 

At Cosmic Noon, we have access to several metal lines in the optical spectra. Since we probe cool photoionized gas, we follow the component structure of \OI\, (if not saturated or contaminated) or low ions (\CII, \SiII, \AlII, \MgII, \FeII) to guide our analysis (see Figure~\ref{fig:absorber}). If no low ions are detected, we adopt the redshift determined by the Lyman limit and report upper limits for metal ions. For 15 absorbers only high ions (\CIV\ and/or \SiIV) are detected. If we have access to Ly$\alpha$ in the optical and its component structure is similar to high ions, we treat the high ions as detections and use them to determine the redshift. If we do not have access to Ly$\alpha$ or if its component structure suggests that the metal lines may be associated with a weaker blended \HI\ absorber, we determine redshift using the high ions, but treat them as upper limits. 

We compare all available metal ion transitions in velocity to determine the appropriate integration range. We integrate the apparent column density profile, $N_a(v)$, within the integration range to estimate the integrated column density, $N_a$, and the equivalent widths, $\rm W_{\lambda}$, of the transitions. Uncertainties on $N_a$ are determined by propagating the uncertainties in the observed flux and the fitted continuum. We use $\rm W_{\lambda}$ to determine whether absorption is detected at the $>2\sigma$ level. If not, we quote a $2\sigma$ upper limit on the column density, which is defined as twice the $1\sigma$ error derived for the column density assuming the absorption line lies on the linear part of the curve of growth (e.g., \ZnII\, in Figure~\ref{fig:absorber}). The $1\sigma$ error is determined by integrating over a velocity interval similar to that of a detected low ion. Our code flags any transition where the flux reaches zero within the integration range as saturated. These cases require special attention and comparison of the $N_a(v)$ profiles for all available transitions for each ion. 

When absorption is detected at the $>2\sigma$ level, we compare the $N_a(v)$ profiles for ions with two or more transitions. At $z=1$--2.2, we have access to a combination of the following ions depending on the redshift: \OI, \OVI, \underline{\CII}, \underline{\CIII}, \underline{\CIV}, \underline{\SiII}, \underline{\SiIII}, \underline{\SiIV}, \underline{\AlII}, \underline{\AlIII}, \NI, \NII, \NV, \underline{\FeII}, \FeIII, \underline{\MgI}, \underline{\MgII}, \SII, \SIII, \ZnII\, (most commonly used ions are underlined).\footnote{For most absorbers \OVI\, and \NV\, show broad velocity profiles that are distinct from the low and intermediate ions, and we therefore exclude them from our photoionization models. Only one absorber in our sample showed narrow \NV\, absorption, with a velocity profile similar to \CIII\, and \CIV. In this case, we included \NV\, in the photoionization model and found the resulting metallicity and density consistent with the model in which \NV\, was excluded.} For high-$z$ absorbers, we may also have access to \HI\, transitions, though these are highly saturated and blended for $\log N_{\HI}\gtrsim16.2$. For ions with multiple transitions where the $N_a(v)$ profiles align well, we calculate the true column density, $N$, as the weighted mean of the integrated apparent column densities, where the weights are the inverse of the uncertainty in $N_a$. If the $N_a(v)$ profiles for transitions of the same ion do not agree closely, we visually inspect them for contamination and exclude contaminated transitions from the weighted mean. If multiple lines for an ion exist, we use only unsaturated, uncontaminated lines for the mean. We can distinguish between contamination and saturation readily by comparing the $N_a(v)$ profiles: if the profiles do not match at peak optical depth and the apparent column density of the stronger transitions is systematically smaller than that of the weaker transitions, the transitions are saturated. If the profiles do not match elsewhere in the profile, or an absorption feature is present only in the $N_a(v)$ profile of the weaker transition, it is a contamination. In Figure~\ref{fig:absorber} we show five \SiII\ transitions: $\lambda1193$ and $\lambda1304$ are contaminated, and $\lambda1260$ is saturated. We therefore take the weighted mean of $\lambda1526$ and $\lambda1808$ alone. 

For mild saturation without contamination in doublets (\OVI\, $\lambda\lambda$1031, 1037; \CIV\, $\lambda\lambda$1548, 1550; \SiIV\, $\lambda\lambda$1393, 1402; \NV\, $\lambda\lambda$1238, 1242; \AlIII\, $\lambda\lambda$1854, 1862; \MgII\, $\lambda\lambda$2796, 2800), where the $N_a(v)$ profiles agree well except at peak optical depth, we apply the correction derived by \cite{savage1991} if two conditions are satisfied: (i) the transitions have $\Delta\log(f\lambda)\simeq0.3$ and, (ii) the difference in apparent column densities of the weak and strong transitions is $\Delta(\log N) = \log N^{\rm{weak}}-\log N^{\rm{strong}} \leq0.13$ dex. If $\Delta(\log N) > 0.13$ dex, we adopt $N^{\rm{weak}}$ as the lower limit (see \citealt{wotta2016}).  

For single transitions like \CIII\, $\lambda$977, \SiIII\, $\lambda$1206, \AlII\, $\lambda$1670, \FeIII\, $\lambda$1122, \NII\, $\lambda1084$, we assess contamination by visual inspection.\footnote{At Cosmic Noon \NII\, and \FeIII\, frequently lie below the atmospheric cutoff (for $z_{\rm abs}\lesssim1.8$). When accessible in the optical band, they lie in the Ly$\alpha$ forest, and are often contaminated.} Strong contamination is easily identified by offsets in mean velocity or peak optical depth relative to similar ions, or by very different velocity structures of the absorption profiles (e.g., \SiIII\, and \FeIII\, are severely contaminated in Figure~\ref{fig:absorber}).  We compare these ions to the velocity structure of low ions like \CII, \SiII, and \OI, rather than high ions like \OVI\, or \NV, which often trace a hotter gas phase. If we cannot confirm that a absorption feature arises solely from an understudied ion (like \CIII\, with no detection of frequently detected ions like \CII), but the feature is plausible given the \HI\, mean velocity, and considering that metal lines are narrower, we treat the measurement as an upper limit.\footnote{We use flags to keep track of our analysis. For each absorber, we assign detection and reliability flags to every ion. Detection flags 0, $-1$, and $-2$ represent measurement, upper limit, and lower limit, respectively. Reliability flags indicate, 1: results based on several transitions or a non-detection, which is always reliable since it would be estimated in an uncontaminated region of the spectrum; 2: results based on at least two transitions, but where only one transition is detected at the 2$\sigma$ level and the upper limits agree with that detection; 3: results based only on a single transition or several saturated transitions.}  

\section{Determining the Metallicities of the Absorbers} \label{sec:metallicity}

\subsection{Case for Photoionization Modeling}\label{subsec:why_photoion}
Empirical studies on the metallicity distribution of pLLSs, LLSs, and SLLSs at both low ($z<1$; \citealt{lehner2013,lehner2016,wotta2016,wotta2019}, hereafter \citetalias{wotta2019}) and high ($z\sim2.2$--3.6;  \citealt{prochaska2015,fumagalli2016}; \citetalias{lehner2022}) redshifts show that the gas probed by these absorbers is cool ($T\sim$ a few $\times 10^4$ K), and predominantly photoionized. This is demonstrated by the low $b$-values for the \HI\, absorption in these samples: for CCC ($z<1$) and KODIAQ-Z ($z\sim2.2$--3.6), $\langle b\rangle=28\pm8$ \kms and $30\pm11$ \kms, respectively. For both samples, nearly 90\% of all absorbers have $b<40$ \kms, which corresponds to $T<4\times10^4$ K, consistent with photoionization (\citealt{prochaska1999,prochaska2015,fumagalli2016,lehner2016,lehner2022,crighton2019}; \citetalias{wotta2019}). In the BRIDGE sample, we do not fit \HI\, absorption profiles directly and rely primarily on the Lyman limit for $N_{\HI}$. The weak constraints on $b$ obtained from fitting the Lyman limit as well as from our limited Voigt profile fits for saturated absorbers are consistent with $b\sim25$--30 \kms.\footnote{We note that for these $N_{\HI}$ values, Ly$\alpha$ is always saturated and almost always composed of more than one component, as seen in metal ions, which means we are more prone to overestimating $b$ and thus gas temperature.} 

As in CCC and KODIAQ-Z, we find that detected low ions (e.g., \OI, \CII, \SiII, \AlII) align well with one another and with \HI\, (whenever accessible in the optical data), consistent with the gas being in a single photoionized phase. At Cosmic Noon, we almost always detect intermediate (\CIII, \SiIII, \AlIII) and high (\CIV\, and \SiIV) ions. If low ions are also detected and have a component structure similar to that of intermediate and high ions, we use them all to constrain photoionization models. If there are no low ions are detected but intermediate and high ions are, we use them to constrain the models. However, if only high ions are detected, we treat them as upper limits in our photoionization models (this is true for 7/105 absorbers or about 7\% of the sample). This is because in the absence of intermediate or low ions, it is challenging to determine if \CIV\, and \SiIV\, are associated with the total estimated $N_{\HI}$ or with a weaker subset of the dominant \HI\, absorber. In the latter case, adopting a higher \HI\, column density while including the high ions would bias models toward lower metallicities and densities (see \S\ref{app:ionsincluded}). Lower densities, in turn, can yield unrealistically large path lengths for the absorbers (see \S\ref{subsec:systematics}). 

For our sample of absorbers with $\log N_{\HI}>16.4$, broad and strong \OVI\, and \NV\, absorption cannot be reproduced by photoionization \citep{fox2013,shull2014,lehner2014}. That is, when we include these ions, our models fail to converge on a single-phase solution. These ions more likely trace a warmer, collisionally ionized phase, or very diffuse photoionized gas (e.g., \citealt{tripp2008}). We therefore exclude \OVI\, and \NV\, from our models (except for one absorber, see 
\S\ref{sec:metal_Cols}).
\subsection{Overview and General Assumptions}\label{subsec:cloudy_basics}

To determine metallicities, we model photoionization using Cloudy (version C13.02; \citealt{ferland2013}), adopting the same methodology and grid of models as \citetalias{wotta2019} and \citetalias{lehner2022}. We assume each absorber can be approximated as a uniform slab in thermal and photoionization equilibrium, illuminated by a Haardt-Madau extragalactic UV background radiation field (HM05; \citealt{hm05} and HM12; \citealt{hm12}). We adopt HM05 to ensure consistent metallicity comparisons across redshifts, and examine the impact of the assumed radiation field on metallicity in \S\ref{subsec:systematics}. There are two main variables in the photoionization models: (i) the ionization parameter, $U\equiv n_{\gamma}/n_{\rm{H}}$, where $n_{\gamma}$ is the density of ionizing photons and $n_{\rm{H}}$ is the total hydrogen number density; and, (ii) the metallicity, $[\rm{X/H}]$. We assume solar relative abundances from \cite{asplund2009}, allowing the ratio of carbon to the $\alpha$ elements ($\rm{O}$ or $\rm{Si}$) to vary. This additional parameter allows non-solar $\left[\rm{C}/\alpha\right]$, accounting for variations in abundance ratios due to nucleosynthesis effects \citep{pettini2008,cescutti2009,cooke2011}. We use the grid of Cloudy models created by \citetalias{wotta2019} and spanning a range of values in $N_{\HI}$, $z_{\rm{abs}}$, $[\rm{X/H}]$, $n_{\rm{H}}$ and $\left[\rm{C}/\alpha\right]$.

\subsection{Bayesian MCMC Technique}\label{subsec:MCMC}

One of the goals of the BRIDGE survey is to estimate robust uncertainties on the derived ionization parameters and metallicities of absorbers. We estimate $\left[\rm{X/H}\right]$ and $\log U$ by comparing the observed $N_{\HI}$ and metal ion column densities against the grid of Cloudy models. Since this is a multi-dimensional optimization problem with non-linear correlations between the various metal ions and model parameters, and since all observed quantities carry uncertainties that propagate into the estimated metallicity, we optimize in a Bayesian context. We use the publicly available code described in \cite{fumagalli2016} and \citetalias{wotta2019} to run an MCMC analysis on each absorber, and obtain posterior distributions for metallicity and $n_{\rm{H}}$ (or $U$).\footnote{This code was packaged in Python as PyIGM and is available at \url{https://github.com/pyigm/pyigm} \citep{pyigm}.} 

\begin{figure*}
\begin{centering}
\includegraphics[width=\textwidth]{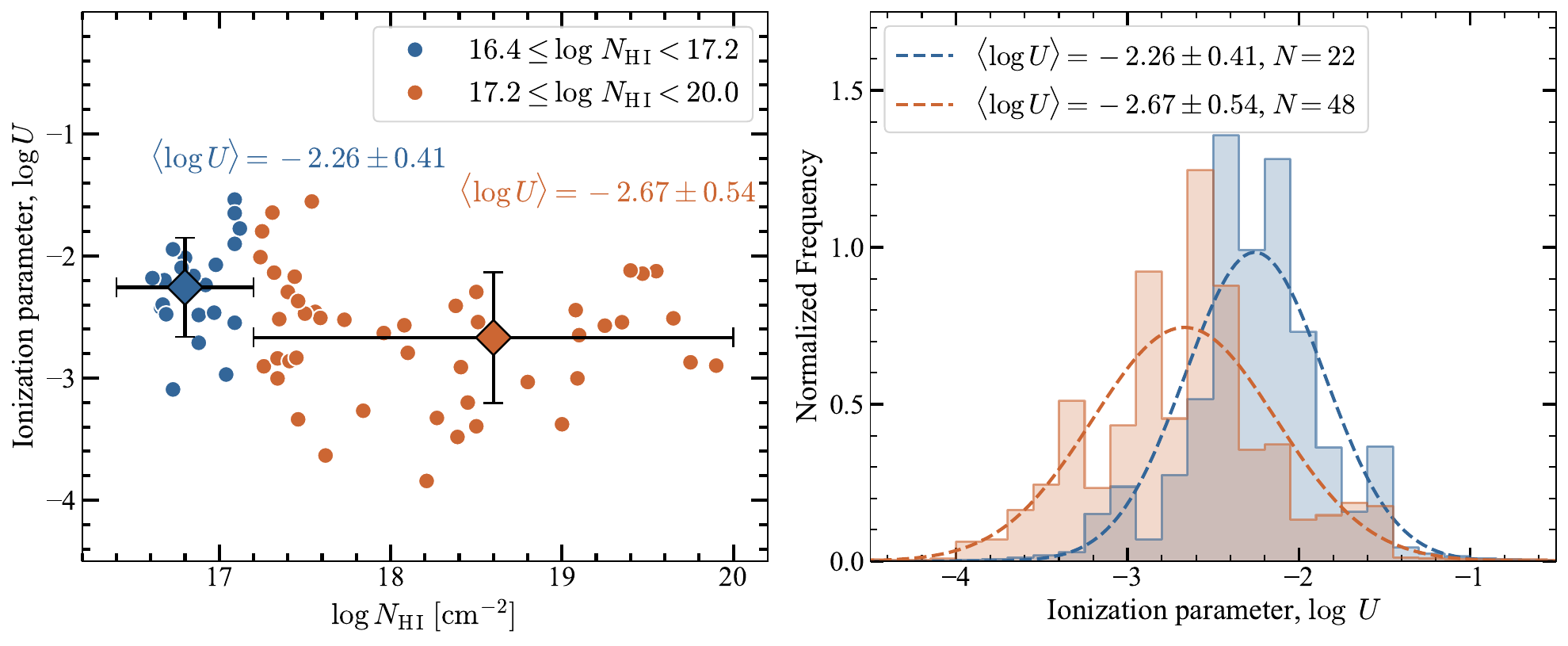}
\par\end{centering}
\begin{centering}
\caption{
{\textit{Left panel:} Best-fit ionization parameter $\log U$ derived from MCMC analysis for 70 well-constrained absorbers in the BRIDGE survey, for which the optimization converged with a flat prior on $\log U$. Blue and orange points represent the two $\log N_{\HI}$ bins (pLLSs and LLSs+SLLSs.). The large diamonds mark the mean of each distribution, with error bars showing the standard deviation derived from a Gaussian. \textit{Right panel:} Gaussian fits to the $\log U$ distributions for pLLSs and LLSs+SLLSs.}
}\label{fig:logU_gauss}
\par\end{centering}
\end{figure*}

\subsection{Priors on Parameters}\label{subsec:priors}
We apply Gaussian priors to the measured parameters: $N_{\HI}$ and $z_{\rm{abs}}$. For all absorbers, we first assumed a flat prior on $\log U$. After the first run, we assessed the model outputs (chain plots, corner plots) to check for convergence. In the BRIDGE sample, $\sim67\%$ of the absorbers converged with a flat prior on $\log U$. We plot our results for $\log U$ as a function of $\log N_{\HI}$ in the left panel of Figure~\ref{fig:logU_gauss}. Similar to the low-$z$ CCC and the high-$z$ KODIAQ-Z sample, we observe an anti-correlation between $\log U$ and $N_{\HI}$ in the BRIDGE sample. The $\log U$ distribution for BRIDGE is more similar to that of KODIAQ-Z (see \S\ref{subsec:UVB}). This is expected, since the ionization conditions of the Universe (flux of ionizing radiation field or the EUVB) peak around $z\sim3$ and do not show considerable evolution between $z\sim1$--3 (\citealt{faucher2008,faucher2020,becker2013}; and see \S\ref{sec:discussion}).

For the remaining 33\% absorbers, the MCMC runs did not converge. This happens when there are fewer metal ion constraints to reliably determine the metallicity and $n_{\rm{H}}$. For these absorbers, we assume a Gaussian prior on $\log U$ to help the models converge. To estimate the appropriate prior, we split our sample in two bins: $16.4\le\log N_{\HI}<17.2$ (pLLSs) and $17.2\le\log N_{\HI}<20$ (LLSs+SLLSs), depicted as blue and orange points in Figure~\ref{fig:logU_gauss}.\footnote{This split is empirically motivated: $\log U$ is roughly constant within each $\log N_{\HI}$ bin but decreases from the lower to the higher bin, so we adopt different priors for each.} We fit the distribution for each bin with a Gaussian to estimate a mean  $\langle \log U\rangle$ and standard deviation. We obtain $\langle \log U\rangle=-2.26\pm0.41$ and $-2.67\pm0.54$ for 22 pLLSs and 48 LLSs+SLLSs, respectively. Our fits are shown in the right panel of Figure~\ref{fig:logU_gauss}. We adopt these means and standard deviations as the Gaussian priors on $\log U$ in each $\log N_{\HI}$ interval.

For all absorbers, we also run MCMC analyses assuming a flat prior on $\left[\rm{C}/\alpha\right]$ between $-1$ and 1. We then divide the sample into absorbers with good constraints on $\left[\rm{C}/\alpha\right]$ (based on corner plots) and those that could benefit from a Gaussian prior on $\left[\rm{C}/\alpha\right]$. For the latter, we rerun the MCMC analyses using a Gaussian prior defined by the sample mean and standard deviation of the ``good" absorbers. We discuss this in detail in \S\ref{subsec:nonsolar}. 

We summarize the results of photoionization modeling in Table \ref{tab:bridge_metallicity} for all absorbers in the BRIDGE sample. For each absorber, we list the sightline, $z_{\rm{abs}}$, $\log N_{\HI}$, metallicity, $\log U$, and $\left[\rm{C}/\alpha\right]$. For each parameter, we use the posterior probability distribution functions (PDFs) to derive the median and 68\% CI, except in the cases where we derived a lower or upper limit, where we list the median and 80\% CI. 

\subsection{The Near-DLAs with $\log N_{\HI}>20.0$}\label{subsec:sampleg20}

The Cloudy grids used in our analysis were constructed by \citetalias{wotta2019}, with $\log N_{\HI}$ ranging between 12.0--20.0. Of the absorbers we draw from the literature, 6 absorbers have $\log N_{\HI}>20.0$. For these SLLSs, we created smaller grids of models at the absorber redshift, with $n_{\rm H}$ and $[\rm{X/H}]$ spanning $-4.5$--$0.0$ and $-5.0$--$2.5$, respectively. Following  \cite{lehner2013}, we obtained the best-fit model by maximizing the likelihood function, which is estimated by comparing the model predictions for column densities to the observed values. The uncertainties in metallicity and $\log U$ are specified using the 68\% confidence interval. Since these absorbers did not undergo the MCMC routine, we do not include them in the statistical sample of 99 absorbers. We list these at the end of Table~\ref{tab:bridge_metallicity}. 

For absorbers drawn from the HD-LLS survey in the literature, the metallicities were estimated by \cite{fumagalli2016} using the same Bayesian framework. However, \cite{fumagalli2016} adopted the HM12 EUVB. We re-estimate metallicities using the HM05 EUVB for consistency across all samples. In Appendix~\ref{app:lit_sample}, we compare the literature-reported metallicities with our own re-estimates for all absorbers drawn from the literature. Table~\ref{tab:lit_met_cols} summarizes the metal ion column densities drawn from the literature and used in our Cloudy models. 

\subsection{Statistical and Systematic Uncertainties}\label{subsec:systematics}

We derive statistical uncertainties on all output parameters ($[\rm{X/H}]$, $\log U$, and $\left[\rm{C}/\alpha\right]$) from Bayesian MCMC ionization modeling. Additionally, there are systematic uncertainties associated with our modeling process due to the assumed physics in the Cloudy models. In this section, we discuss the systematic impact of the assumed ionizing background radiation, the selection of ions used for photoionization modeling, and the relative solar abundances; we present these tests in full in Appendix~\ref{app:systematics}.

The shape of the assumed EUVB is one of the largest systematic uncertainties in ionization modeling. Adopting the harder HM12 EUVB instead of HM05 systematically raises the estimated metallicities by $0.37\pm0.19$ dex at $z<1$ \citep{wotta2019} and $0.10\pm0.17$ dex at $2.2<z\le3.6$ \citepalias{lehner2022}. For a subsample of 70 BRIDGE absorbers with well-constrained densities, we find $\langle\rm{[X/H]_{\rm{HM12}}}-\rm{[X/H]_{\rm{HM05}}}\rangle=+0.16\pm0.13$ dex, intermediate between the $z<1$ and $2.2<z\le3.6$ offsets, as expected. We discuss our analysis in detail in Appendix~\ref{app:euvb} and revisit the evolution of the UV
background in \S\ref{subsec:UVB}.

The set of ions available to constrain each model also varies with redshift and wavelength coverage. Repeating the photoionization modeling with and without the high ions \CIV\ and \SiIV\ for the 70 absorbers with such coverage, we find that metallicities are largely unaffected (mean absolute difference $0.13\pm0.04$ dex, with only 7\% of absorbers differing by $>0.4$ dex and no systematic trend), whereas the ionization parameter is systematically higher when high ions are included (mean difference $0.33\pm0.04$ dex, with 25\% exceeding $0.4$ dex). Including high ions therefore biases the inferred densities (and therefore the estimated physical scale and overdensity) but not the metallicities, consistent with results at lower \citep{lehner2019} and higher \citepalias{lehner2022} redshifts We discuss the effect of this systematic on different quantities in detail in Appendix~\ref{app:ionsincluded}.

Finally, dust depletion and nucleosynthesis effects can alter abundance ratios. Since we typically use a combination of C, Si, and O ions to constrain metallicity via photoionization modeling, deviations from solar $\left[\rm{C}/\alpha\right]$ ratios may bias our results. To quantify the impact of non-solar abundances, we allow the $\left[\rm{C}/\alpha\right]$ ratio to vary in the range $-1\leq\left[\rm{C}/\alpha\right]\le1$ using a flat prior. We obtain $\langle\left[\rm{C}/\alpha\right]\rangle=0.08\pm0.35$ for the 23 absorbers in the BRIDGE redshift range with sufficient data to constrain $\left[\rm{C}/\alpha\right]$, consistent with solar. The assumed $\left[\rm{C}/\alpha\right]$ therefore does not introduce a significant systematic in the estimated metallicities. We discuss the trends between $\left[\rm{C}/\alpha\right]$ and $N_{\HI}$, metallicity, and neutral fraction in \S\ref{subsec:nonsolar}.

\section{Results}\label{sec:results}

\subsection{Metallicity of CGM-like Absorbers at Cosmic Noon}\label{subsec:met_results}

From this point forward, we use CCC, BRIDGE and KODIAQ-Z ($z<1$, $1\le z\le 2.2$, and $2.2<z\le3.6$, respectively) to denote redshift-based classifications rather than the survey-of-origin, with absorbers reassigned accordingly. 

\begin{figure}
\begin{centering}
\includegraphics[width=\columnwidth]{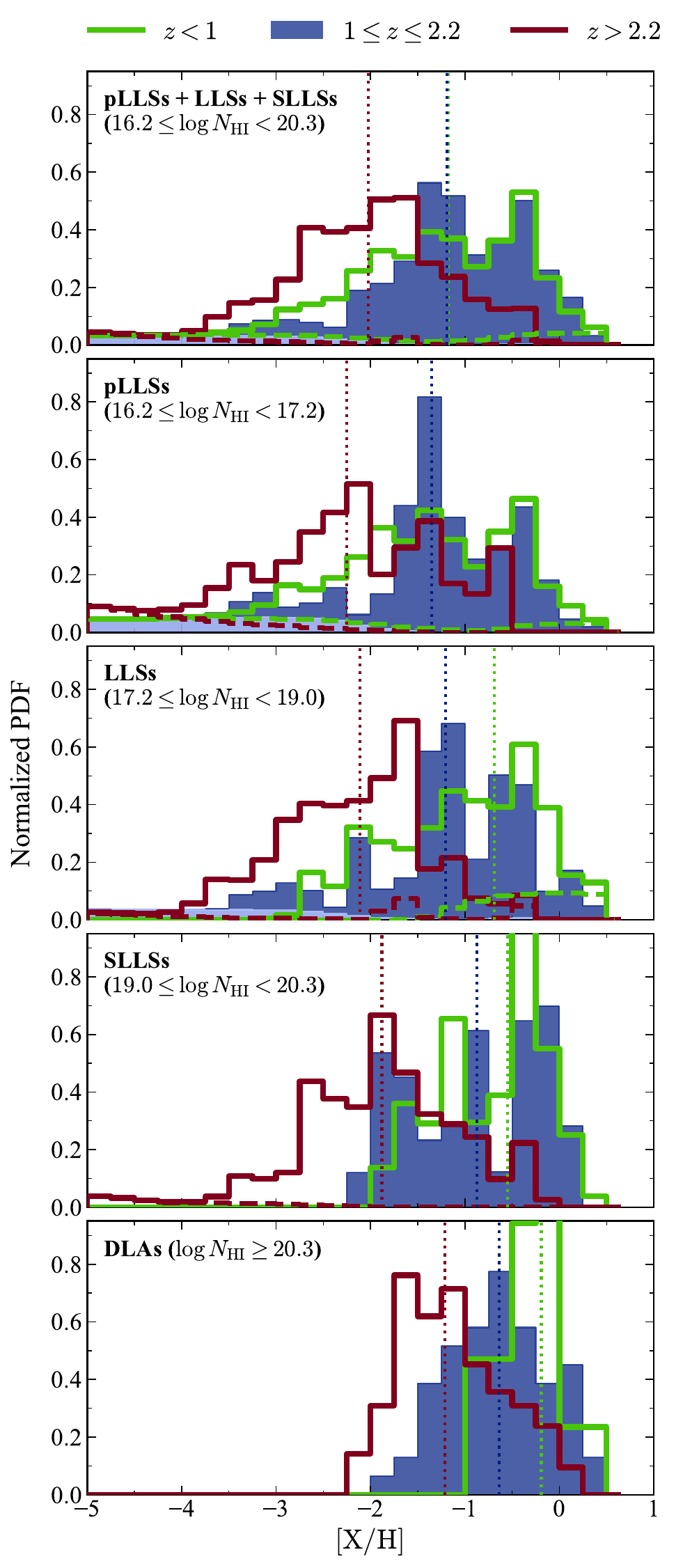}
\par\end{centering}
\caption{{\textit{Top panel:} Metallicity PDF for ionized CGM-like absorbers (pLLSs+LLSs+SLLSs) in the CCC ($z<1$; green), BRIDGE ($1\le z \le2.2$; blue), and KODIAQ-Z ($z>2.2$; red) samples. The lighter colored region and dashed lines represent contributions from limits. \textit{Middle panels:} Metallicity PDFs for each absorber type for the 3 samples. In each panel, the vertical dotted lines represent median metallicities for the respective samples. \textit{Bottom panel:} Metallicity PDFs for DLAs derived from point estimates drawn from \citetalias{peroux2021}.
}}\label{fig:met_pdf}
\end{figure}

\subsubsection{Metallicity Distributions}\label{subsub:met_pdf}

For each absorber in our sample, the Bayesian ionization modeling framework yields a posterior metallicity distribution. We use the MCMC walkers of each absorber to construct a combined metallicity probability distribution function for the sample. Figure~\ref{fig:met_pdf} shows the metallicity PDF for the BRIDGE survey ($1\le z\le2.2$; blue) of ionized CGM-like absorbers (pLLSs + LLSs + SLLSs), as well as the distribution for each absorber subclass, defined by its $N_{\HI}$ range. The shaded light region represents limits. We also plot the corresponding metallicity PDFs for the CCC ($z<1$; green) and KODIAQ-Z ($2.2<z\le3.6$; red) samples for comparison. In each panel, we plot the median metallicities for the samples as dotted lines. Among pLLSs and LLSs, several absorbers have no metal-ion detections, or only \CIV\, and/or \SiIV\, were detected (7 absorbers) with component structures distinct from Ly$\alpha$; for these, we report upper limits on metallicity. The low-$z$ CCC sample has several lower limits resulting from saturated metal-ion absorption, shown in the pLLSs and LLSs panels (dashed lines). Since there are only 8 SLLSs in the CCC sample, we adopt wider bins for the histogram. We also include a fifth panel showing the metallicity PDF for DLAs drawn from \citetalias{peroux2021} and classified into the same three redshift bins. Unlike the ionized absorbers, DLA metallicities are point estimates so the PDF is built directly using point values rather than walker chains. For the full BRIDGE sample ($16.4\le\log N_{\HI}<20.3$), we estimate the mean metallicity as $\langle\left[\rm{X/H}\right]\rangle=-1.27\pm1.10$, with a median of $-1.19$ and interquartile range (IQR) of $[-1.70$, $-0.49]$. The mean, median, and IQR for each absorber subclass across all three samples are listed in Table~\ref{tab:meanZ}. 

As shown in all four panels of Figure~\ref{fig:met_pdf}, BRIDGE ($1\le z\le2.2$) metallicity PDFs are more similar to CCC ($z<1$) than to KODIAQ-Z ($z>2.2$): by Cosmic Noon, most CGM-like absorbers are already enriched to the level seen at low $z$, consistent with peak metal production during this epoch. The median metallicities of the CCC ($-1.18$) and BRIDGE ($-1.19$) samples are nearly identical. At the same time, the peak of the BRIDGE metallicity PDF lies between those of the high-$z$ and low-$z$ comparison samples, consistent with the gradual increase in the mean CGM metallicity across cosmic time: $\langle[\rm{X/H}]\rangle=-2.11\pm0.93$, $-1.27\pm1.10$, and $-1.20\pm1.31$ for KODIAQ-Z, BRIDGE, and CCC, respectively. The DLAs, on the other hand, show substantial evolution between the three samples, with the mean metallicity increasing from $\langle[\rm{X/H}]\rangle=-1.14\pm0.56$ at $z=2.2$--3.6, to $-0.63\pm0.54$ at $z=1$--2.2, to $-0.30\pm0.31$ at $z<1$. These correspond to a three-fold increase between KODIAQ-Z and BRIDGE, and a two-fold increase from BRIDGE to CCC. 

The DLA metallicity PDF at Cosmic Noon is nearly Gaussian with a median of $-0.64$. By contrast, the metallicity PDFs of pLLSs, LLSs, and the combined pLLSs+LLSs+SLLSs sample for BRIDGE are negatively skewed, with the median of each distribution higher than its mean. This offset between mean and median is likely driven by the presence of limits in the sample, and may indicate a non-Gaussian underlying distribution. We therefore test the overall shape of each PDF statistically, rather than relying just on summary statistics. We apply Hartigan's \citep{hartigan1985} dip test to the per-system metallicities (medians and 80\% CI for limits) in each $N_{\HI}$ subclass; a significant dip statistic indicates two peaks in the PDF. For all absorber types in the BRIDGE sample, we fail to reject the null hypothesis for unimodality, finding no statistical evidence for bimodality (the dip statistic $D$ and $p$-values for all absorbers: $D=0.04$, $p=0.41$; pLLS: $D=0.07$, $p=0.58$; LLS: $D$=0.06, $p=0.39$; SLLS: $D=0.08$, $p=0.40$). We also perform a Gaussian mixture model (GMM) comparison, and find that a single-component model is preferred over a two-component mixture for every absorber type. Applying the same tests to CCC and KODIAQ-Z gives the same result for every subclass but one:  for the CCC sample of ionized absorbers, the GMM comparison prefers a two-component Gaussian fit with means at $-1.80$ and $-0.38$, consistent with \citetalias{wotta2019}. We also note that the CCC metallicity distribution is higher than that of BRIDGE in the range $-3.0<[\rm X/H]<-1.5$. This reflects the incompleteness of the BRIDGE-I pLLS sample, since we cannot access the absorbers with $\log N_{\HI}\lesssim16.4$ in the low resolution WFC3 spectra (see \S\ref{sec:HI}). 

Although statistically insignificant, visual inspection of the metallicity distributions of pLLSs and LLSs in the BRIDGE sample shows apparent double peaks. The pLLS distribution has peaks near $-1.3$ and $-0.4$, while the LLS metallicity PDF peaks near $[\rm{X/H}]\sim-1.2$, with a secondary peak at $\sim-0.5$, strikingly reminiscent of the bimodality observed at $z<1$ in the CCC sample \citep{lehner2013,lehner2019}. There is also a minor peak around $-3.0$, driven by a population of extremely metal-poor absorbers (EMPs), defined following \citetalias{lehner2022} as systems with $[\rm{X/H}]<-3.0$, which overlaps with the low metallicity tail in the high-$z$ PDF. The metallicity PDF of LLSs in BRIDGE is thus similar to that of CCC, with the addition of a small EMP population also seen in the high-$z$ sample. The SLLSs show significant metallicity evolution: while $\sim6\%$ of the high-$z$ SLLSs are EMPs, none of the SLLSs in our Cosmic Noon sample reach such low metallicities. The DLA sample does not show EMPs even at $z>3$, and the mean metallicity increases steadily over cosmic time ($\langle[\rm{X/H}]\rangle=-1.14$, $-0.63$, and $-0.30$ for KODIAQ-Z, BRIDGE, and CCC, respectively). 

These trends suggest that lower-$N_{\HI}$ absorbers (pLLSs, LLSs), which trace CGM-like gas, retain a substantial low-metallicity population ($[\rm{X/H}]\lesssim-2.0$) from Cosmic Noon through $z<1$, chemically decoupled from the central galaxy. This is consistent with continued accretion of metal-poor gas onto the CGM \citep{crighton2013,crighton2016,hafen2019,sultan2026}. The relatively larger IQRs of pLLSs and LLSs, compared to SLLSs and DLAs, are consistent with CGM-like gas originating from a mix of pristine inflows and previously enriched material, while remaining chemically decoupled from ISM metallicities \citep{kacprzak2019,kacprzak2026,nunez2024}. In contrast, the higher-$N_{\HI}$ SLLSs and DLAs, more closely associated with ISM-like gas, are enriched more rapidly. This likely reflects more efficient mixing and shorter enrichment timescales in the ISM compared to the diffuse CGM, which appears to be poorly mixed even at low $z$.

\input{table2_metpdf}

\begin{figure*}
\begin{centering}
\includegraphics[trim = 85 0 85 10, clip, width=\textwidth]{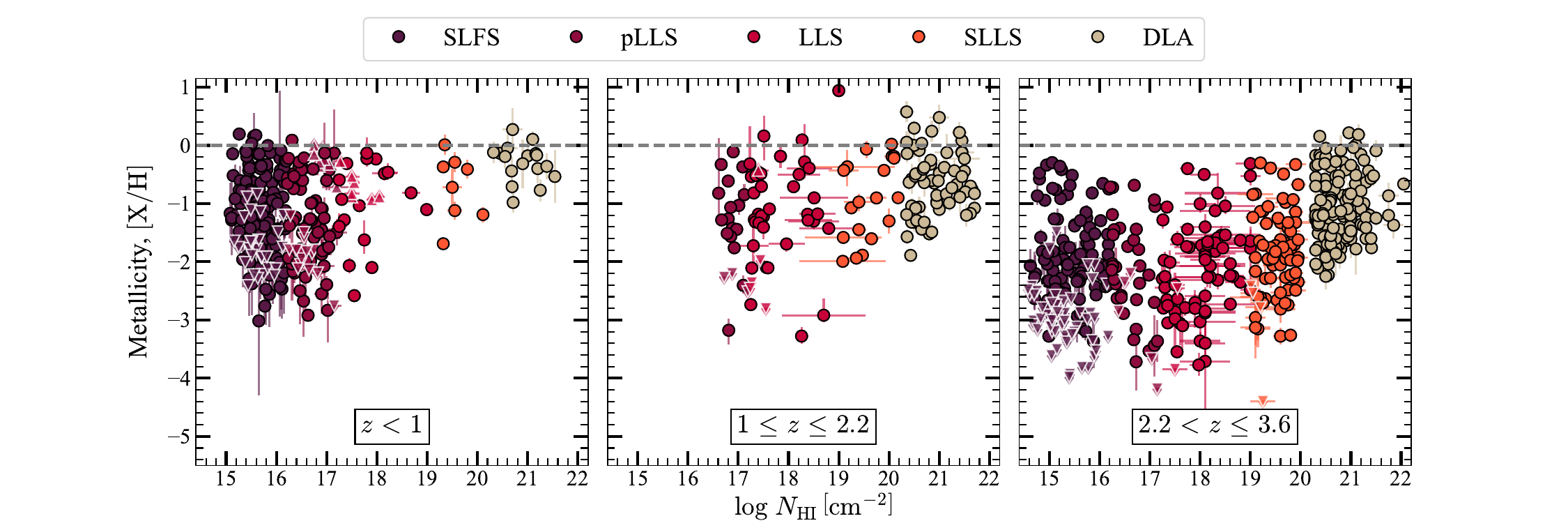}
\par\end{centering}
\begin{centering}
\caption{
{Metallicity as a function of $N_{\HI}$ for the CCC ($z<1$), BRIDGE ($1\le z\le2.2$), and the KODIAQ-Z ($z>2.2$) samples. The absorbers are color-coded based on $N_{\HI}$; each sample is reassigned to lie strictly within its defined redshift range. We plot the 68\% confidence interval for all measurements, except for the upper and lower limits (triangles), where we plot the 90th and 10th percentiles, respectively.}
} \label{fig:3_panel}
\par\end{centering}
\end{figure*}

\subsubsection{Metallicity vs $N_{\HI}$}\label{subsub:met_v_nhi}

Figure~\ref{fig:3_panel} shows the estimated metallicities as a function of \HI\ column density for CCC, BRIDGE, and KODIAQ-Z. Across all three samples, metallicity increases with increasing $N_{\HI}$, consistent with denser, cooler gas in galactic environments (traced by DLAs and SLLSs) being enriched more rapidly and retaining more metals than lower $N_{\HI}$ gas. 

At all redshifts, the metallicities of lower column density systems (SLFSs, pLLSs, LLSs) show considerably larger scatter, illustrating the diversity of environments traced by these absorbers. For all absorber types, the mean metallicity remains sub-solar, though some CGM-like absorbers reach super-solar values at Cosmic Noon (2/80) and $z<1$ (14/133), consistent with metal build-up due to continued deposition over cosmic time. Only $3\%$ ($1\%$--$7\%$ CI) or 2/80 CGM-like absorbers (pLLSs, LLSs, SLLSs) in BRIDGE have super-solar metallicities. In fact, such absorbers remain rare at any redshift, though several lower limits at $z<1$ leave this uncertain. We find one super metal-rich ($[\rm{X/H}]\sim1.0$) absorber with $\log N_{\HI}=19.00\pm0.10$, along J092705+562114 at $z_{\rm{abs}}=1.779$, previously reported by \cite{prochaska2006}. With our larger sample, we now find that such super-enriched absorbers are very rare at Cosmic Noon.

At Cosmic Noon, we define very metal-poor (VMP) absorbers as those with $[\rm{X/H}]<-1.7$ (the 2$\sigma$ lower limit of DLA metallicities).\footnote{\citetalias{lehner2022} defined VMP systems for the KODIAQ-Z sample as those with $[\rm{X/H}]<-2.4$, which is the 2$\sigma$ lower limit for DLA metallicities at $2.2<z\lesssim3.6$. Similarly, for CCC ($z<1$) VMPs are absorbers with $[\rm{X/H}]<-1.4$ \citepalias{wotta2019}.} For each absorber type, we report the fraction of $[\rm{X/H}]<-1.7$ (VMP), $[\rm{X/H}]<-2.4$, and $[\rm{X/H}]<-3.0$ (EMP) in Table~\ref{tab:meanZ}. To quantify the fraction of absorbers below a given metallicity threshold, we compute binomial proportions and with uncertainties from the Wilson score 90\% CIs \citep{wilson1927}. The Wilson score gives accurate coverage for binomial proportions even at sample sizes or when the proportion is near 0 or 1. In BRIDGE, 30\% of pLLSs, 27\% of LLSs, and 16\% of SLLSs are VMP; EMP absorbers comprise pLLS ($\sim1\%$--$20\%$) and LLS ($\sim1\%$--$10\%$), and no absorbers have $[\rm{X/H}]<-3.5$. That is, we see no evidence for pristine gas at Cosmic Noon. By contrast, $10\%$ of pLLSs (4/40) and $\sim6\%$ of LLSs (4/68) in the higher-$z$ KODIAQ-Z sample have $[\rm{X/H}]<-3.5$.

As shown in Table~\ref{tab:meanZ}, the upper quartiles of the metallicity distributions evolve weakly across $\log N_{\HI}$, whereas the lower quartiles decrease substantially with decreasing $\log N_{\HI}$, indicating that the spread in the metallicity is driven mainly by these metal-poor systems with $\log N_{\HI}<19.0$. 
\begin{figure}
\begin{centering}
\includegraphics[width=\columnwidth]{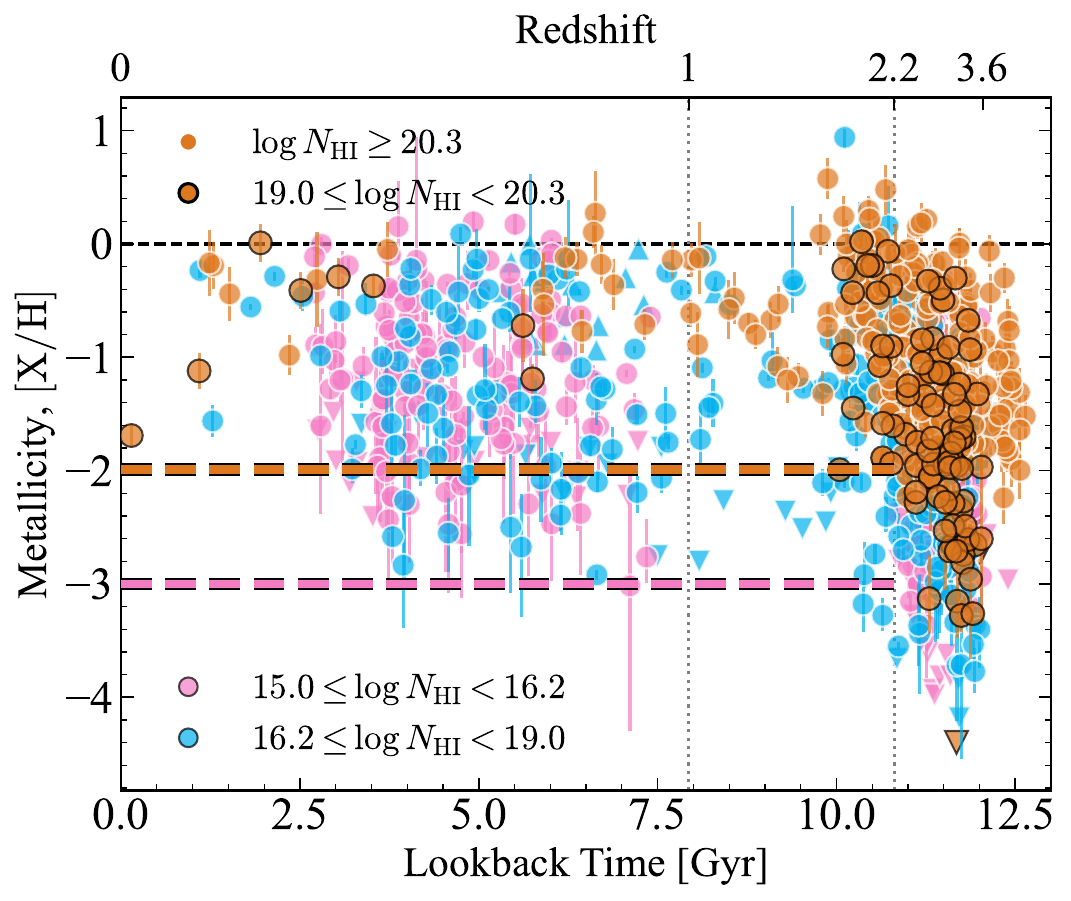}
\par\end{centering}
\begin{centering}
\caption{
{Metallicity as a function of lookback time for the CCC ($z<1$, BRIDGE ($1\le z\le2.2$), and KODIAQ-Z ($2.2<z\lesssim3.6$) samples. The top axis shows the corresponding redshift. Absorbers are color-coded by $N_{\HI}$; black marker edges denote SLLSs and white marker edges denote DLAs. The horizontal dashed black line marks solar metallicity, and the horizontal dashed orange line marks the approximate metallicity ``floor" observed for SLLSs and DLAs at $z\lesssim2.2$. The dashed pink line marks the ``floor" for all absorbers at $z\lesssim 2.2$. Vertical gray lines mark the redshift boundaries separating the three samples.}
} \label{fig:LBT}
\par\end{centering}
\end{figure}

\subsubsection{Evolution of Metallicity Distribution}\label{subsub:lbt}
The overarching goal of the BRIDGE survey is to study the evolution of the metallicity distribution of absorbers probing CGM- and IGM-like gas with cosmic time. In lookback time, CCC, BRIDGE and KODIAQ-Z span $\Delta t_{\rm LB}\sim8$ Gyr, $\sim3$ Gyr, and $\sim1.2$ Gyr, respectively. We show estimated metallicities as a function of lookback time in Figure~\ref{fig:LBT}, with absorbers color-coded according to $N_{\HI}$: pLLSs (pink), LLSs (blue), and SLLSs/DLAs (orange). Gray dotted lines mark the boundaries between the CCC, BRIDGE, and KODIAQ-Z samples. The metallicity distribution evolves significantly around $z\sim2.2$, over a period of just 1 Gyr. Lower metallicity absorbers, often with $\log N_{\HI}<19.0$, are preferentially found at higher-$z$ within BRIDGE ($1\le z \le2.2$).

We note a metallicity ``floor" of $[\rm{X/H}]\sim-2.0$ (orange dashed line in Figure~\ref{fig:LBT}) at $z\le2.2$. This is set by two independent statistical estimates from the combined DLA+SLLS sample (108 absorbers): a CDF-inversion estimate which gives the metallicity under which only 2\% of absorbers fall ($[\rm{X/H}]=-1.94$) and an order-statistic estimate of where the lowest-metallicity absorber in a sample of this size is expected to fall ($[\rm{X/H}]=-1.99$). These agree within 0.05 dex. Given the finite sample size, a non-parametric Dvoretzky-Kiefer-Wolfowitz \citep{dkw,massart1990} 68\% confidence interval allows for up to $9\%$ of the population to lie below $[\rm{X/H}]\sim-2.0$ despite none being observed, a limit set by sampling uncertainty rather than empirical evidence. While $25\%$ of pLLSs and $24\%$ of LLSs in BRIDGE fall below $[\rm{X/H}]=-2.0$ (and $30\%$ and $13\%$ respectively, in CCC), no SLLSs or DLAs in our statistical sample reach such low metallicities at these redshifts. At $z>2.2$, 4\% of DLAs and 40\% of SLLSs in our sample have $[\rm{X/H}]<-2.0$. Even at high $z$, metal-poor DLAs are rare in statistical surveys; dedicated surveys of metal-poor DLAs find several low metallicity ($[\rm{X/H}]\lesssim-2.0$) DLAs, though still at $z\gtrsim2.2$ \citep{cooke2011,cooke2017,welsh2022,welsh2024,berg2025}.

At $z<2.2$, the most metal-poor SLLSs occupy $-2.0<[\rm{X/H}]<-1.7$ ($16\%$ in BRIDGE). By contrast, $z>2.2$, $40\%$ SLLSs have $[\rm{X/H}]<-2.0$. At these redshifts, SLLSs probe much smaller overdensities in the Universe, and these metal-poor SLLSs may be tracing denser patches in the IGM or the diffuse CGM. We discuss overdensities of absorbers is more detail in \S\ref{subsec:bov_origin}.

Considering diffuse ionized absorbers with $16.2<\log N_{\HI}<20.3$ (since BRIDGE-I does not include SLFSs) across all redshifts, 34\% and 13\% of absorbers in the KODIAQ-Z sample have $[\rm{X/H}]<-2.4$ and $[\rm{X/H}]<-3.0$, respectively. BRIDGE has fewer metal-poor absorbers (14\% and 3\%), and CCC fewer still, with only 8\% below $[\rm{X/H}]<-2.4$, and none that are EMP. This ``floor" for all CGM-like absorbers is shown by the dashed pink line at $[\rm{X/H}]=-3.0$ in Figure~\ref{fig:LBT}. Even the lowest-$N_{\HI}$ gas at $z<1$, traced by SLFSs, have $[\rm{X/H}]\gtrsim-3.0$ (see Figure~\ref{fig:3_panel}), though most of them are upper limits. The decline in the fraction of metal-poor absorbers from high to low redshift suggests that diffuse CGM- and IGM-like gas undergoes sustained metal enrichment from $z\sim3$ to $z<1$, with the most metal-poor gas (potentially enriched by Population III stars) becoming progressively rarer with cosmic time. 

From KODIAQ-Z to BRIDGE to CCC, the IQR for all ``CGM-like" absorbers (pLLSs+LLSs+SLLSs) changes from [$-2.67, -1.54$] to [$-1.70,-0.49$] to [$-1.95,-0.39$]. Thus, the IQRs widens, the span increasing from 1.13 dex to 1.21 dex to 1.56 dex from higher to lower redshift. This is driven primarily by the ionized $N_{\HI}$ absorbers: considered separately, the DLA IQR does not widen, remaining similar between KODIAQ-Z and BRIDGE ($\sim0.8$ dex) and narrowing between BRIDGE and CCC ($0.3$ dex). The truncation of the metal-poor tail in the denser SLLSs and DLAs, alongside the persistence of VMP pLLSs and LLSs in the BRIDGE sample, indicates that global metal mixing throughout the population at Cosmic Noon is well advanced in ISM-like/denser CGM-like gas but relatively incomplete in the more diffuse CGM. 

The large metallicity scatter among ionized absorbers points to diverse physical origins: enriched outflows likely elevate the metallicities of many CGM-like absorbers, while continued accretion of metal-poor gas sustains a metal-poor population with $[\rm{X/H}]<-2.4$ and even $[\rm{X/H}]\sim-3.0$, at $1\le z\le2.2$. The BRIDGE LLS PDF illustrates this behavior, with features of both high- and low-$z$ populations: its median matches CCC, but the low metallicity tail seen in the KODIAQ-Z persists. By $z<1$, all absorbers have $[\rm{X/H}]>-3.0$, with no evidence of pristine gas among CGM-like absorbers. 

Another independent assessment of metal mixing comes from the large metallicity variations seen across small velocity separations within individual absorbers in both the KODIAQ-Z \citepalias{lehner2022} and CCC \citep{lehner2019} samples. \citetalias{lehner2022} show that the mean metallicity difference between paired absorbers ($|\Delta| v<150$ \kms) is $\Delta[\rm{X/H}]=0.68\pm0.14$ at $2.2< z\le3.6$. Similarly, \cite{lehner2019} show that $75\%$--$96\%$ of paired absorbers in the CCC sample ($|\Delta| v<500$ \kms) can have mean metallicity difference as high as $\Delta[\rm{X/H}]=1.7$. Complementary evidence comes from the CUBS survey: \cite{zahedy2021} resolve individual velocity components within four $z<1$ LLSs and find a median metallicity of $-0.7_{-0.2}^{+0.1}$, with a 16--84 percentile range spanning $(-1.3,\,-0.1)$, a wide spread in enrichment among components in the same absorption complex. Similarly, \cite{cooper2021} find component-to-component metallicity variations of $>0.8$ dex for a pLLS within a single halo, evidence that distinct components within the same halo can have disparate physical origins and are not well mixed. Together, these results indicate that even at small velocity scales, ionized CGM-like gas is poorly mixed. 

\begin{figure}
\begin{centering}
\includegraphics[width=\columnwidth]{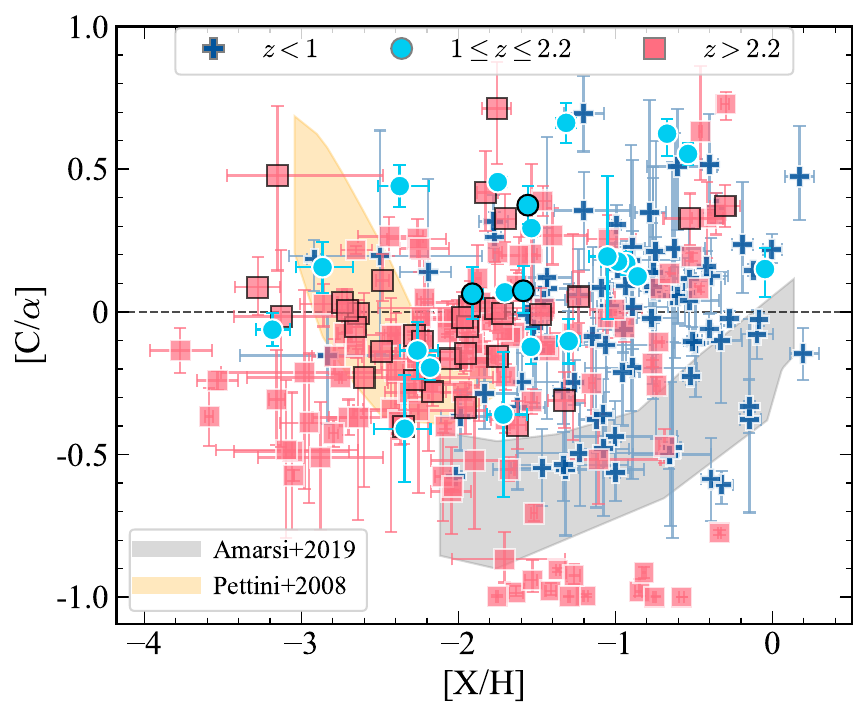}
\par\end{centering}
\begin{centering}
\caption{
{The $[\rm{C}/\alpha]$ ratio versus metallicity $[\rm{X/H}]$ for the KODIAQ-Z (pink; $2.2<z\le3.6$), BRIDGE (cyan; $1\le z\le2.2$), and the CCC (blue; $z<1$) samples. The points with black edges are SLLSs. The gray and orange shaded regions show the expected $[\rm{C}/\rm{O}]$ for stars \citep{amarsi2019} and SLLSs/DLAs \citep{pettini2008}.}
}\label{fig:ca_met}
\par\end{centering}
\end{figure}

\begin{figure*}
\begin{centering}
\begin{minipage}{0.48\textwidth}
\begin{centering}
\includegraphics[width=\columnwidth]{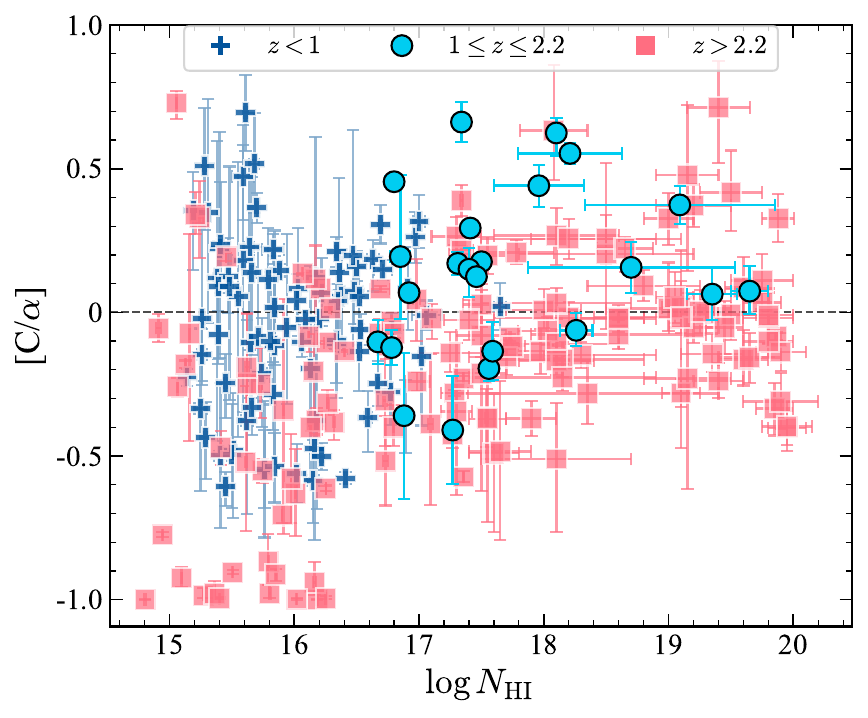}
\par\end{centering}
\end{minipage}
\hfill
\begin{minipage}{0.48\textwidth}
\begin{centering}
\includegraphics[width=\columnwidth]{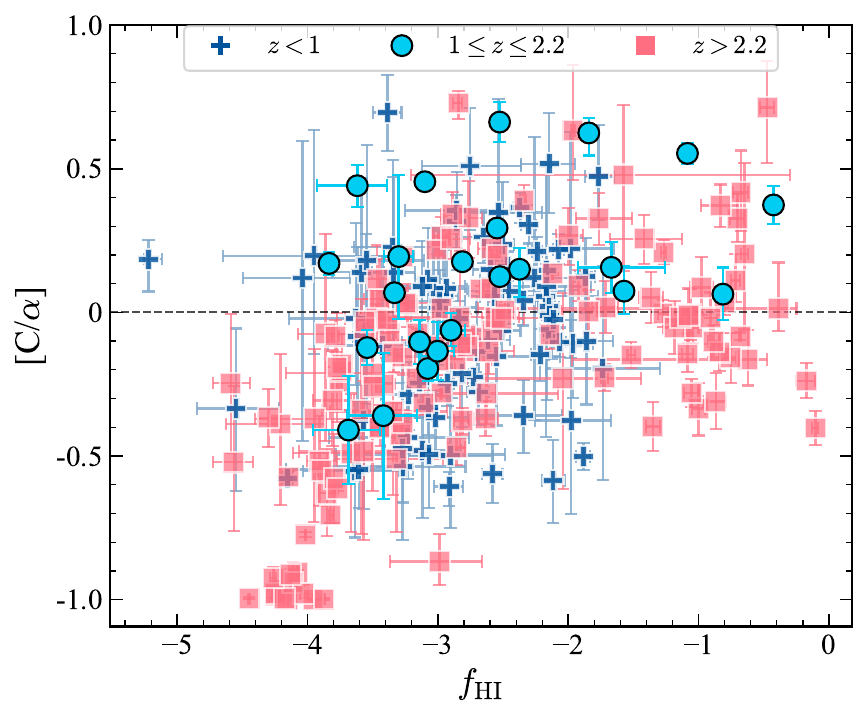}
\par\end{centering}
\end{minipage}
\par\end{centering}
\caption{
\textit{Left panel:} The $[\rm{C}/\alpha]$ ratio versus $N_{\HI}$ for the KODIAQ-Z (pink; $2.2<z\le3.6$), BRIDGE (cyan; $1\le z\le2.2$), and the CCC (blue; $z<1$) samples.
\textit{Right panel:} The $[\rm{C}/\alpha]$ ratio versus $f_{\HI}$ for the three surveys.
}\label{fig:carbalpha}
\end{figure*} 

\subsection{Non-solar Relative Abundances}\label{subsec:nonsolar}

Relative abundances trace the enrichment history of the absorbing gas independently of its total metallicity, since carbon and the $\alpha$-elements are produced on different timescales by different stellar populations. Our photoionization models constrain [C$/\alpha$] directly for 23 absorbers in the BRIDGE sample. We plot the estimated $[\rm C/\alpha]$ ratio as a function of metallicity in Figure~\ref{fig:ca_met}, and perform the Spearman rank order test in each redshift bin to test for a correlation. We find a significant positive correlation only at $1\le z \le 2.2$ ($r_s=0.44$, $p=0.04$); neither  $z<1$ ($r_s=0.14$, $p=0.20$) nor $z>2.2$ ($r_s=0.08$, $p=0.34$) shows a statistically significant correlation. Within the KODIAQ-Z sample, excluding absorbers with $\left[\rm{C}/\alpha\right]<-0.9$ (which are more strongly affected by ionization corrections, see \S6.5 in \citetalias{lehner2022}), reveals a statistically significant positive correlation with metallicity ($r_s = 0.28$, $p=0.002$). 

In Figure~\ref{fig:ca_met}, we also show  the $[\rm{C}/\alpha]$--$[\alpha/\rm{H}]$ trend seen in stars (shaded orange; \citealt{amarsi2019}) and in DLAs/SLLSs (shaded gray; \citealt{pettini2008}). These studies find that $[\rm{C}/\alpha]$ decreases from $[\rm{X/H}]\sim0$ to $-2$, with an upturn below $[\rm{X/H}]=-2$. \citetalias{lehner2022} categorized their sample by $N_{\HI}$ and found that SLLSs in the KODIAQ-Z sample show a similar trend to that seen in stars and DLA/SLLS \citep{akerman2004,fabbian2009,pettini2008,cooke2011}, including the upturn in $[\rm{C}/\alpha]$ below $[\rm{X/H}]=-2$. By contrast, absorbers with $\log N_{\HI}<19$ do not show this upturn. With our larger sample, we revisit this trend. While BRIDGE shows an overall positive correlation between $[\rm{C}/\alpha]$ and $[\rm{X/H}]$, we do not see the specific decline-then-upturn trend reported for stars and DLAs/SLLSs. We note that only 6 of the 23 well-constrained BRIDGE absorbers have metallicities $[\rm{X/H}]<-2$, and only 3 are SLLSs, too few to test this trend independently for higher-$N_{\HI}$ absorbers. 

We also plot the estimated $\left[\rm{C}/\alpha\right]$ as a function of $N_{\HI}$ and the neutral fraction $f_{\HI}=N_{\HI}/N_{\rm{H}}$ in Figure~\ref{fig:carbalpha}, for the CCC, BRIDGE, and KODIAQ-Z samples, and perform the same Spearman test for $\left[\rm{C}/\alpha\right]$ against $N_{\HI}$ and $f_{\HI}$ in each sample. For CCC, we find no statistically significant correlation between $\left[\rm{C}/\alpha\right]$ and $N_{\HI}$ ($r_s = -0.01$, $p = 0.91$) or $f_{\HI}$ ($r_s = 0.16$, $p = 0.14$). BRIDGE likewise shows no significant correlation with $N_{\HI}$ ($r_s = 0.19$, $p = 0.40$) and only a marginal positive correlation with $f_{\HI}$ ($r_s = 0.37$, $p = 0.08$). In contrast, the KODIAQ-Z sample exhibits a strong positive correlation with both $N_{\HI}$ ($r_s = 0.40$, $p = 1.3 \times 10^{-6}$) and $f_{\HI}$ ($r_s = 0.61$, $p = 3.0 \times 10^{-15}$), consistent with \citetalias{lehner2022}. These correlations are not driven by the small number of extremely carbon-poor absorbers: excluding systems with $\left[\rm{C}/\alpha\right]<-0.9$, the KODIAQ-Z trends remain significant with both $N_{\HI}$ ($r_s = 0.27$, $p = 2.2 \times 10^{-3}$) and $f_{\HI}$ ($r_s = 0.50$, $p = 4.9\times 10^{-9}$). 

The carbon abundance in the CCC sample is constrained primarily by the lower ionization states \CII\ and \CIII, whereas BRIDGE and KODIAQ-Z also use \CIV. Because ionization corrections are largest in the low $N_{\HI}$, low $f_{\HI}$ systems, which also have lower $\left[\rm{C}/\alpha\right]$ values, we attribute the observed positive correlations of $\left[\rm{C}/\alpha\right]$ with $N_{\HI}$ and $f_{\HI}$ at $z>1$ to the large ionization corrections applied to absorbers, rather than to nucleosynthetic effects. 

By contrast, we find that the $\left[\rm{C}/\alpha\right]$--$[\rm{X/H}]$ correlation is less affected by this systematic. Excluding the most ionization-affected KODIAQ-Z absorbers with $\left[\rm{C}/\alpha\right]<-0.9$, the scatter in $\left[\rm{C}/\alpha\right]$ is nearly identical across all three samples ($\sigma=0.30$, $0.29$, and $0.29$ for CCC, BRIDGE, and KODIAQ-Z, respectively). This is despite the varying levels of ionization corrections applied to the three samples. This indicates that excess scatter in the entire KODIAQ-Z sample ($\sigma=0.37$) is concentrated only among the absorbers that have the largest ionization correction. The positive $\left[\rm{C}/\alpha\right]$--$[\rm{X/H}]$ correlations in BRIDGE and KODIAQ therefore may be tracing nucleosynthetic effects, though with substantial added noise from ionization corrections \citepalias{lehner2022}.

\input{table3_physicspdf}
\subsection{Physical Properties of Absorbers at Cosmic Noon}\label{subsec:phys_prop}

The photoionization models discussed in \S\ref{sec:metallicity} also predict the physical properties of the gas. Specifically, we derive the gas density $n_{\rm{H}}$, the neutral fraction $f_{\HI}$, the total hydrogen column density $N_{\rm{H}}=N_{\HI}+N_{\HII}$, and the line-of-sight length scale ($l=N_{\rm{H}}/n_{\rm{H}}$) of the absorbers. However, there is a known degeneracy between the density $n_{\rm{H}}$ and the intensity of the ionizing radiation field when constraining the ionization parameter (since $U=n_{\gamma}/n_{\H}$), making the inferred density sensitive to the assumed EUVB \citep{fumagalli2016,lehner2022,gibson2022}. In contrast, metallicity depends on the relative abundances of metal and hydrogen ions, which are set by the shape of the ionizing spectrum rather than its absolute normalization, and is therefore more robustly constrained. Additionally, our models assume a uniform EUVB and purely photoionized gas; in practice, local ionizing sources (e.g., galaxies or AGN) may contribute to the ionization, further affecting the inferred densities and length scales. This effect should be minimal for pLLSs and LLSs that dominate our sample and are typically found at larger impact parameters \citep{berg2023}. Of the 105 absorbers in the BRIDGE sample, only 5 lie within 3000 km s$-1$ of the quasar and 3 show \OVI\, and/or \NV\, absorption. This includes the absorber for which the \NV\, absorption is consistent with photoionization (see \S\ref{sec:metal_Cols}). Contamination from local ionizing sources associated with the quasar host is therefore unlikely to affect our sample.

Photoionization models also assume uniform density along the line of sight, and combining them with the observed integrated column densities effectively collapses the three-dimensional structure of the gas into a single, path-length-averaged dimension. The inferred densities and physical scales therefore represent averaged quantities for an absorption system that may comprise several inhomogeneous components. This limitation is most severe for BRIDGE. At $1\le z\le2.2$, the higher-order Lyman series lines fall in the near-UV, where the available HST spectra ($R\sim200$ for WFC3) cannot resolve the \HI\, velocity structure. By contrast, in CCC and KODIAQ-Z the weaker Lyman series lines fall in the far UV ($R\sim20000$) and optical ($R\sim45000$), respectively, where \HI\, can be resolved into individual components.

\begin{figure*}
\begin{centering}
\includegraphics[trim = 0 5 0 0, clip, width=\textwidth]{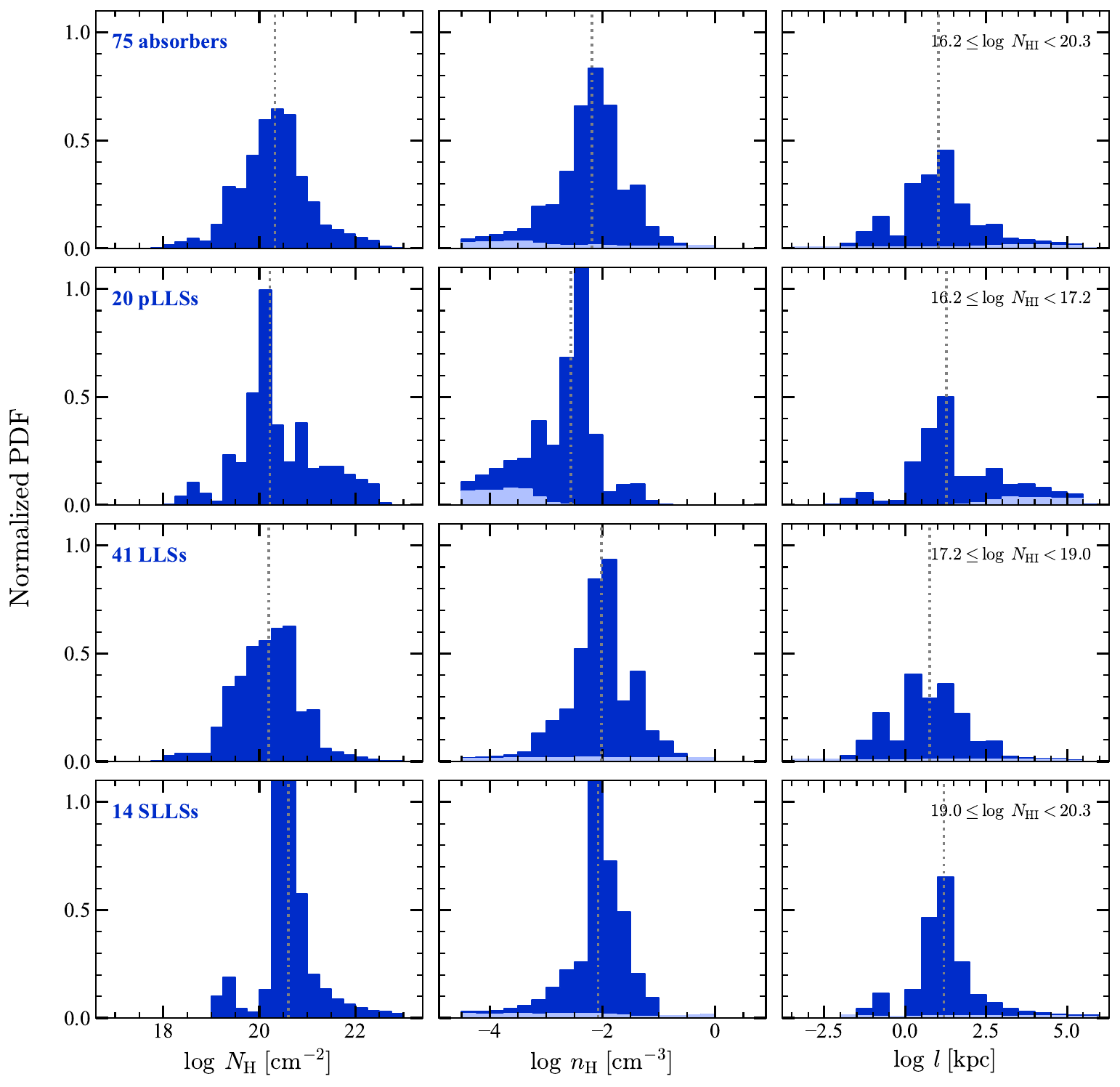}
\par\end{centering}
\caption{Posterior PDFs of the total hydrogen column density $N_{\rm H}$, hydrogen number density $n_{\rm H}$, and line-of-sight length scale $l$ for all (pLLSs+LLSs+SLLSs) (top), pLLSs (second), LLSs (third), and SLLSs (bottom), derived using walkers from our photoionization models. The light-blue shaded region represents lower and upper limits. The dotted gray lines represent the medians for each distribution.}
\label{fig:physics}
\end{figure*}

In Figure~\ref{fig:physics}, we plot the combined posterior PDFs (derived directly using MCMC walkers) of $N_{\rm{H}}$, $n_{\rm{H}}$, and $l$ for each absorber type in the BRIDGE sample. The corresponding mean, standard deviation, and median (dotted gray lines in Figure~\ref{fig:physics}) for all quantities are listed in Table~\ref{tab:physprop}. We also report the summary statistics using only those systems for which the density $n_{\rm H}$ is well constrained, that is, those for which we did not impose a Gaussian prior on $\log U$. We find that excluding the absorbers with a Gaussian $\log U$ prior leaves the statistics unchanged. This is not surprising, since only 12 absorbers required the prior ( 4 pLLSs and 8 LLSs), corresponding to 16\% of the entire sample and 20\% of each of the pLLS and LLS subsamples. We therefore adopt the entire sample for all subsequent statistics. 

The PDFs shown in Figure~\ref{fig:physics} are much narrower than the metallicity PDFs in Figure~\ref{fig:met_pdf}, suggesting that gas clouds of similar physical properties, such as density and size, can nonetheless span a wide range of metallicities. We summarize the physical quantities ($N_{\rm{H}}$, $n_{\rm{H}}$, $l$) derived from our photoionization modeling of the BRIDGE-I sample for each absorber in Table~\ref{tab:bridge_physical}. This includes absorbers which were analyzed in this paper but lie in the CCC and KODIAQ-Z redshift ranges. In this section we analyze the redshift evolution of physical properties across absorber types.\footnote{The CCC ($z<1$) SLLSs are drawn from the literature, and their physical properties were not derived; we therefore compare only the KODIAQ-Z and BRIDGE SLLS samples.}

\subsubsection{Hydrogen Column Densities, Densities, and Physical Scales}\label{subsub:physprop}

For the BRIDGE sample, the mean $\log N_{\rm H}$ shows no significant trend with $N_{\HI}$: $\overline{\log N_{\rm H}}$ is 20.42, 20.17, and 20.61 for pLLSs, LLSs, and SLLSs, respectively, though the large standard deviations mean the three PDFs overlap substantially. The PDF for the full sample is unimodal and nearly Gaussian, with $\overline{\log N_{\rm H}}=20.32\pm0.74$ and a median of 20.33, remarkably similar to DLA \HI\, column densities and $N_{\rm H}$, assuming DLAs are fully neutral, i.e., $N_{\rm H}=N_{\HI}$ (although see \citealt{fox2007,fox2011,lehner2014}). We note that all absorber types have similar $N_{\rm H}$ irrespective of $N_{\HI}$. Moreover, the \HI\, column density distribution functions of \citetalias{ribaudo2011} (their Figure 9) and \citetalias{omeara2013} (their Figure 14), show that lower \HI\, column density systems (LyAF, SLFSs, pLLSs, LLSs) far outnumber SLLSs and DLAs, at all redshifts. This implies that low-$N_{\HI}$ systems contribute more to the total baryon budget at Cosmic Noon compared to high-$N_{\HI}$ systems. We also note that $\overline{\log N_{\rm H}}$ decreases from high to low redshift: the evolution between KODIAQ-Z ($20.12$, $20.75$, $20.84$ for pLLSs, LLSs, and SLLSs) and BRIDGE is modest, but $\overline{\log N_{\rm H}}$ is higher at Cosmic Noon than at $z<1$ ($19.21$ and $19.53$ for pLLSs and LLSs) by factors of $\sim16$ and $\sim5$, respectively. This points to an evolving baryon budget over cosmic time: at high $z$ more baryons reside in low-$N_{\HI}$ (and thus, low overdensity) gas, whereas at low redshift most of the gas lies in the more overdense regions of the Universe, following structure formation expectations (\citetalias{peroux2021}; \citealt{deepak2025}, hereafter \citetalias{deepak2025}).

We caution that densities, overdensities, and length scales systematically depend on the treatment of high ions (\CIV, \SiIV). At Cosmic Noon, including high ions in our models results in a systematically higher $\log U$ and correspondingly a lower $n_{\rm H}$ by $\sim0.3$ dex. This propagates to a $\sim0.7$ dex offset in $\log l\,\rm{[kpc]}$ (see Appendix~\ref{app:ionsincluded}). The trends discussed below should therefore be treated as statistical results particular to our model assumptions, and not be over-interpreted.  

The estimated hydrogen densities $n_{\rm{H}}$ broadly increase with $N_{\HI}$ as expected for self-shielding gas, though as with $N_{\rm{H}}$ the dispersion in the density PDFs is large, and the distributions for the three absorber types overlap. The mean $\log n_{\rm H}$ is $-2.75$, $-2.04$, and $-2.12$ for pLLSs, LLSs, and SLLSs, respectively. For some absorbers, $\log U$ is not well constrained, yielding only a limit on $n_{\rm{H}}$ (3 pLLSs have lower limits, and 4 LLSs and 2 SLLSs have upper limits on $\log n_{\rm H}$), shown as the light-colored regions of the PDFs in Figure~\ref{fig:physics}. At $z\sim1.81$ (the mean redshift of the BRIDGE sample), the mean cosmic density is $\overline{n}_{\rm H}\sim10^{-5.42}$ cm$^{-3}$ \citep{schaye2001a}. At Cosmic Noon, then, pLLSs, LLSs, and SLLSs trace overdensities $\delta_b=n_{\rm H}/\overline{n}_{\rm H}$ of $\sim400$--2200, with a broader range of $\sim100$--9000 if we account for the dispersion. Such overdensities ($200\lesssim\delta\lesssim10^4$) are generally associated with the CGM of galaxies \citep{schaye2001a,schaye2001b,steidel2010,pallottini2014,stern2016}. Compared to the high redshift ($z\sim3$) KODIAQ-Z sample where $\overline{\log n_{\rm H}}=-2.70$, $-2.50$, and $-2.29$ for pLLSs, LLSs and SLLSs, respectively ($\delta_b\sim180$--500), BRIDGE absorbers probe higher overdensities. This results from both the mean cosmic density decreasing and the $\overline{n_{\rm H}}$ of absorbers increasing with cosmic time. For the CCC sample, $\overline{\log n_{\rm H}}=-2.50$ and $-2.22$ for pLLSs and LLSs. Since the mean densities for BRIDGE and CCC samples are comparable, while the mean density of the Universe continues to decrease, pLLSs and LLSs in the CCC sample probe even higher overdensities ($\delta_b\sim4500$ and 8500).

The line-of-sight length scales are directly related to $n_{\rm H}$ and $N_{\rm H}$ and show the largest dispersion among all derived physical parameters. Upper and lower limits in $n_{\rm H}$ propagate to lower and upper limits in length scale, respectively. Since the PDFs in Figure~\ref{fig:physics} are estimated from MCMC walkers in our photoionization models, limits on parameters produce extended tails, as seen in the $\log l$ PDFs; these show unphysical values of $\log l>2.5$ that do not represent actual measurements but are instead upper limits, indicated by the light-colored regions. For pLLSs, LLSs, and SLLSs, the length-scale IQRs are 6--410 kpc, 1--30 kpc, and 4--36 kpc, respectively. The corresponding values for KODIAQ-Z are 4--112 kpc, 12--238 kpc, and 10--175 kpc. For CCC, we do not have estimates for SLLSs; pLLSs and LLSs both have IQRs $\sim0.3$--7 kpc. High-$z$ pLLSs+LLSs+SLLSs therefore probe larger structures (CGM, filaments, intra-group gas), whereas at low $z$ the same absorbers trace much smaller clouds within galactic halos or overdense patches in the IGM \citep{crighton2015,rubin2018,berg2023,butsky2024,richter2025}.

\begin{figure}
\begin{centering}
\includegraphics[width=\columnwidth]{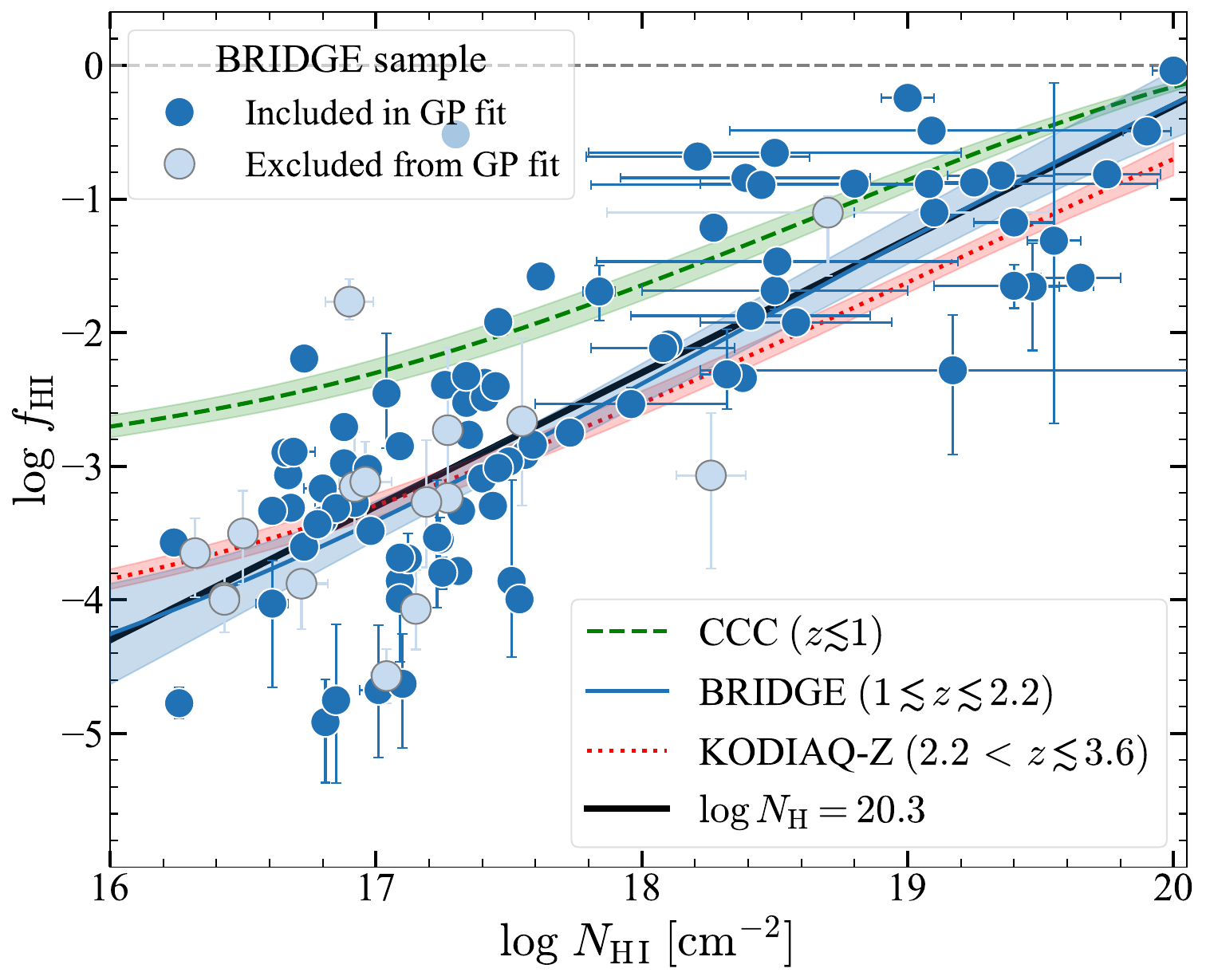}
\par\end{centering}
\begin{centering}
\caption{
{Neutral fractions $f_{\HI}$ as a function of $\log N_{\HI}$ for the BRIDGE sample. The solid blue curve is the resulting GP model, and the shaded area around it represents the standard deviation predicted by the fit. Blue circles represent absorbers with a flat prior on $\log\,{U}$, whereas light-blue circles are absorbers for which we adopted a Gaussian prior on $\log\,{U}$ and are excluded from the Gaussian process fit. We also plot the neutral fraction fits for the CCC (dashed green) and KODIAQ-Z (dotted red) surveys for comparison. The black line represents $f_{\HI}$ values for $\log N_{\rm{H}}=20.3$.}
}\label{fig:neutral_frac}
\par\end{centering}
\end{figure}

\subsubsection{Neutral Fractions}\label{subsub:neut_frac}

Estimating the metallicity of an absorber requires ionization corrections to both the metal-ion and neutral hydrogen column densities. Using observed column densities and Cloudy models within our Bayesian framework, we derive the neutral fraction $f_{\HI}$. The correlation between $N_{\HI}$ and $f_{\HI}$ at any redshift, along with the column density distribution function, can be used to derive the cosmic mass density of cool, diffuse gas. Both at high $z$ \citepalias{lehner2022} and low $z$ \citep{lehner2019}, $f_{\HI}$ and $N_{\HI}$ are positively correlated. Following \citepalias{lehner2022}, we use a Gaussian process (GP) model for nonparametric regression to describe the mean $f_{\HI}$ as a function of $N_{\HI}$, allowing probabilistic fitting without specifying a functional form. The resulting predictions from the GP model can be approximated by simpler polynomials while retaining the full posterior distribution. We use the empirical confidence intervals from the GP model to characterize uncertainties in the fit. 

We use the Gaussian Process Regression package in scikit-learn \citep{scikit-learn}, adopting a Matern kernel with parameters specifying the variance, the length scale of influence, and the scatter (approximated by the mean standard deviation of the absorbers in the neutral fraction space). We include only absorbers whose models converged assuming a flat prior on $\log U$. To better constrain the fit at higher column densities, we include a few anchor points at $\log N_{\HI}>20.3$, for which we adopt $f_{\HI}=0.96\pm0.04$, consistent with DLAs being predominantly neutral gas. The resulting fit is shown in Figure~\ref{fig:neutral_frac}, alongside GP fits from \citetalias{lehner2022} and \citetalias{deepak2025} for the KODIAQ-Z and CCC surveys, respectively. Shaded regions show the 68\% confidence intervals for each GP model. We also plot a constant $\log N_{\rm H}=20.3$ line, and find that the BRIDGE absorbers show total hydrogen column densities similar to DLA \HI\, (or \H) column densities across the full $N_{\HI}$ range. We note that the high- and low-$z$ GP fits include neutral fractions for SLFSs, which are absent from our survey; this likely explains the larger uncertainties at low $N_{\HI}$ in our fit, where the absence of data leaves the model poorly constrained. 

\begin{figure}
\begin{centering}
\includegraphics[width=\columnwidth]{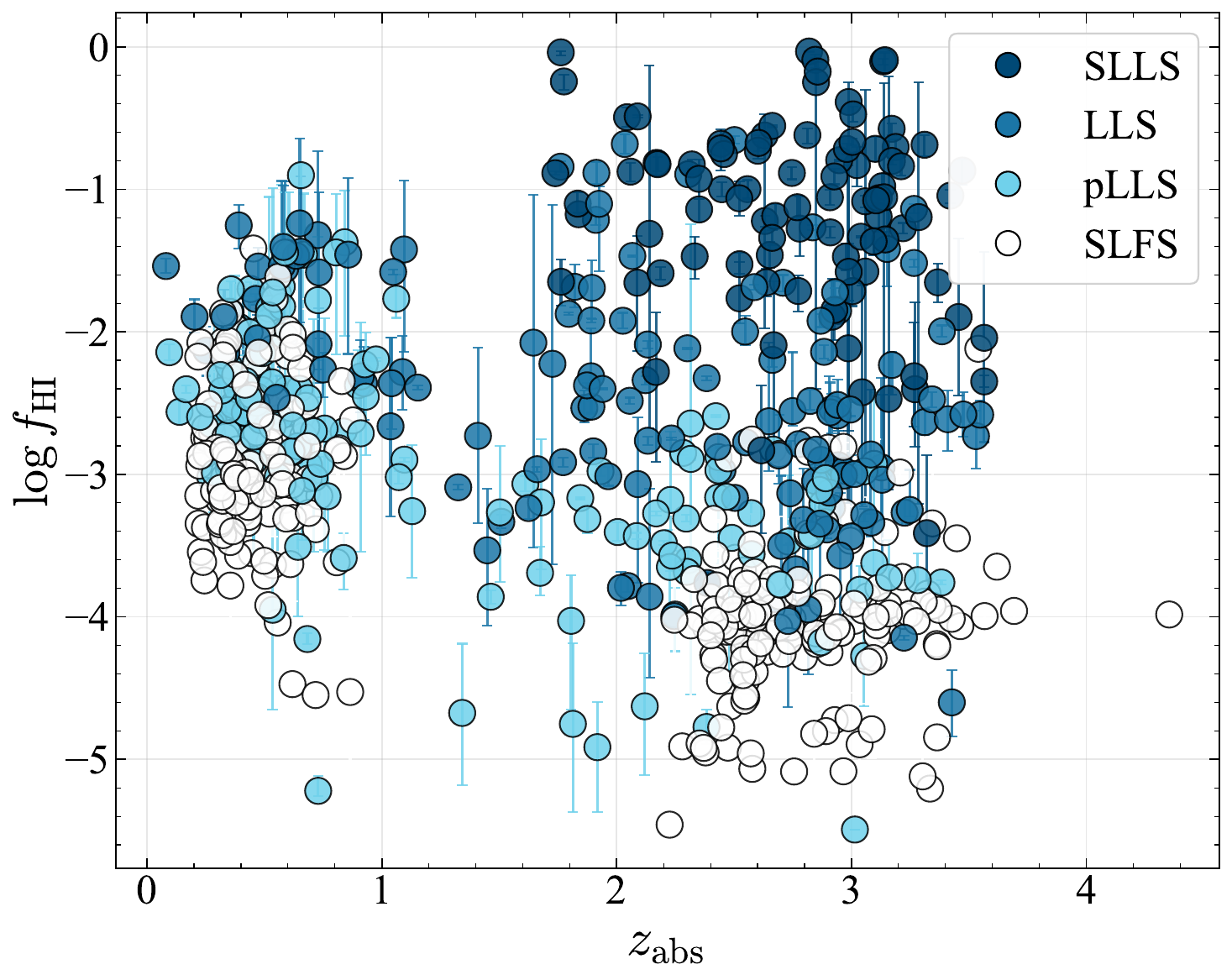}
\par\end{centering}
\begin{centering}
\caption{
{Neutral fraction $f_{\HI}$ as a function of redshift. Across all absorber types, the mean neutral fraction increases from high to low $z$. At Cosmic Noon, $z\sim1$--2.2, pLLSs have lower neutral fractions compared to both higher and lower redshifts.}
}\label{fig:fHIvz}
\par\end{centering}
\end{figure}

The overall $f_{\HI}$--$N_{\HI}$ trend at Cosmic Noon falls between those of the CCC ($z<1$) and KODIAQ-Z ($z>2.2$) surveys, though the slope is steeper than both comparison samples across the full $N_{\HI}$ range. Two effects may contribute to this. First, the UV background peaks near Cosmic Noon due to increased AGN activity, which would keep low-$N_{\HI}$ gas more highly-ionized (lower $f_{\HI}$) while having comparatively less effect on the denser, self-shielded high $N_{\HI}$ gas, naturally steepening the slope. This is shown in Figure~\ref{fig:fHIvz} where we plot the neutral fraction of absorbers as a function of redshift. At $z\sim1$--2.2, pLLSs have relatively lower neutral fractions, while LLSs and SLLSs show trends similar to $z>2.2$ absorbers, steepening the $f_{\HI}$--$N_{\HI}$ relation at Cosmic Noon. Second, the absence of SLFSs may cause the GP model to extrapolate into an unconstrained regime at low $N_{\HI}$, artificially steepening the fit. A better determination of the slope will be possible once SLFSs are included in the GP model. 

\section{Discussion}\label{sec:discussion}

\subsection{Enrichment Timescales in the CGM}\label{subsec:cdf}

\begin{figure}
\begin{centering}
\includegraphics[width=\columnwidth]{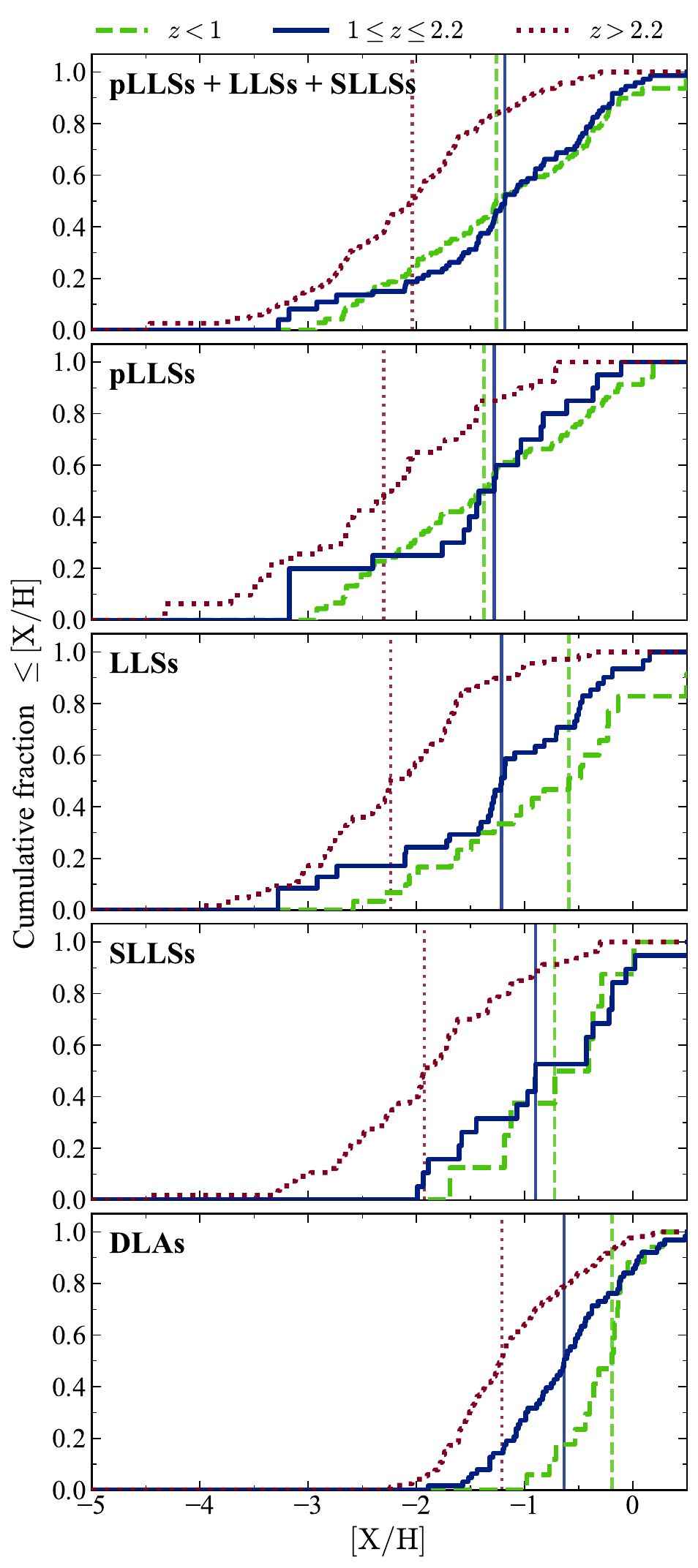}
\par\end{centering}
\begin{centering}
\caption{
{Cumulative distributions functions of metallicity for CCC ($z<1$; dashed green), BRIDGE ($1\le z\le 2.2$; solid blue), and KODIAQ-Z ($z>2.2$; dotted red). From top to bottom, we show CDFs for the pLLSs+LLSs+SLLSs sample, and the individual subsamples. Vertical lines mark the medians for each CDF. The CCC and BRIDGE distributions are statistically indistinguishable in all except LLSs and DLAs panels; all absorber types differ from KODIAQ-Z at high significance in every panel (see \S\ref{subsec:cdf}).}
}\label{fig:cdf}
\par\end{centering}
\end{figure}

One of the key questions we set out to answer in the BRIDGE survey was the timescales of enrichment of the ISM, CGM and IGM. We show in \S\ref{subsub:lbt} that the metallicity distribution of the ISM- and CGM-like gas evolved significantly over a relatively short period of $\sim1$ Gyr, starting around $z\sim2$. This is also seen Figure~\ref{fig:met_pdf}, where we see that the CCC and BRIDGE metallicity distributions are very similar in all except the LLS panel. Some metallicity estimates in each sample are upper and/or lower limits; to account for their non-uniform weighting at the tails of the metallicity PDFs, we estimate each sample's cumulative distribution function (CDF) using the Turnbull nonparametric maximum-likelihood estimator (TNMLE; \citealt{turnbull1976}), which handles data censored in both directions. The TNMLE reduces to the Kaplan–Meier estimator when only upper limits are present, and to the empirical distribution function when no censoring is present. Figure~\ref{fig:cdf} shows the estimated CDFs for all ionized absorbers (pLLSs+LLSs+SLLSs) and each absorber type, for the CCC (green), BRIDGE (blue), and KODIAQ-Z (red) samples, with dotted vertical lines marking the median of each distribution. The evolution in the CDFs for pLLSs and SLLSs between BRIDGE and CCC is discernible but marginal. The sample size for SLLSs is very small at $z<2.2$, so these trends are only suggestive. Most of the evolution in the metallicity distribution is instead driven by LLSs and DLAs at these redshifts. In all panels, we see significantly more evolution between KODIAQ-Z and BRIDGE/CCC than between BRIDGE and CCC.

To quantify the apparent redshift evolution of the metallicity distributions, we compare the samples using the Kolmogorov-Smirnov (KS; \citealt{hodges1958}) and Anderson-Darling (AD; \citealt{scholz1987}) tests on the estimated CDFs. The KS test is most sensitive near the distribution medians, whereas the AD statistic is more sensitive to differences in the tails. Since the tabulated null distributions for these statistics assume fully observed data and do not apply to censored estimates, so we obtain $p$-values by permutation: we randomly reassign sample labels (CCC, BRIDGE, or KODIAQ-Z) to absorbers 10,000 times with fixed sample sizes, recompute the estimator and statistic for each realization, and estimate the $p$-value as the fraction of permutations reaching or exceeding the observed statistic. The null hypothesis is that the two samples share both the same underlying metallicity distribution and the same censoring pattern. The CCC and BRIDGE metallicity CDFs are statistically indistinguishable in the combined absorber, pLLSs, and SLLS samples ($p\ge 0.23$), with median offsets below $0.18$ dex, while LLSs show a marginal difference (median offsets of 0.62 dex;  $p=0.038$ AD, $0.092$ for KS). Both lower-redshift samples differ from KODIAQ-Z at high significance in every subsample ($p\le0.022$), with median offsets across all absorber types between $0.79$--$1.65$ dex. The DLAs on the other hand show statistically significant evolution between KODIAQ-Z and BRIDGE as well as BRIDGE and CCC, with median offsets of $0.58$ dex and $0.44$ dex, respectively. 

The range of $N_{\HI}$ covered by CCC, BRIDGE, and KODIAQ-Z differs due to observational constraints (redshift, wavelength coverage, and instrument sensitivity). To eliminate this systematic, we repeat the statistical tests on restricted samples, defining pLLSs as $16.6\le\log N_{\HI}<17.2$ and LLSs as $17.2\le\log N_{\HI}\le18.2$, the ranges set by BRIDGE and CCC coverages, respectively. The observed trends remain unchanged, with larger median differences between the CCC/BRIDGE and KODIAQ-Z samples ($0.85$--$2.11$ dex). LLSs again stand out as the only population differing between CCC and BRIDGE, with a median offset of $0.72$ dex (although at $p=0.08$ for AD, this is marginal). 

The metallicity evolution, therefore, occurs predominantly between BRIDGE and KODIAQ-Z. Between BRIDGE and CCC, LLSs (among the ionized absorbers) are the only population showing evidence of evolution. This suggests that Cosmic Noon marks a transitional phase in which the enrichment of low-$N_{\HI}$ CGM-like gas is largely complete: pLLSs reach their present-day metallicity distribution by $1\le z\le2.2$, whereas the denser gas traced by LLSs continues to be enriched down to $z<1$. We note that the SLLSs show no comparable low-redshift evolution. This is unlikely to be physical: DLAs show continued enrichment down to $z<1$, with the mean DLA metallicity increasing by $0.24$ dex between $z\sim1.6$ and $z<1$ \citepalias{peroux2021}. The CCC SLLS sample contains only 8 systems, for which the two-sample test has little statistical power, so we regard the null SLLS result as a limitation of the sample rather than evidence against evolution.

\begin{figure}
\begin{centering}
\includegraphics[width=\columnwidth]{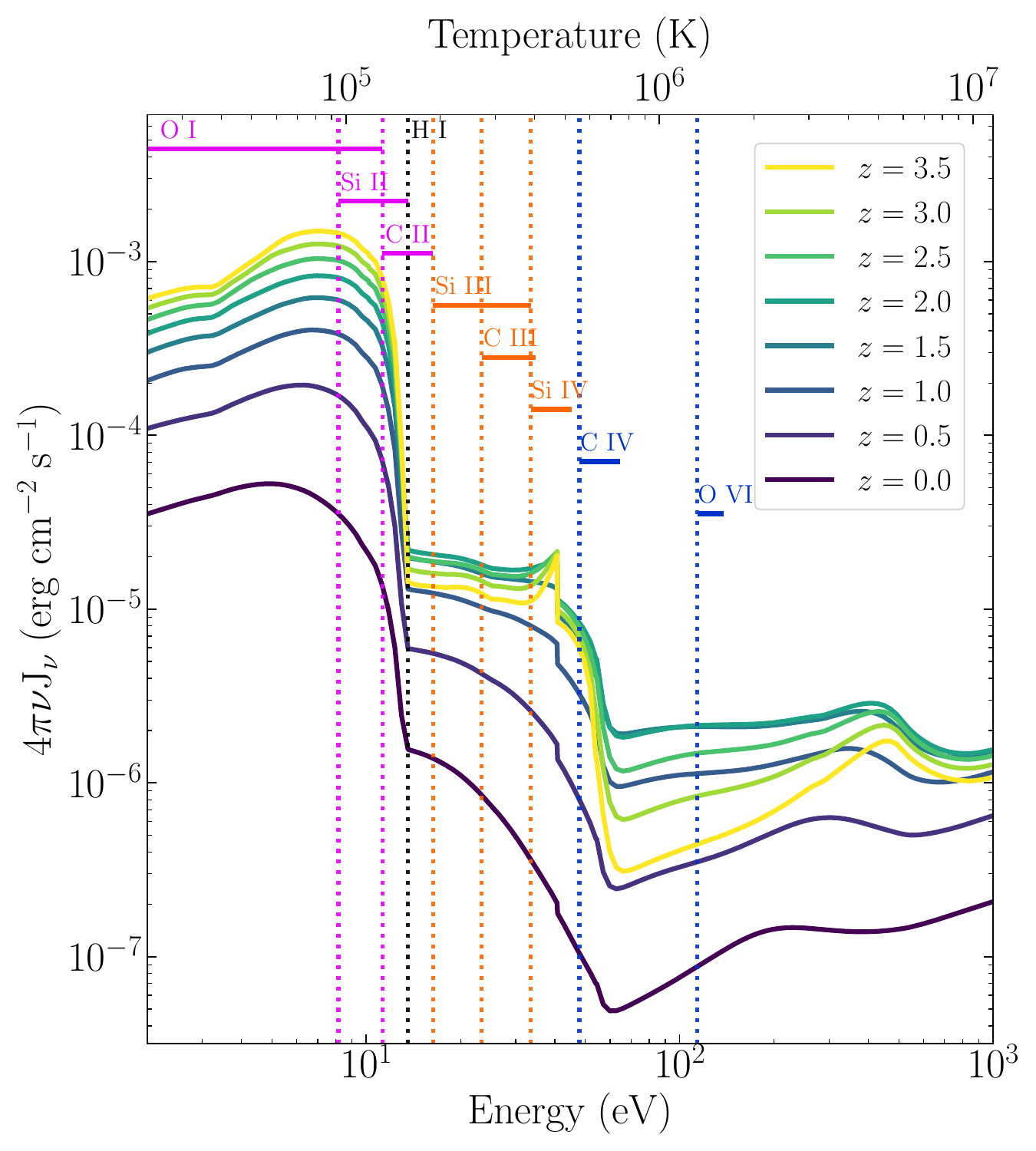}
\par\end{centering}
\begin{centering}
\caption{
{The evolution of the spectral energy distributions for the \citetalias{hm05} UV background. The key ions used to constrain metallicities are shown, with dotted vertical lines marking the minimum ionizing energy and horizontal bars showing their range of ionization energy.}
}\label{fig:UVB}
\par\end{centering}
\end{figure}

\begin{figure}
\begin{centering}
\includegraphics[width=\columnwidth]{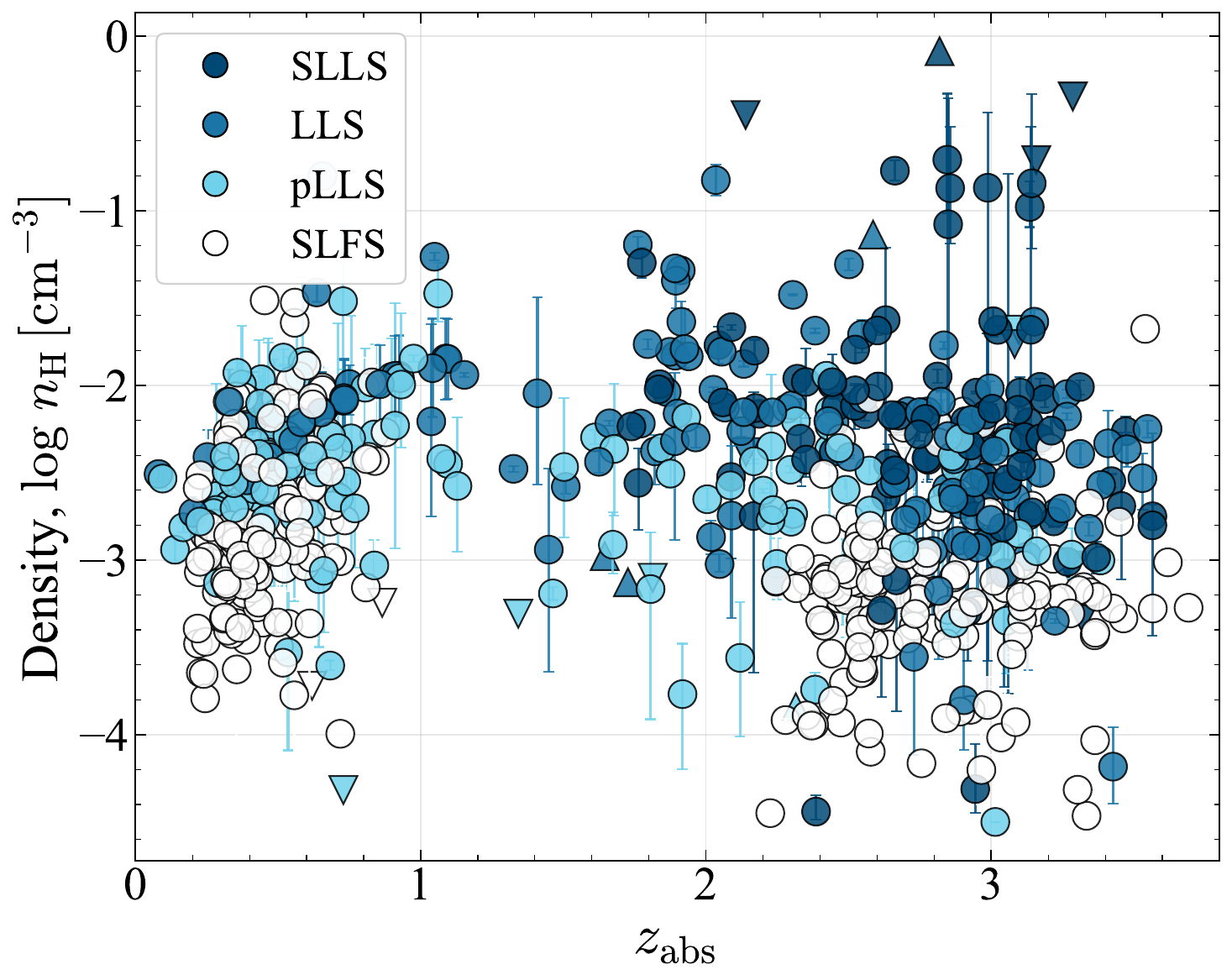}
\par\end{centering}
\caption{ The hydrogen number density of gas, $n_{\rm{H}}$, as a function of redshift. The different absorber types are color-coded. We note that densities are estimated via ionization modeling and scale with the intensity of the radiation field.}
\label{fig:logU_dens}
\end{figure} 

\begin{figure}
\begin{centering}
\includegraphics[width=\columnwidth]{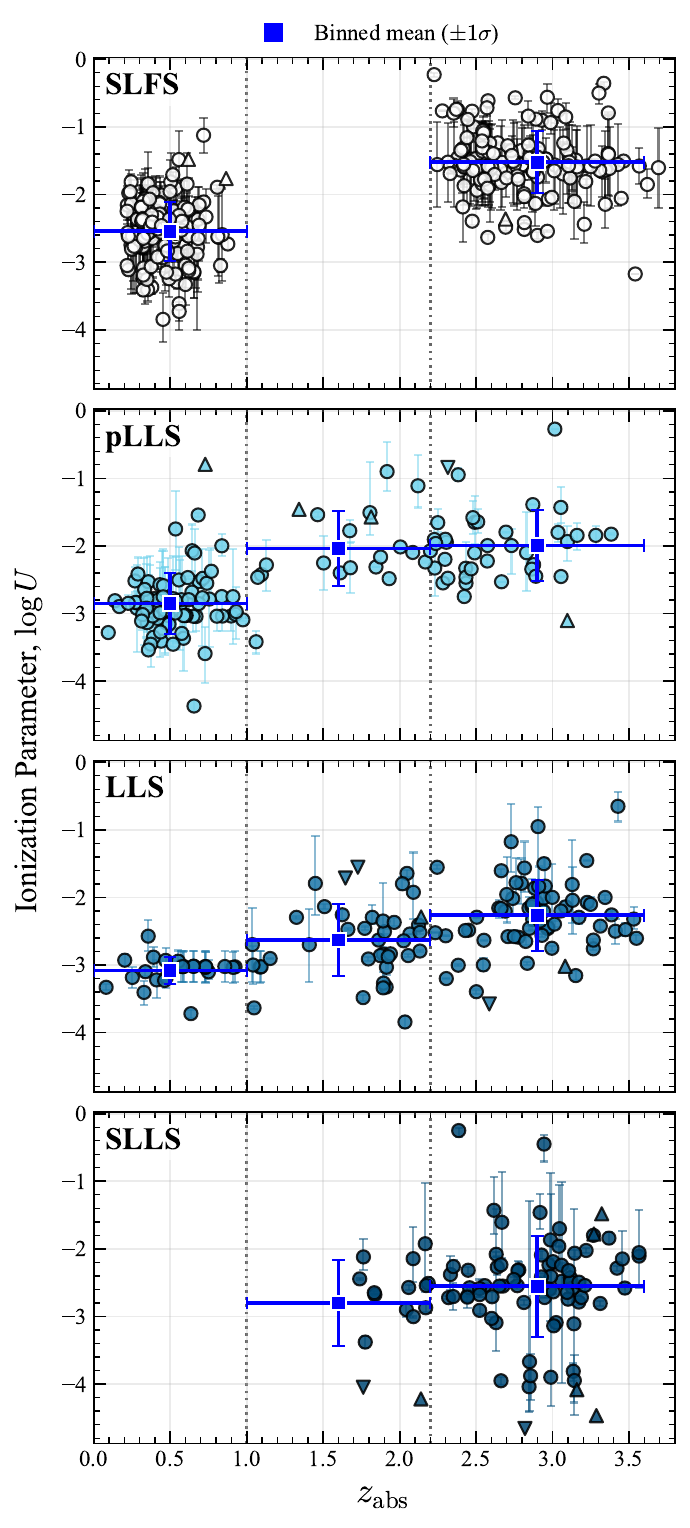}
\par\end{centering}
\begin{centering}
\caption{
{Ionization parameter as a function of $z_{\rm{abs}}$ for each absorber type. The blue squares represent the redshift-binned mean ionization parameter ($\langle\log U\rangle$) and standard deviation. The blue horizontal bars represent the redshift covered by each redshift sample (CCC, BRIDGE, KODIAQ-Z), which are also demarcated by dotted gray lines. For SLLSs and SLFSs, we do not have $\log U$ in the CCC and BRIDGE samples, respectively. For each absorber class, $\log U$ decreases with $z$.}
}\label{fig:logUzstat}
\par\end{centering}
\end{figure}

\subsection{Evolution of the Ionization Properties of Cool, Diffuse Gas}\label{subsec:UVB}

The physical properties of the cool, diffuse, CGM-like gas (neutral fraction, ionization parameter) are driven by the spectral shape and intensity of the extragalactic UV background illuminating it. Figure~\ref{fig:UVB} shows the \citetalias{hm05} EUVB spectral energy distribution at several redshifts spanning the CCC, BRIDGE, and KODIAQ-Z samples, along with some of the key ions used to constrain our photoionization models. The background evolves substantially both in normalization and shape over $0\le z\le3.5$: intensity increases from low to high $z$, with most of the distribution (energy $\gtrsim11$ eV), including the \HI\, ionizing radiation at 13.6 eV, peaking near $z\sim1.5$--2.5, before declining toward higher redshift. This peak tracks the evolution in the cosmic star formation and AGN activity. 

The evolution in the EUVB informs the neutral fraction trends in \S\ref{subsub:neut_frac}. A harder, more intense EUVB during Cosmic Noon photoionizes and heats low-density gas, consistent with the lower $f_{\HI}$ found in pLLSs (and SLFSs in CCC and KODIAQ-Z). Denser, self-shielded gas in LLSs and SLLSs, shows a more gradual evolution. To understand the combined effect of the evolving EUVB and gas density on the ionization state of the gas, we examine the evolution of $n_{\rm H}$ in Figure~\ref{fig:logU_dens} and the ionization parameter, $U$, in Figure~\ref{fig:logUzstat}, color-coding absorbers by their \HI\ column densities. Densities increase with decreasing redshift across all absorber types, as expected from large-scale structure formation. At all redshifts, SLLSs and LLSs probe higher densities compared to pLLSs and SLFSs. For all absorber types, the ionization parameter decreases from high to low $z$. \\
\input{table4_meanlogU}
We tabulate the mean $\log U$ for each absorber class in Table~\ref{tab:meanlogU}. Between KODIAQ-Z and BRIDGE, $\overline{\log U}$ decreases by a factor of 1.2 (0.09 dex) for SLLSs and 3 (0.48 dex) for LLSs, but stays nearly constant for pLLSs. Between BRIDGE and CCC, $\overline{\log U}$ decreases further, by a factor of $\sim6$ (0.80 dex) for pLLSs and $2.5$ (0.40 dex) for LLSs. BRIDGE lacks SLFSs; comparing KODIAQ-Z to CCC directly, $\overline{\log U}$ for SLFSs decreases by a factor of $\sim11$ (1.04 dex). This behavior suggests that SLFSs evolve similarly to pLLSs, with the ionization parameters for these lower-$N_{\HI}$ absorbers remaining nearly constant from $z\sim3$ to Cosmic Noon---as the rising photon density, $n_{\gamma}$, is offset by the concurrent rise in $n_{\rm{H}}$---before dropping more steeply toward $z<1$. The higher-$N_{\HI}$ absorbers are less affected by the increase in $n_{\gamma}$ due to self-shielding, so the rise in $n_{\rm{H}}$ dominates their $\log U$ trend. 
To quantify this trend, we compute the Spearman rank-order coefficient ($r$) between $\log U$ and $z_{\rm{abs}}$ for each absorber type in Figure~\ref{fig:logUzstat}. We find strong, statistically significant positive correlations with redshift for SLFSs ($r=0.65$, $p<0.0001$), pLLSs ($r=0.69$, $p<0.0001$), and LLSs ($r=0.62$, $p<0.0001$). For SLLSs, we do not have $\log U$ values at $z<1$, and observe no significant evolution between $1<z<3.6$ ($r=0.13$, $p=0.22$). 
We find that at any redshift, denser gas is the least ionized and shows the smallest evolution among all absorber types (due to the evolving EUVB), consistent with increased self-shielding.

\begin{figure}
\begin{centering}
\includegraphics[width=\columnwidth]{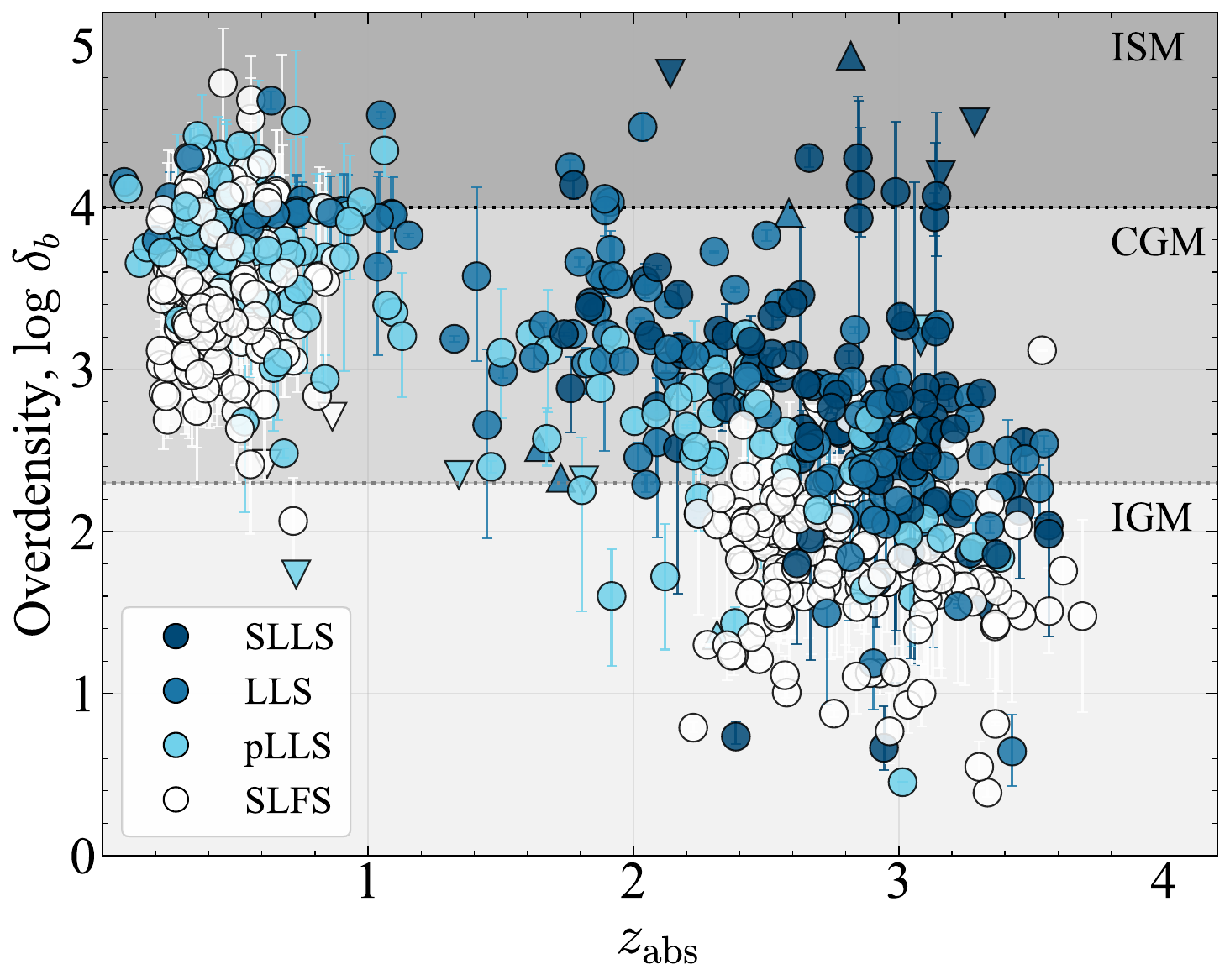}
\par\end{centering}
\begin{centering}
\caption{
{Overdensity as a function of absorber redshift. We color-code absorbers by their $N_{\HI}$. Dotted lines at $\delta_b=200$ and $\delta_b=10^4$ mark transitions from the IGM to CGM, and CGM to ISM, respectively.}
}\label{fig:bov_z}
\par\end{centering}
\end{figure}

\subsection{Overdensity and the Physical Origin of ``CGM-like" Gas}\label{subsec:bov_origin}

\begin{figure*}
\begin{centering}
\begin{minipage}{0.48\textwidth}
\begin{centering}
\includegraphics[width=\columnwidth]{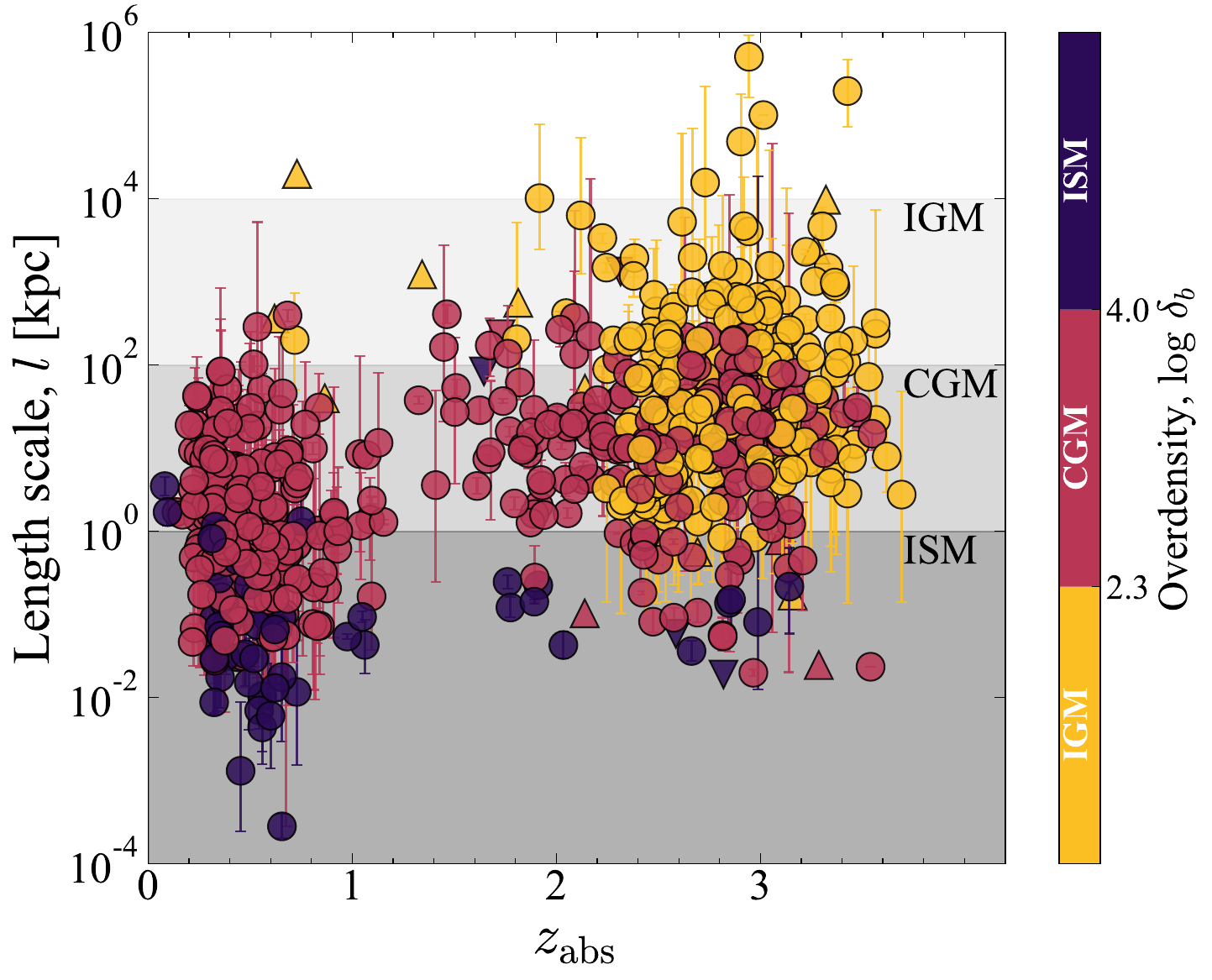}
\par\end{centering}
\end{minipage}
\hfill
\begin{minipage}{0.48\textwidth}
\begin{centering}
\includegraphics[width=\columnwidth]{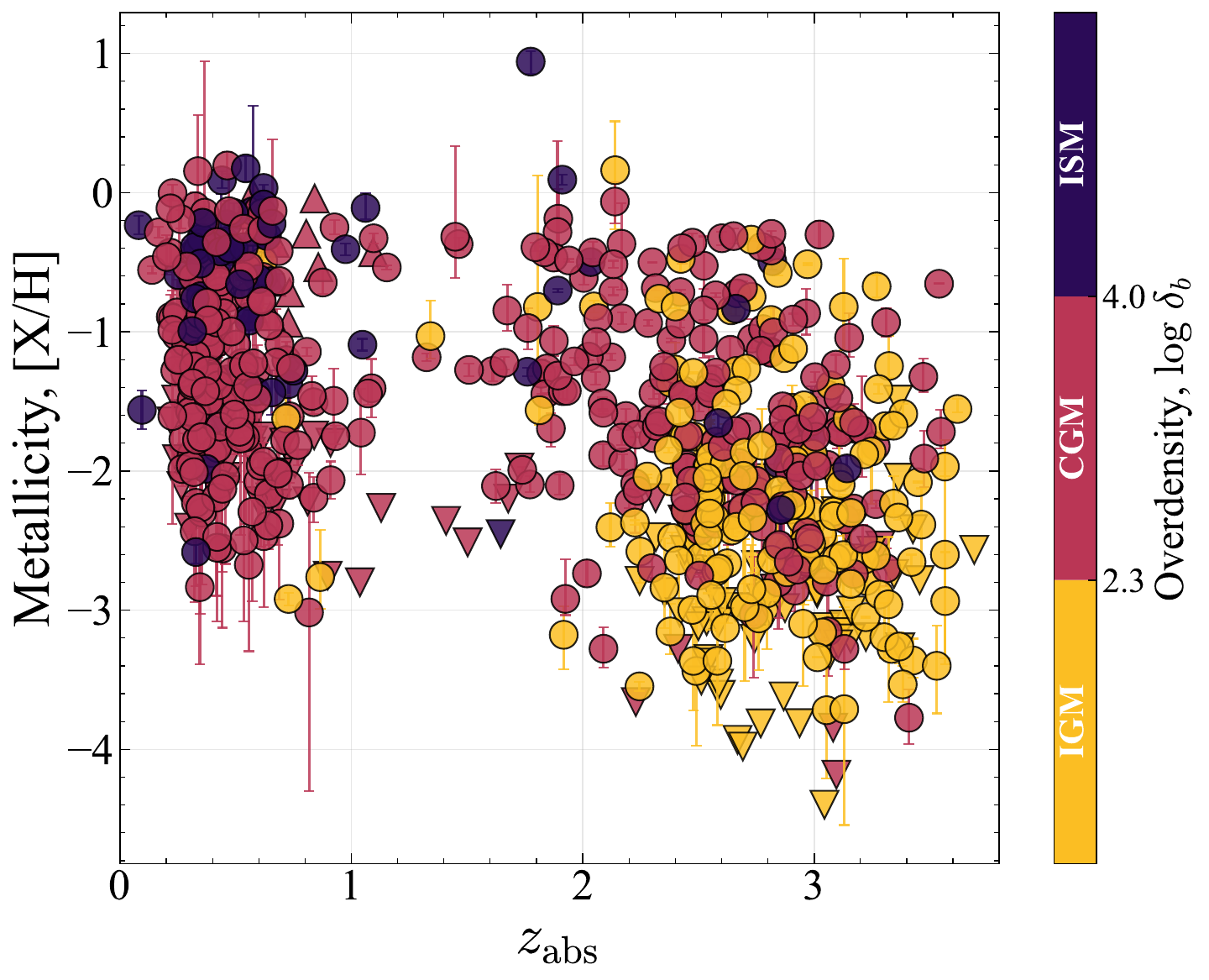}
\par\end{centering}
\end{minipage}
\par\end{centering}
\caption{
\textit{Left panel:} The line-of-sight length scale $l$ of absorbers as a function of redshift for the CCC, BRIDGE, and KODIAQ-Z surveys. We color-code absorbers by their overdensity $\delta_b$. We also shade regions with length-scales characteristic of ISM-, CGM-, and IGM-like clouds. 
\textit{Right panel:} Metallicity as a function of redshift, color-coded by $\delta_b$.
}\label{fig:color_code_bov}
\end{figure*} 

The baryonic overdensity $\delta_b$ provides a redshift-independent measure of the absorber density relative to the cosmic mean baryon density at that epoch. This is important because a given $N_{\HI}$ traces gas of different overdensity (and thus physical structures) at different redshifts: at higher redshift, the same $N_{\HI}$ corresponds to lower $\delta_b$. Motivated by simulations, we define gas with $\delta_b<200$ as IGM-like,  $200\lesssim\delta_b<10^4$ as CGM-like, and $\delta_b>10^4$ as ISM-like \citep{schaye2001a,schaye2001b,ferriere2001,klessen2016,mcquinn2016,tumlinson2017}. We compute $\delta_b$ for each absorber using the hydrogen density, $n_{\rm H}$, derived from our ionization models and the mean cosmic hydrogen number density $\overline{n_{\rm{H}}}$ at the absorber's redshift \citep{schaye2001a}. 

Figure~\ref{fig:bov_z} shows $\delta_b$ as a function of redshift, color-coded by $N_{\HI}$ class, with shaded regions marking the ISM-, CGM- or IGM-like overdensity ranges. For all absorber classes, overdensity increases from high to low $z$, and more steeply than $n_{\rm H}$, since $\overline{n_{\rm H}}$ is simultaneously decreasing. We attribute this trend to structure formation and the progressive collapse of gas into denser structures. Notably, we do not find many absorbers with metal detections in the IGM-like regime at low $z$, consistent with large-scale structure funneling most of the metal budget into collapsed, overdense gas \citepalias{deepak2025}. A given absorber class does not occupy a fixed overdensity regime across cosmic time: at $z>2.2$, SLLSs lie predominantly in the CGM regime (with some in the ISM-like regime), while SLFSs trace IGM-like gas. At $z\lesssim1$, SLLSs are more ISM-like and SLFSs enter the CGM-like regime; pLLSs and LLSs show some overlap with IGM and ISM regimes, but are predominantly CGM-like at all redshifts. Most of the BRIDGE survey---pLLSs, LLSs, and SLLSs at Cosmic Noon---falls in the CGM band. 

The physical scale and metallicity of an absorber offer give key insights into the structures it traces. To examine how these properties evolve over cosmic time and physical environment, Figure~\ref{fig:color_code_bov} shows the line-of-sight length scale (left panel) and metallicity (right panel) as functions of redshift, color-coded by $\delta_b$. Also shown in the left panel are shaded regions representing characteristic length scales for ISM ($\lesssim1$ kpc), CGM ($\sim1$--$10^{2}$ kpc), and IGM ($\sim10^2$--$10^4$ kpc) clouds. We emphasize that the derived length scales carry significant uncertainty from the assumptions in the photoionization modeling, but remain useful for identifying statistical trends across redshift and overdensity. We also note that the length scale and overdensity are not independent quantities: both are derived from $n_{\H}$, and we may express $l=N_{\H}/(\overline{n_{\H}}\delta_b)$. Since $N_{\H}$ remains nearly constant for all absorber types at a given redshift, the trends between $\log l$ and $\delta_b$ are expected from definitions. 

As shown in the left panel of Figure~\ref{fig:color_code_bov}, at $z>2$ the length scales probed by a given overdensity are diverse. Absorbers with $\delta_b<200$ typically span $l\sim100$--1000 kpc and may trace diffuse IGM-like gas in filaments or outer regions of intra-group and intra-cluster media \citep{mackenzie2019,manuwal2019}. Absorbers with $\log\delta_b\sim2.5$--3 typically span $l\sim10$--100 kpc, consistent with CGM scales, while absorbers with even higher $\delta_b$ span $l\sim0.01$--1 kpc. Denser absorbers therefore generally probe smaller clouds. At $z<2$, nearly all absorbers have $l<100$ kpc, with the densest tracing structures as compact as 10 pc. 

The lowest-overdensity, largest-scale filamentary absorbers on average the most metal-poor, as shown in the right panel of Figure~\ref{fig:color_code_bov}. Several of these have only upper limits on metallicity, since no metal ions were detected. Such absorbers are more common in the KODIAQ-Z sample ($z>2.2$). Even so, gas at the lower end of CGM-like $\delta_b$ spans a wide range of metallicities. At $z<2.2$, no clear trend emerges: absorbers with relatively high overdensity ($\log\delta_b\sim4$) can be very metal poor ($[\rm{X/H}]<-2.4$) while others with $\log\delta_b\sim2$ reach high metallicities ($[\rm{X/H}]>-1$), reflecting a large range in metallicities traced by both CGM- and IGM-like gas at Cosmic Noon. At $z<1$, both the mean overdensity and the mean metallicity of the sample are higher, suggesting that almost all gas has been enriched. 

\subsection{Comparison with Galaxy Surveys}\label{subsec:compare_lit}

As we attempt to interpret our results in the context of galactic environment, it is useful to compare our \HI-selected sample to galaxy surveys. We compare our metallicities and overdensities to the \HI-selected, galaxy-associated absorbers in the BASIC survey \citep{berg2023}, which characterizes the associated galaxies at $z<1$, using a sample of 19 pLLSs and LLSs drawn from the CCC sample. \citeauthor{berg2023} found that pLLSs and LLSs with $[\rm{X/H}]<-1.4$ have a probability of $0.39$ of being associated with a $\log (M_*/M_{\odot})\ge9$ galaxy (within $\rho/R_{\rm vir}\le1.5$), while higher-metallicity absorbers have a probability of $0.78$. In our sample, $\sim60\%$ of all $z<2.2$ (CCC+BRIDGE, including lower limits) absorbers have $[\rm{X/H}]\ge-1.4$. 
These higher-metallicity absorbers are also more overdense, and correspondingly probe shorter length scales (Figure~\ref{fig:color_code_bov}). Both are expected if these absorbers are associated with galaxies and trace enriched outflowing or recycling gas.This is supported by surveys of galactic outflows at Cosmic Noon \citep{rupke2018,rudie2019,pointon2019,thompson2024}. For example, \cite{liou2026} studied 98 galaxies at $z=2$--3 with outflow signatures from the Keck Baryonic Structure Survey (KBSS; \citealt{steidel2014,strom2017}). They find that up to 30\% of ionized absorbers with $T\sim10^4$ K may trace galactic winds.

The lower-metallicity absorbers ($[\rm{X/H}]<-1.4$) are harder to place in a galactic context. \cite{berg2023} distinguish two populations of metal-poor absorbers: those associated with galaxy halos ($\delta_b=100$--$10^4$), and those not associated with galaxies, which lie at lower overdensities ($\delta_b=10$--100). Almost all of our low-metallicity absorbers fall in the former range. Only 2\% of the CCC+BRIDGE sample are both low-overdensity and low-metallicity, and this conclusion is unchanged whether we adopt their boundary at $\delta_b=100$ or our own at $\delta_b=200$. We illustrate this in the right panel of Figure~\ref{fig:color_code_bov}: absorbers with $\delta_b<200$ and metal detections are found almost exclusively at $z\gtrsim2.2$, with very few remaining in this IGM-like regime by $z<1$, as $\delta_b$ increases with cosmic time (\S\ref{subsec:bov_origin}). Using their absorbers not associated with galaxies, \cite{berg2023} further place an upper limit on the unweighted geometric mean metallicity of the IGM, $[\rm{X/H}]<-2.1$. About 23\% of CCC and BRIDGE absorbers fall below this threshold, but given their higher overdensities, they are more likely to trace low-metallicity CGM-like gas than IGM-like gas.

At $z\sim2$, where we do have IGM-like absorbers, galaxy searches around metal-poor absorbers are rare, since the faint Ly$\alpha$ emitters used to trace these environments are difficult to detect at blue wavelengths. At higher redshift, the picture is mixed. \cite{fumagalli2016b} found no galaxies associated with a metal-poor LLS at $z\sim3.2$ ($[\rm{X/H}]=-2.90$; \citealt{robert2022}), yet found five Ly$\alpha$ emitters coincident with an even more metal-poor LLS ($[\rm{X/H}]<-3.8$) at $z\sim3.1$. The first is likely probing the IGM, while the second more plausibly traces near-pristine filamentary gas feeding one or more galaxies (see also \citealt{lofthouse2020}). That the more metal-poor of the two is the one with galaxies makes the point: neither metallicity nor overdensity alone identifies an absorber's origin, though the two together are informative. A definitive classification at Cosmic Noon will require galaxy--absorber association surveys like BASIC, which have yet to be carried out at these redshifts.

\begin{figure}
\begin{centering}
\includegraphics[width=\columnwidth]{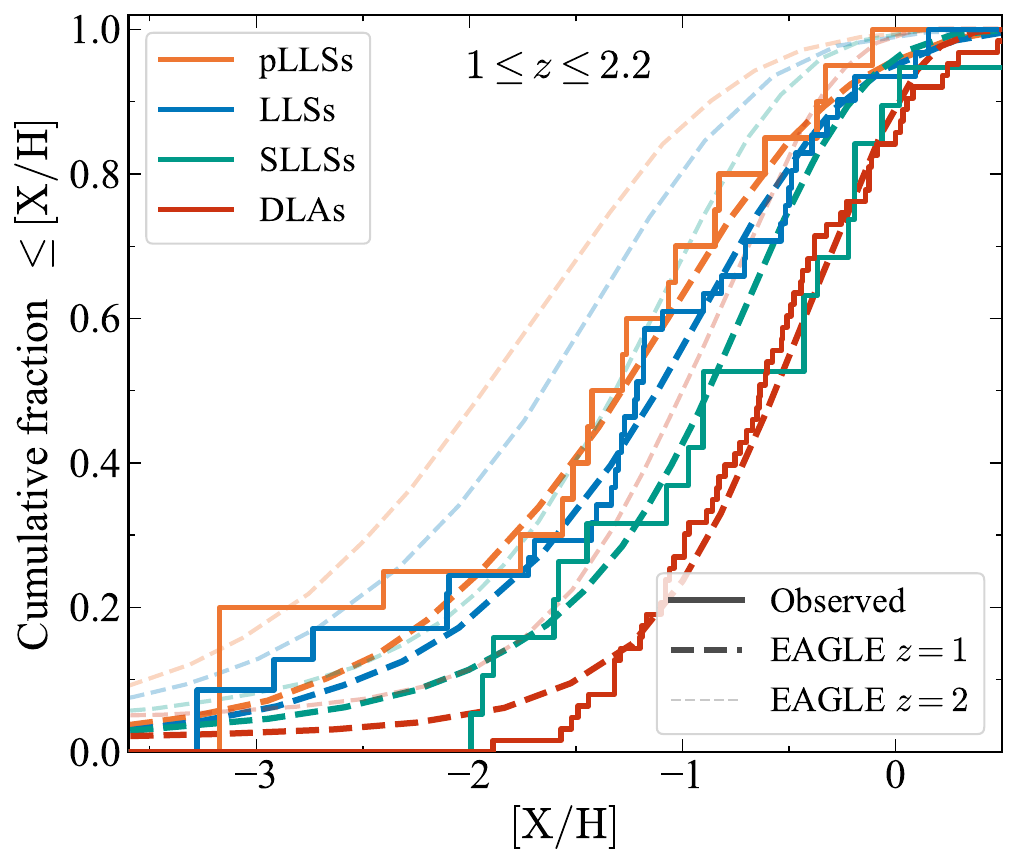}
\par\end{centering}
\begin{centering}
\caption{
{The observed CDFs (solid lines) for different absorber types in the BRIDGE samples, and the predicted CDFs (dashed lines) from the EAGLE simulation \citep{rahmati2018} at $z=1.0$ and 2.0 (transparent).}
}\label{fig:eagle}
\par\end{centering}
\end{figure}

\subsection{Comparison with Simulations}\label{subsec:compare_sim}

We show in \S\ref{subsec:bov_origin} that metallicity alone is a poor predictor of gas origin in observations \citep{hafen2017,weng2024}. Cosmological simulations offer a unique opportunity to disentangle origin and metallicity of gas through particle tracking. It is therefore useful to compare global metallicity surveys with mock absorbers drawn from cosmological simulations. Such comparisons can help test feedback models, especially at Cosmic Noon, the peak of AGN activity \citep{strawn2024}. 

In Figure~\ref{fig:eagle}, we compare the estimated metallicity CDFs for each absorber type in the BRIDGE sample with predictions from the EAGLE (Evolution and Assembly of GaLaxies and their Environments) simulation \citep{springel2005,schaye2015,crain2015}. The simulation CDFs are drawn from \cite{rahmati2018}, who use the higher-resolution version of the simulation, \textit{Recal}-L025N0752 EAGLE HiRes, which includes subgrid prescriptions for radiative cooling, star formation, stellar and chemical enrichment, and superwind feedback associated with star formation, features essential for comparison at Cosmic Noon. \cite{rahmati2018} also use the HM05 EUVB, which removes that systematic from the comparison with our sample. 

The CDFs for our sample at $1\le z\le 2.2$ are more consistent with the EAGLE predictions at $z=1$ than at $z=2$, despite the mean redshift of BRIDGE being $\langle z_{\rm abs}\rangle=1.81\pm0.31$; that is, the simulations predict delayed enrichment relative to the observations. A similar trend was noted by \cite{wotta2019} (their Figure 21 and 22), who compared the CCC sample with EAGLE and found that simulations predict strong metallicity evolution between $z\le0.3$ and $z>0.6$, a trend not seen in observations. For both CCC and BRIDGE samples, the EAGLE simulation underproduces low-metallicity pLLSs ($[\rm{X/H}]<-1.4$ for CCC, $[\rm{X/H}]<-2.0$ for BRIDGE). The simulation overproduces low-metallicity SLLSs and DLAs ($[\rm{X/H}]<-2.0$) even at $z=1$, which, as noted in \S\ref{subsub:lbt}, are absent from the BRIDGE sample below $z=2.2$. The higher-metallicity ends of the pLLS and LLS CDFs agree well with our empirical CDFs, whereas high-metallicity SLLSs and DLAs are overproduced. \cite{wotta2019} attribute this over-abundance of high-metallicity denser gas, along with the underprediction of low-metallicity, more diffuse absorbers, to too strong feedback in the simulations that can cause over-mixing. We note the same trend at Cosmic Noon. The EAGLE simulation, therefore, predicts stronger redshift evolution and adopts more aggressive feedback than is needed to reproduce the observed CDFs. 

Disagreements between simulations and observed metallicity trends may also arise from modeling assumptions and limited resolution. Using the Feedback in Realistic Environments (FIRE) simulations \cite{hafen2019} show that simulated CGM metallicities depend heavily on the subgrid metal diffusion and the halo mass of the central galaxy. Using the TNG50 hydrodynamical simulation, \cite{weng2024} find that the stellar mass of the galaxy is similarly important. Notably, they find that inflowing and outflowing gas differ in mean metallicity by only $\lesssim0.5$ dex at $z=0.5$, an offset that decreases further for higher stellar mass central galaxies, consistent with FIRE \citep{muratov2017}. \cite{wotta2019} also compared CCC CDFs with FIRE and found several inconsistencies driven by strong feedback and selection effects (simulated absorbers are identified along sightlines through individual halos biasing metallicity PDFs). Together, these results suggest that metallicity-based origin classification, in both simulations and observations, is inherently limited by the mixed, multi-origin nature of CGM- and IGM-like gas.

\subsection{Future directions}\label{subsec:future}

This paper (BRIDGE-I) focuses on CGM-like gas ($16.4\le\log N_{\HI}<20.3$) at Cosmic Noon ($1\le z\le2.2$). The second paper in this series (BRIDGE-II) will extend this work to the lower-$N_{\HI}$, IGM-like population ($\log N_{\HI}<16.2$): SLFSs and the Ly$\alpha$ forest. As shown in Figure~\ref{fig:bov_z}, SLFSs occupy a shifting position in the overdensity space across cosmic time, tracing IGM-like gas at $z>2.2$, moving into the CGM-like regime at $z<1$; at Cosmic Noon, these likely span both regimes. A dedicated metallicity survey of this population at Cosmic Noon is therefore essential for completing the picture across the full range of gas phases at this epoch. This regime presents distinct observational challenges, including weaker metal-line absorption and increased line blending in the Ly$\alpha$ forest. 

Together, BRIDGE-I and II, combined with surveys at high redshift (KODIAQ-Z) and low redshift (CCC) will provide the first comprehensive \HI-selected metallicity survey spanning $z=0$--3.6. This survey will cover the full range of cool diffuse gas, from ISM-like to IGM-like, enabling a more complete accounting of the baryon and metal budget presented in \citetalias{deepak2025}. While BRIDGE-I contains no pristine CGM-like absorbers, pristine IGM-like absorbers may still exist at Cosmic Noon. The low-$N_{\HI}$ sample from BRIDGE-II will let us test this directly and constrain the enrichment timescale of IGM-like gas. In this work, we find that the large-scale collapse of baryons together with continued deposition of metals in dense gas, drives the evolution of the global metallicity distribution of CGM-like gas. Because the IGM is more decoupled from galactic environments than the CGM, it offers a cleaner probe of large-scale enrichment via outflows. Simulations suggest that enriched outflows are diluted by the time they reach the CGM, more so in massive halos. Estimating the metallicity of IGM-like gas at Cosmic Noon, the epoch of peak feedback activity, will help constrain the extent to which metals are diluted or transported beyond the CGM into the IGM.

\section{Summary and Conclusions}\label{sec:summary}

In this paper, the first in the BRIDGE series, we use archival UV spectra from HST combined with archival high-resolution, 56 optical quasar spectra from Keck/HIRES and VLT/UVES to construct a sample of \numberabs\, \HI-selected absorbers with $16.4<\log N_{\HI}<20.3$ (40 pLLSs, 45 LLSs, 20 SLLSs) at Cosmic Noon. In this regime, absorbers trace cool, diffuse, ionized CGM-like gas with overdensities $200<\delta_b\lesssim10^4$. Our sample is selected ``blindly" in terms of metallicity: quasar sightlines were not pre-selected based on the presence of metal absorbers or proximity to a galaxy, enabling an unbiased metallicity survey. The high resolution and signal-to-noise of the archival Keck/HIRES and VLT/UVES spectra allow us to sensitively probe (very) metal-poor gas, with typical detection limits (or $2\sigma$ upper limits) reaching $[\rm{X/H}]\lesssim-3.5$. The sample is drawn primarily from \citetalias{ribaudo2011} and \citetalias{omeara2013}, supplemented with SLLSs and DLAs from the literature (\citealt{prochaska2015,decia2018}; \citetalias{peroux2021}). For each absorber, we estimate metallicity, ionization parameter, density, $[\rm{C/\alpha}]$ ratio, neutral fraction, total hydrogen column density, and line-of-sight length scale using a Bayesian MCMC framework applied to a grid of Cloudy photoionization models \citepalias{wotta2019,lehner2022}. We compare the metallicity distribution and physical properties for our sample at Cosmic Noon ($1\le z\le2.2$) to the CCC ($z<1$) and KODIAQ-Z ($2.2<z\le3.6$) samples. Our main results are summarized below.
\begin{enumerate}
    \item For ``CGM-like" absorbers at Cosmic Noon, we find that several low (\OI, \MgI, \MgII, \CII, \SiII, \AlII, \FeII, \SII, \ZnII), intermediate (\CIII, \SiIII, \AlIII, \FeIII), and high ions (\CIV\, and \SiIV) probe the same gas phase. In all cases except one (see \S\ref{sec:metal_Cols}), \OVI\, and \NV\, have broader absorption profiles, more characteristic of collisional ionization, and are excluded from our photoionization models. Metal ion absorption typically lies within $\pm250$ \kms, which we adopt as our default absorber definition. We estimate column densities for 88 absorbers and adopt estimates for 17 from the literature (see \S\ref{subsec:lit_sample}). 
    \item We quantify two systematic uncertainties in estimated metallicities: the assumed EUVB, and the inclusion of high ions in the models. We find that assuming the HM12 \citep{hm12} EUVB instead of our fiducial HM05 \citep{hm05} results in systematically higher metallicities, a trend also observed at high and low $z$. The mean difference in metallicities is $\langle[\rm{X/H}]_{\rm{HM12}}-[\rm{X/H}]_{\rm{HM05}}\rangle=0.16\pm0.13$ at Cosmic Noon (compared to $0.37\pm0.19$ for CCC and $0.10\pm0.17$ for KODIAQ-Z). Including high ions (\CIV, \SiIV) leaves metallicities largely unchanged but yields systematically lower densities and higher length scales (mean offset of $0.3$ dex and $0.7$ dex, respectively).
    \item For 23 BRIDGE absorbers, the $[\rm{C}/\alpha]$ ratio is well-constrained with a mean of $0.08\pm0.35$. We find a significant positive correlation between $[\rm{C}/\alpha]$ and $[\rm{X/H}]$ and a marginal positive correlation between $[\rm{C}/\alpha]$ and $f_{\HI}$ at Cosmic Noon. Both correlations are stronger in the higher-$z$ KODIAQ-Z sample, which also shows a positive correlation with $N_{\HI}$. We argue that the $N_{\HI}$/$f_{\HI}$ trends are likely driven by large ionization corrections among low-$N_{\HI}$ KODIAQ-Z absorbers, while the $[\rm{C}/\alpha]$--$[\rm{X/H}]$ correlation, which are more robust to ionization effects, may have nucleosynthetic origins.
    \item The BRIDGE metallicity PDF is statistically indistinguishable from CCC, and the marginal evolution between the two is dominated by LLSs. The main qualitative differences in the two samples are the persistence of extremely metal-poor pLLSs and LLSs at Cosmic Noon, which are absent in CCC. By contrast, we find a significant evolution between KODIAQ-Z and BRIDGE over a period $\sim1$ Gyr at $z\sim2$. This suggests that most of the metal enrichment of the CGM-like gas is complete by $z\sim1$, and peaks during Cosmic Noon.
    \item A ``metallicity floor" of $[\rm{X/H}]\sim-2.0$ emerges at Cosmic Noon for high-$N_{\HI}$ absorbers (SLLSs and DLAs) and persists to $z<1$. There is no evidence of pristine CGM-like gas at $z<2.2$; though a substantial fraction of absorbers have only upper limits. The dispersion in the metallicity PDF for BRIDGE is large, suggesting a range of physical origins; combined with the large metallicity variations seen across small velocity separations within individual absorbers at both higher and lower redshift (\citetalias{lehner2022}; \citealt{lehner2019}), this points to incomplete metal mixing in the diffuse CGM at Cosmic Noon, while denser ISM-like gas appears more thoroughly mixed.
    \item At any given redshift $N_{\H}$ is nearly constant across all $\log N_{\HI}$ classes. The absorber types thus differ mainly in ionization state and metallicity. The mean $\log N_{\rm H}$ decreases from high to low $z$, indicating more baryons reside in the cool, diffuse CGM/IGM phase at high $z$. Densities and overdensities increase toward low $z$ ($\delta_b=180$--500, $400$--2200, and $4500$--8500 for KODIAQ-Z, BRIDGE, and CCC, respectively. For all three samples, pLLSs and LLSs roughly fall within our CGM-like definition ($200<\delta_b\lesssim10^4$).
    \item The $f_{\HI}$--$N_{\HI}$ relation is steeper at Cosmic Noon than at $z<1$ and $z>2.2$. We attribute this to the increased strength of the EUVB at Cosmic Noon or insufficient data in the low-$N_{\HI}$ regime. Correspondingly, the ionization parameter $ U=n_{\gamma}/n_{\H}$ increases with redshift, with statistically significant positive correlations for SLFSs, pLLSs, and LLSs. This is driven by the combined evolution of the EUVB ($n_{\gamma}$ increases from $z>3.5$, peaks at $z\sim2$, and then declines toward $z=0$) and the decline of $n_{\H}$ toward lower redshift. 
    \item Line-of-sight length scales span $\sim10$ pc in the densest systems to $\sim1000$ kpc in the most diffuse (more likely tracing IGM-like gas in filaments), though the largest values are upper limits. Because $N_{\rm H}$ is nearly constant across absorber types at a given redshift, length scale and overdensity are not independent: denser absorbers necessarily probe smaller structures. High metallicity absorbers ($[\rm{X/H}]\ge-1.4$; $\sim60\%$ of the $z<2.2$ sample) are on average more overdense, consistent with simulation predictions that global metal enrichment of CGM-like gas tracks structure formation, and with a galaxy--absorber association study in an \HI-selected samples at $z<1$ \citep{berg2023}. 

    \item Comparing our metallicity CDFs to predictions from the EAGLE cosmological simulation \citep{rahmati2018}, we find that the high-metallicity ends agree well across most absorber types, though EAGLE overproduces low-metallicity SLLSs and DLAs ($[\rm{X/H}]<-2$) and underproduces low-metallicity pLLSs and LLSs. At Cosmic Noon, our empirical CDFs resemble simulation predictions at $z=1$ even though our mean sample redshift is $\langle z\rangle=1.81$, indicating that the simulation predicts delayed enrichment relative to the observations.
\end{enumerate}

We will upload the 26 Keck/HIRES spectra used in this project to the KODIAQ database. All tables presented in this paper are available in machine-readable format, providing, for each of the \numberabs\, absorbers, the metal ion column densities, and the estimated metallicities, densities, and other physical properties.

\section*{acknowledgements}

We are grateful to Sebastián López and Hugo Cortés for providing us with the extracted and coadded UVES spectra for the lensed quasar system HE1104-1805.
This project was supported by NASA through the Astrophysics Data Analysis Program (ADAP) grant 80NSSC25K7587. Additional support was provided by NASA through grants HST-AR-17862 from the Space Telescope Science Institute, which is operated by the Association of Universities for Research in Astronomy, Incorporated, under NASA contract NAS5-26555. S.D. also acknowledges support for this project from the 2024 Future Investigators in NASA Earth and Space Science and Technology program (NASA grant 80NSSC25K7302).
This research made use of the supercomputing facility at the Notre Dame Center for Research Computing. We expressly acknowledge the assistance of Dodi Heryadi.
This research has made use of NASA's Astrophysics Data System.

\software{Astropy \citep{astropy2022},
  Matplotlib \citep{hunter2007}, Cloudy \citep{ferland2013}, pyigm \citep{pyigm}, rbcodes \citep{rbcodes}, $\textsc{scikit-learn}$ \citep{scikit-learn,sklearn_api}.}

\bibliography{references}

\appendix
We discuss our approach to fitting the Lyman limit for the low-resolution WFC3 data in \S\ref{app:hicols} and our treatment of upper and lower limits on $N_{\HI}$ in \S\ref{app:vpfit}. In \S\ref{app:systematics} we discuss the systematic uncertainties in metallicity estimates due to the assumed ionizing background, and the ions included in our photoionization models. Tables summarizing our sample, along with estimated column densities for different ions, Cloudy results for metallicities, ionization parameters, [C/$\alpha$], and physical properties, are provided in \S\ref{appendix:tables}. We also provide detailed results from our analysis in \S\ref{appendix:figs}, including refitted models of the WFC3 and ACS spectra. 
\\

\section{Modeling \HI\, Absorption in UV Data}\label{app:hicols}

The optical depth at any wavelength is:
\begin{equation}
    \tau\left(\lambda\right) = -\ln \left[F_{\rm obs}(\lambda)/F_{\rm cont}(\lambda)\right],
\end{equation}
where $F_{\rm obs}$ is the observed spectra and $F_{\rm cont}$ is the unabsorbed quasar continuum.

We determine the quasar continuum by fitting the unabsorbed regions of the observed spectrum to the \cite{telfer2002} UV quasar template allowing for a scaled normalization $C$ and a power law tilt $\alpha$ \citepalias{omeara2013}:

\begin{equation}
    F_{\rm cont} = C\,F_{\rm telf}\left[\frac{\lambda_{\rm obs}}{950\,\text{\AA}\,(1+z_{\rm qso})}\right]^\alpha.
\end{equation}

where $F_{\rm telf}$ is the Telfer template flux, and $\lambda_{\rm obs}$ is the observed wavelength, and $z_{\rm qso}$ is the quasar emission redshift. We perform the fitting over the rest wavelength range: $950\,\text{\AA}<\lambda_r<1150\,\text{\AA},\,\lambda_r>1280\,\text{\AA}$. This ensures that we do not include the quasar Ly$\alpha$ emission in the continuum fitting since the emission strengths can vary for different quasars and may bias the template toward a higher normalization parameter. The total optical depth of an absorber at redshift $z_i$ can be estimated using:
\begin{equation}
    \tau_i(\lambda) = \tau_{\rm LL}(\lambda) + \tau_{\rm LS}(\lambda).
\end{equation}
where $\lambda = \lambda_r(1+z_i)$. 
The first term is the optical depth at the Lyman limit ($\lambda_{\rm{LL}}=911.75\,\text{\AA}$):
\begin{equation}
    \tau_{\rm LL} = \sigma N_{\HI} = \sigma_0 \left( \frac{\lambda_r}{\lambda_{\rm{LL}}} \right)^{2.75} N_{\HI},
\end{equation}
\\
where $\sigma_0=6.3\times10^{-18} \rm{cm}^{-2}$ is the absorption cross section of hydrogen at the Lyman limit. Here we have used the power law coefficient 2.75 derived using equation (2.4) from \cite{osterbrock2006}.
The second term is the optical depth due to the Lyman series absorption which encompasses information about all the transitions. For any transition (e.g., Ly$\alpha$, Ly$\beta$, and so on):
\begin{equation}
    \tau_{\rm LS} = \frac{\pi e^2}{m_e c}f\lambda_0 N_{\HI}.
\end{equation}
where $f$ is the oscillator strength of the transition centered at the wavelength $\lambda_0$ \citep{draine2011}. The electronic charge and mass are specified by $e$ and $m_e$, respectively, and $c$ is the speed of light. We model the Lyman series lines as Voigt profiles with a Doppler parameter $b=25$ \kms, consistent with values reported in the literature \citep{werk2016,tumlinson2017,lehner2022}. This corresponds to a gas temperature of $\sim10^4$ K, which is typical for photoionized gas \citep{ferland2003}. The two terms ($\tau_{\rm LL}$ and $\tau_{\rm LS}$) overlap near the Lyman limit, since thermal broadening blends the Lyman series lines together as their spacing shrinks with $n\rightarrow\infty$. For each absorber, we construct the optical depth model $\tau_i$ in the absorber's rest frame and then redshift it to obtain the contribution from the absorber to the total optical depth at the corresponding observed wavelength. For multiple absorbers then, we obtain:
\begin{equation}
    F_{\rm model} = F_{\rm cont}\,\rm{exp} \left[-\sum^N_i \tau_i\right].
\end{equation}

We convolve the generated model with the instrumental line spread function (LSF). We approximate the LSF as a Gaussian with the full width at half maxima set to twice the wavelength dispersion ($\delta\lambda$) for the instrument. Both WFC3 and ACS spectra have $\delta\lambda$ that changes with wavelength. Thus, we define a wavelength-dependent LSF at each pixel to account for the varying dispersion. Finally, we rebin the spectra to the observed wavelength grid. We then employ the {\tt{emcee}} \citep{emcee} package in Python to perform a MCMC optimization of the model to simultaneously solve for the normalization and tilt of the continuum fit, and the redshifts and optical depths of the intervening absorbers. For better fitting, we allow the b-values of the Lyman series lines to vary within $15$--$30$ \kms, which improves the fit near the Lyman limit where the merging of Lyman series lines at limited instrumental resolution causes a smooth flux drop-off.

\section{Treatment of lower limits on \texorpdfstring{$N_{\HI}$}{NHI}}\label{app:vpfit}

For sightlines with higher $N_{\HI}$ ($\log N_{\HI} \gtrsim17.8$) absorbers, the flux below the Lyman limit is often completely absorbed, and we obtain only a lower limit on the $N_{\HI}$. If there are multiple absorbers along a sightline, we perform two rounds of fitting. The first round fits the unsaturated absorbers and gives a lower limit estimate for the saturated absorbers. Because the model cannot converge for the saturated absorbers, this lower limit is unreliable and often overestimated. Thus, we perform a second fitting, where we fix the parameters for the unsaturated absorbers, and use a statistical test to determine the best lower limit for the saturated absorber. Our method leverages the fact that the difference in chi-square ($\Delta\chi^2$) between models and the best-fit model follows a chi-square distribution. Since this second round derives only a statistical lower limit, and does not refit the continuum or absorber redshift, we do not reintroduce systematic errors such as the wavelength calibration uncertainty that was already accounted for in the first round. We construct the distribution by estimating $\chi^2$ for different values of $N_{\HI}$, and determine $\chi_{\rm{crit}}^2=3.842$, which when added to the best-fitting (minimum $\chi^2$) model, corresponds to a $N_{\HI}$ value that represents the 95th percentile of the asymmetric lower limit tail. We use this as the minimum value of $N_{\HI}$ that completely absorbs the continuum flux at the Lyman limit. 

\begin{figure}
\begin{centering}
\includegraphics[trim= 0 10  0 20,clip,scale=0.6]{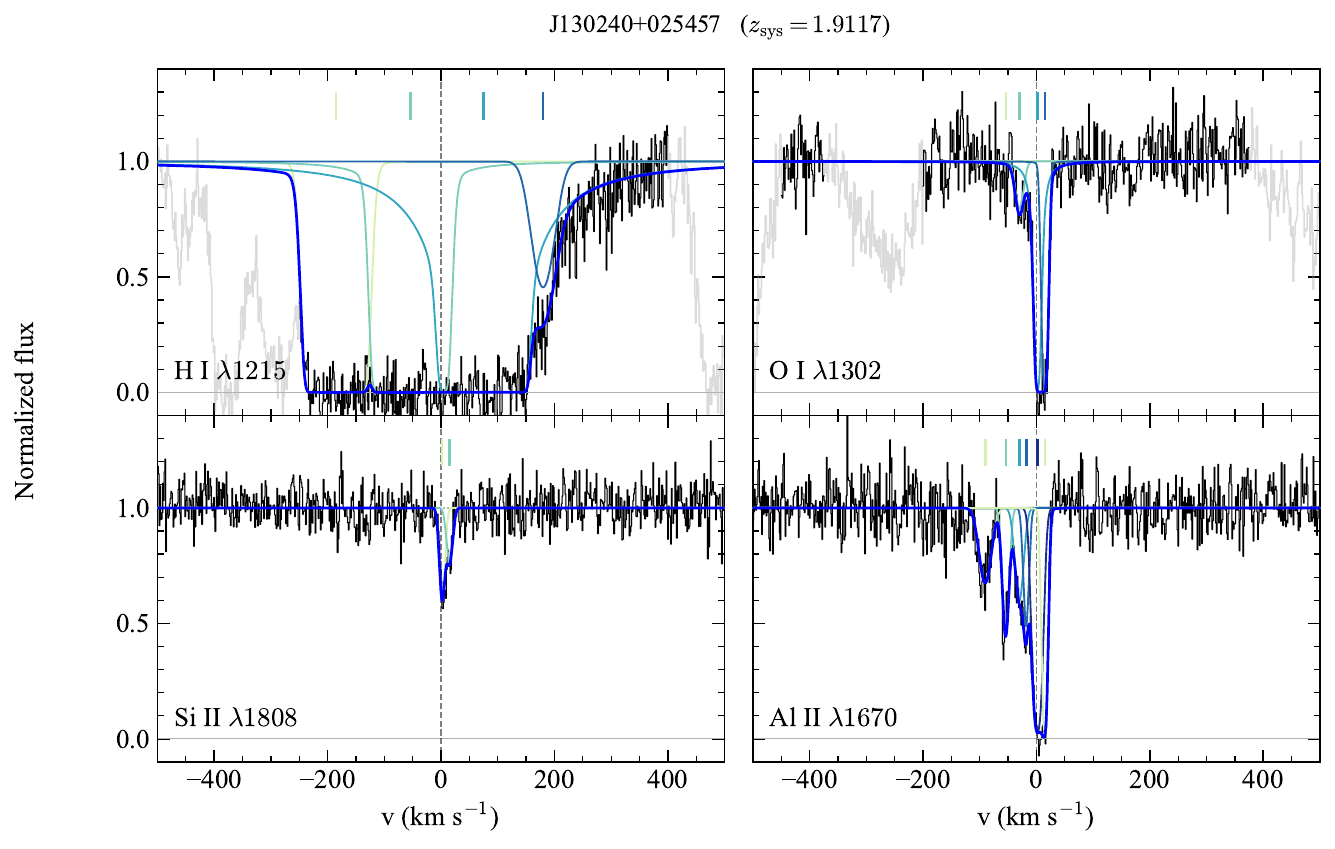}
\par\end{centering}
\begin{centering}
\caption{
{Voigt profile fitting of Ly$\alpha$, \OI\, 1302 and \SiII\, 1808 lines for an absorber at $z=1.9117$ along J130240+025457. We note the \HI\, fit requires only 4 components (as opposed to 6 in \AlII), since severe blending makes it impossible to fit lower $N_{\HI}$ components. This preliminary fitting gives $\log N_{\HI}=18.28\pm 0.05$.}}\label{fig:voigt}
\par\end{centering}
\end{figure}

\begin{figure*}
\begin{centering}
\begin{minipage}{0.49\textwidth}
\begin{centering}
\includegraphics[trim= 33 9 10 20,clip,scale=0.31]{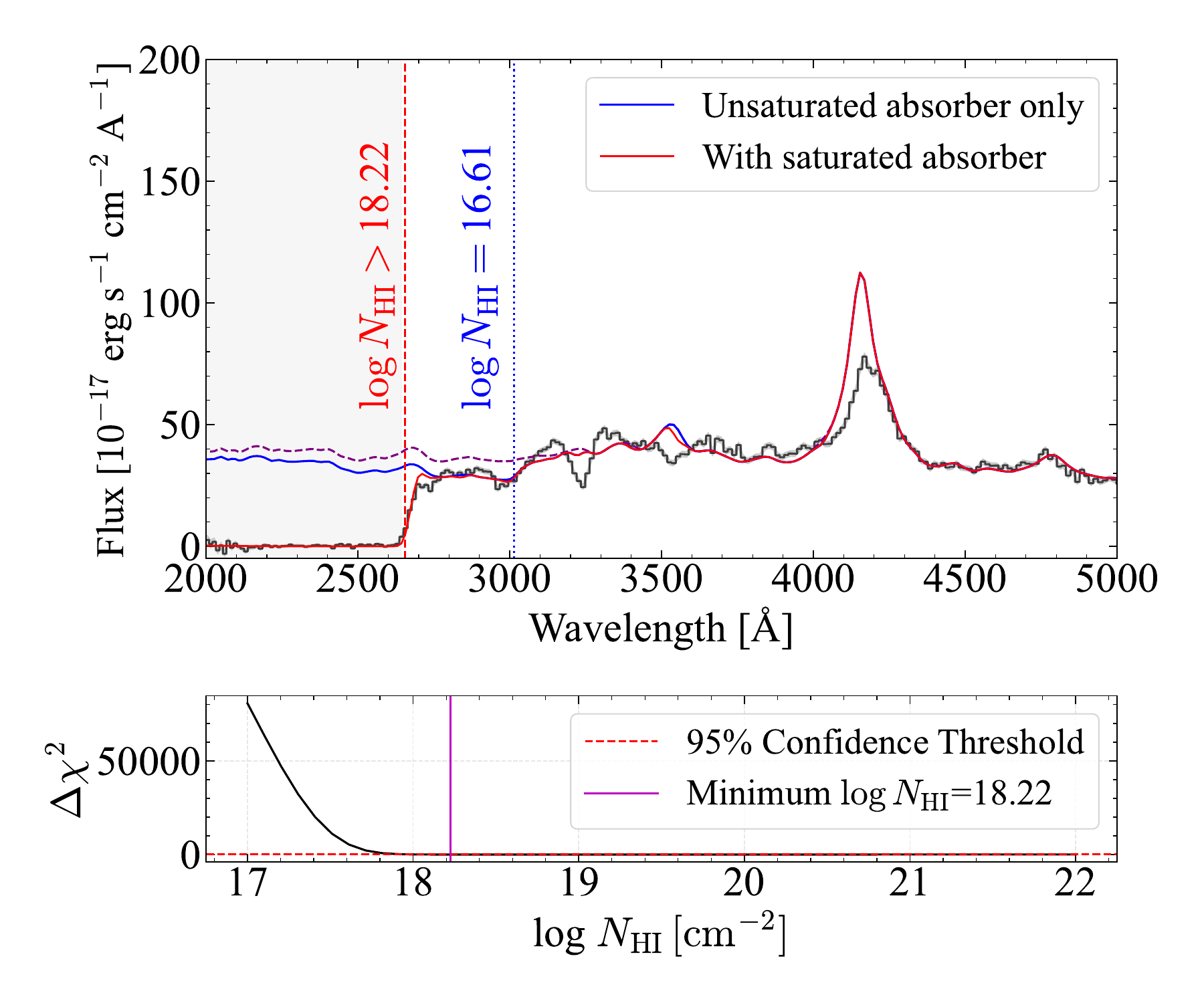}
\par\end{centering}
\end{minipage}
\hfill
\begin{minipage}{0.49\textwidth}
\begin{centering}
\includegraphics[trim= 8 9 0 5,clip,scale=0.31]{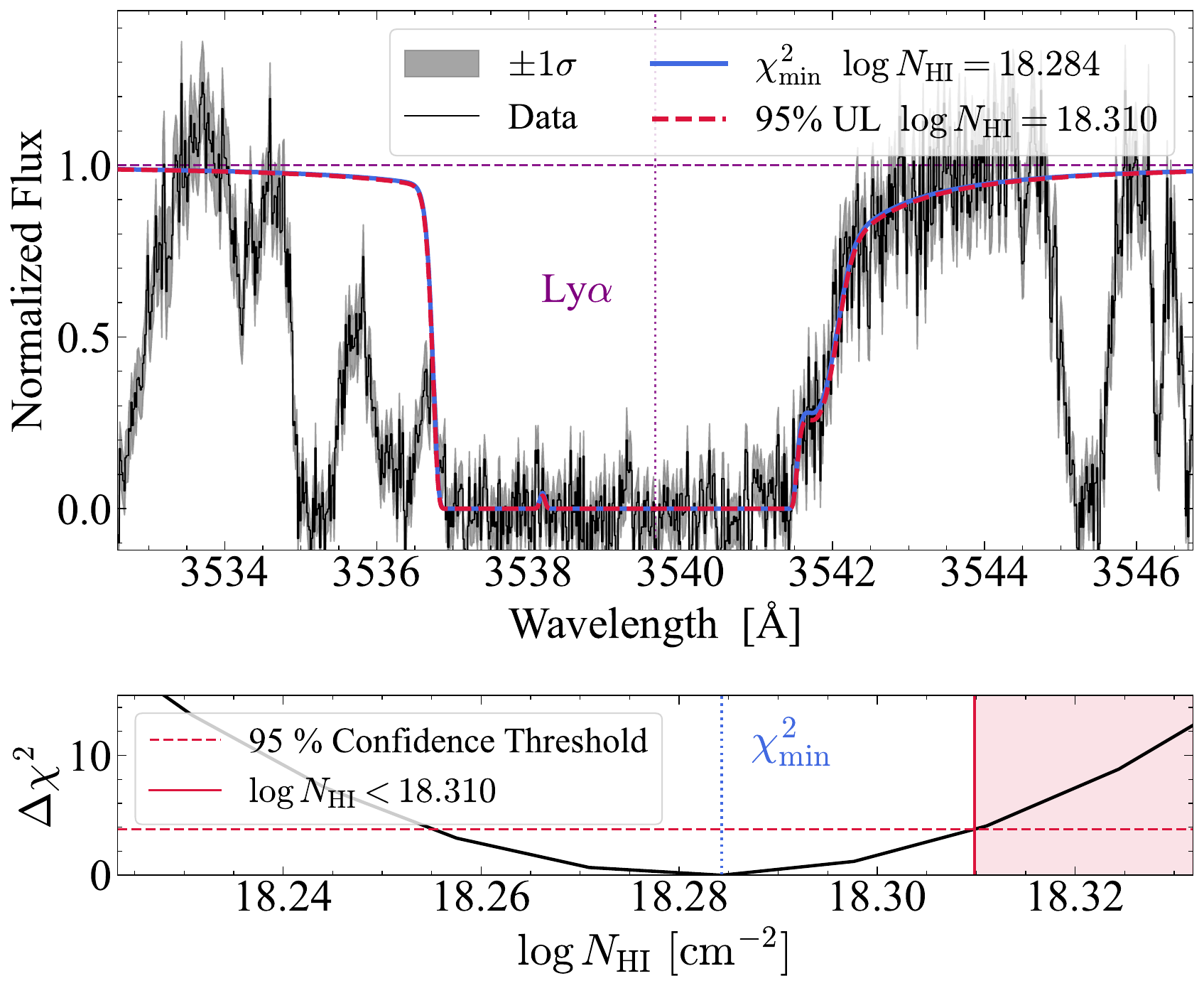}
\par\end{centering}
\end{minipage}
\par\end{centering}
\caption{
\textit{Left panels:} Determining robust lower limits for a Lyman-limit-saturating absorber along the quasar J130240+025457 using WFC3/UVIS-G280 spectra. In the top panel, we plot the modeled quasar continuum (dashed purple line), the model with an unsaturated absorber at $z_{\rm{abs}}=2.31$ (solid blue line), and the statistically determined minimum-$N_{\HI}$ model for the saturated absorber at $z_{\rm{abs}}=1.91$ (red line). The bottom panel shows $\Delta\chi^2$ as a function of $\log N_{\HI}$ used to determine the minimum $N_{\HI}$ that satisfies the observed flux at the 95\% confidence level.
\textit{Right panels:} Upper limit analysis for the same absorber. In the top panel, we plot the normalized HIRES spectrum of J130240+025457 for Ly$\alpha$, the best-fit from preliminary Voigt profile fitting (blue), and the statistically determined maximum-$N_{\HI}$ model (dashed red). The bottom panel shows $\Delta\chi^2$ as a function of $\log N_{\HI}$ used to determine the maximum $N_{\HI}$ consistent with the observed flux at the 95\% confidence level. For this absorber, the lower and upper bounds on $\log N_{\HI}$ are 18.22 and 18.31, respectively.
}\label{fig:lowlim}
\end{figure*}

For the saturated absorbers at a redshift $z_{\rm{abs}}\gtrsim2$ we sometimes have access to lower-order Lyman series lines in the high-resolution optical spectra. These lines are highly saturated for the range of $N_{\HI}$ we are probe, but the presence or absence of damping wings can still constrain $N_{\HI}$ within reasonable bounds. Damping wings appear in the absorption profile of the Ly$\alpha$ line for $\log N_{\HI}\gtrsim19$; for smaller $\log N_{\HI}$, we use their absence to estimate upper bounds on $N_{\HI}$\citep{cooper2015}. Often \HI\, absorbers are composed of several blended components, which are separable in the metal lines owing to their smaller $b$-values, lower abundances and weaker line strengths of metal ions. We therefore use low ions (e.g. \OI, \CII, \SiII, \AlII) to identify the components and simultaneously fit Ly$\alpha$ and metal-ion line profiles with VoigtFit \citep{krogager2018}. For cool, photoionized gas, surveys of \HI-selected absorbers find $b<40$ \kms with, $\langle b \rangle=30\pm11$ \kms for KODIAQ-Z absorbers \citepalias{lehner2022}, and $28\pm8$ \kms for CCC absorbers \citep{lehner2018}, corresponding to a temperature $T<4\times10^4$ K. Motivated by this, we fix the $b$-value for \HI\, transitions to 25 \kms. We find that a few dominant components suffice to reproduce the absorption in \HI\, and we can fit Lyman series lines with fewer components than metal lines. Since our goal is to obtain an upper limit on the total \HI\, column density, the exact component structure is not crucial. Figure~\ref{fig:voigt} shows one such case for J130240+025457. We perform a similar analysis with $\chi^2$ as we did for lower limits and use the highest value of $N_{\HI}$ that satisfies the profile. The left and right panel in Figure~\ref{fig:lowlim} demonstrate our approach to estimating the lower and upper limits to $N_{\HI}$, respectively. We use this method for 19 absorbers and find our results consistent with the lower limits derived from the lower resolution WFC3 data. 

\newpage

\section{Systematic Uncertainties in Ionization Modeling}\label{app:systematics}

Here we present in full the tests summarized in \S\ref{subsec:systematics}: the impact of the assumed ionizing background, the choice of ions used to constrain the photoionization models, and the assumption of solar relative abundances.
\begin{figure}
\begin{centering}
\includegraphics[trim= 0 0 0 20,clip,scale=0.45]{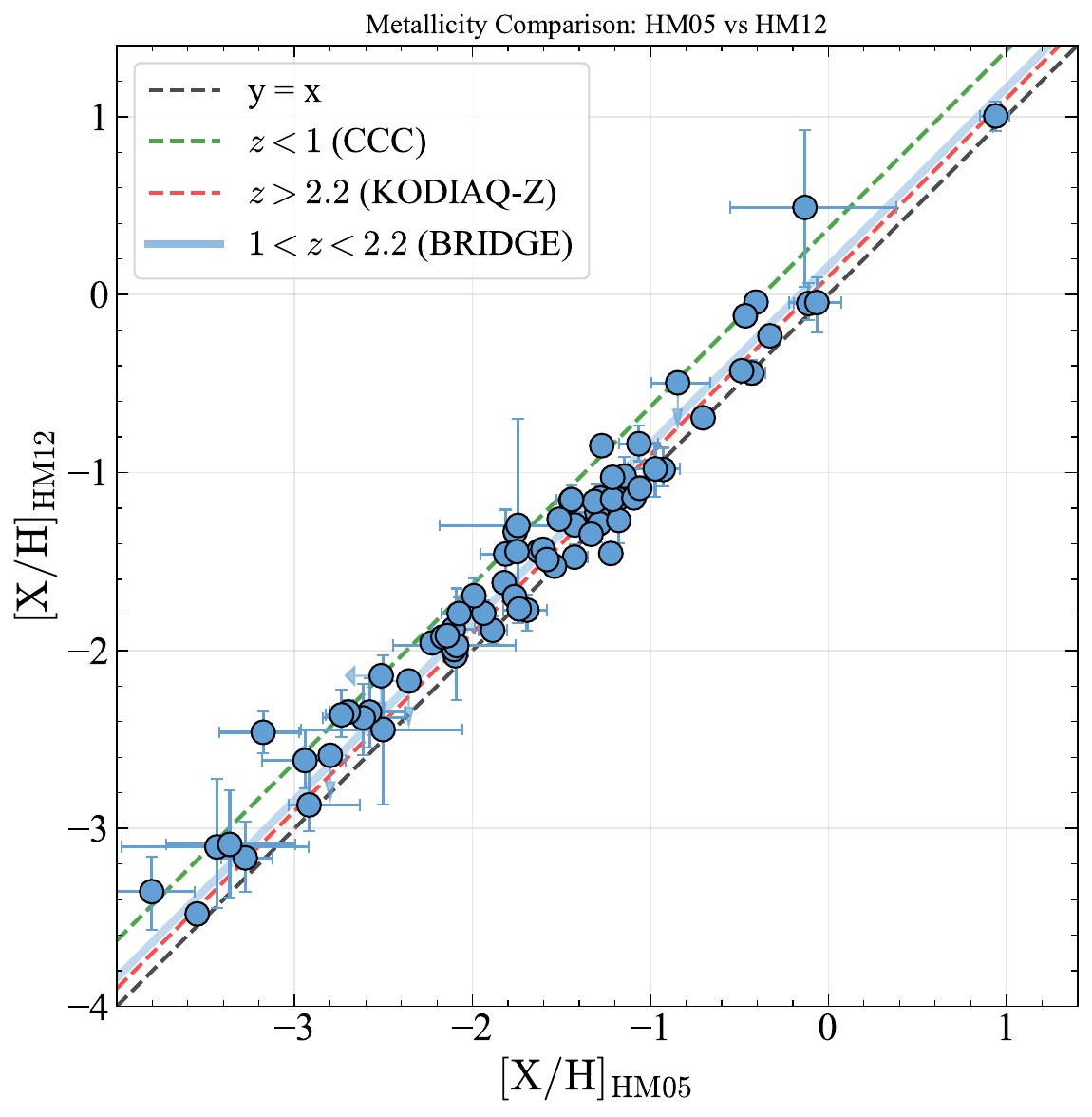}
\par\end{centering}
\begin{centering}
\caption{
{Comparison of median metallicities (within 68\% CI) of BRIDGE derived using the HM05 and HM12 EUVBs. The black dashed line is the 1:1 relationship. We also plot the mean offset between metallicities derived from the two EUVBs ($\langle\rm{[X/H]_{\rm{HM12}}}-\rm{[X/H]_{\rm{HM05}}}\rangle$) for the CCC, KODIAQ-Z, and BRIDGE surveys.}
}\label{fig:compare_radiations}
\par\end{centering}
\end{figure}

\subsection{Ionizing Background}\label{app:euvb}

One of the largest uncertainties in the determination of metallicities via ionization modeling is the shape of the assumed EUVB. Two widely used EUVBs are the HM05 and HM12 EUVBs. Due to the greatly reduced radiation escape fraction in the HM12 EUVB, the metallicities resulting from these models are generally higher. \citetalias{wotta2019} and \citetalias{lehner2022} explored the impact of changing the EUVB from HM05 to the harder HM12 model. At both $z<1$ and $z=2.2$--3.6, metallicities are systematically higher for models with HM12 EUVB. The mean difference between the metallicities was $\langle\rm{[X/H]_{\rm{HM12}}}-\rm{[X/H]_{\rm{HM05}}}\rangle=+0.37\pm0.19$ for $z<1$ \citep{gibson2022} and $0.10\pm0.17$ for $z=2.2$--3.6 \citepalias{lehner2022}. 

Using a sample of 70 absorbers for which the densities were well-constrained (flat prior on $\log U$), we study the impact in the intermediate redshift range covered by BRIDGE. The models for 8 absorbers did not converge when HM12 was used. We used the same ions for both EUVBs. Figure~\ref{fig:compare_radiations} shows our comparison. We find $\langle\rm{[X/H]_{\rm{HM12}}}-\rm{[X/H]_{\rm{HM05}}}\rangle=+0.16\pm0.13$ for this sample. We also plot the mean offset in metallicities derived from the two EUVBs for the CCC and KODIAQ-Z sample. Our results for the BRIDGE survey lie between the two as anticipated. 

For the CCC survey ($z<1$), \cite{gibson2022} derived a systematic increase of 0.35 dex in the derived metallicities if one switched from HM05 to the \cite{ks19} (KS19) EUVB. This is very similar to the 0.37 dex difference between HM12 and HM05. We expect a similar trend at Cosmic Noon. Thus, compared to CCC, the systematic uncertainty in metallicity due to the assumed EUVB is much smaller for the BRIDGE survey. We revisit the evolution in the UV background in \S\ref{subsec:UVB}.

\begin{figure*}
\begin{centering}
\begin{minipage}{0.48\textwidth}
\begin{centering}
\includegraphics[width=\columnwidth]{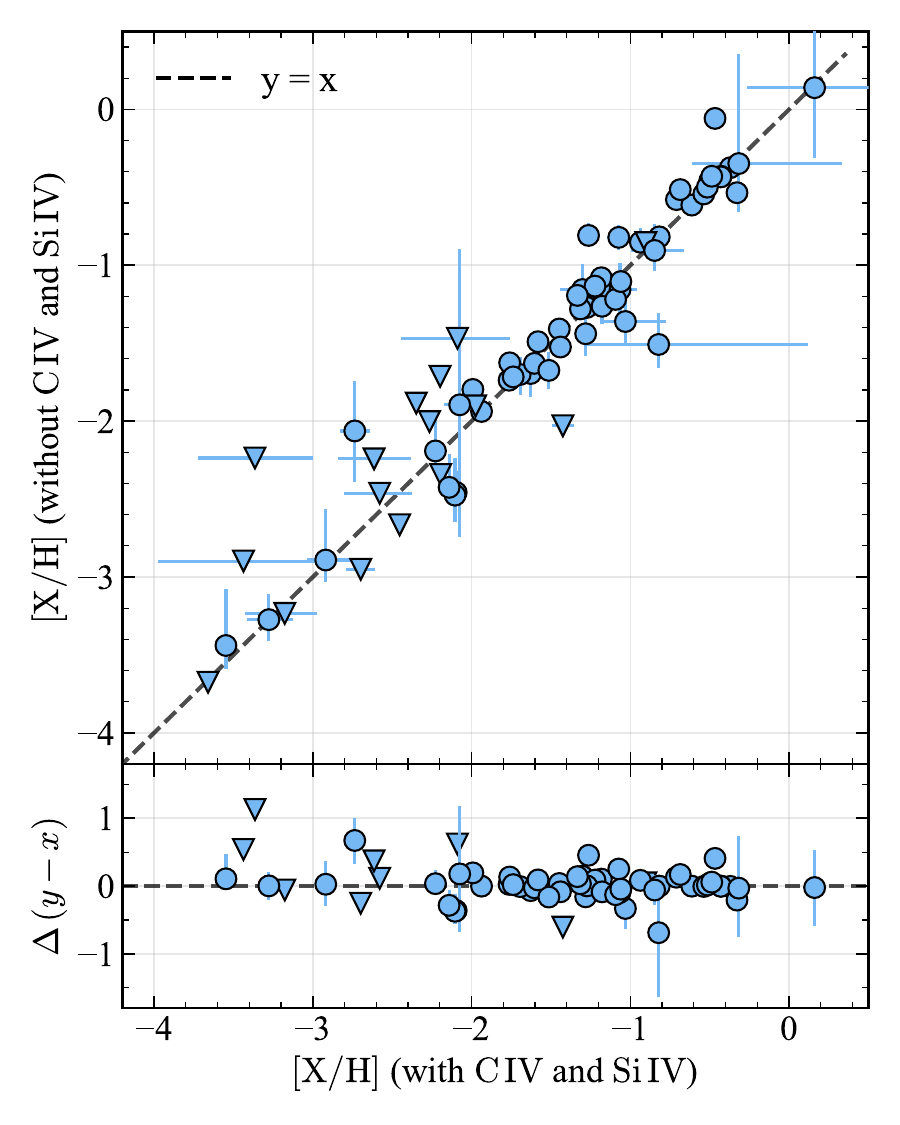}
\par\end{centering}
\end{minipage}
\hfill
\begin{minipage}{0.48\textwidth}
\begin{centering}
\includegraphics[width=\columnwidth]{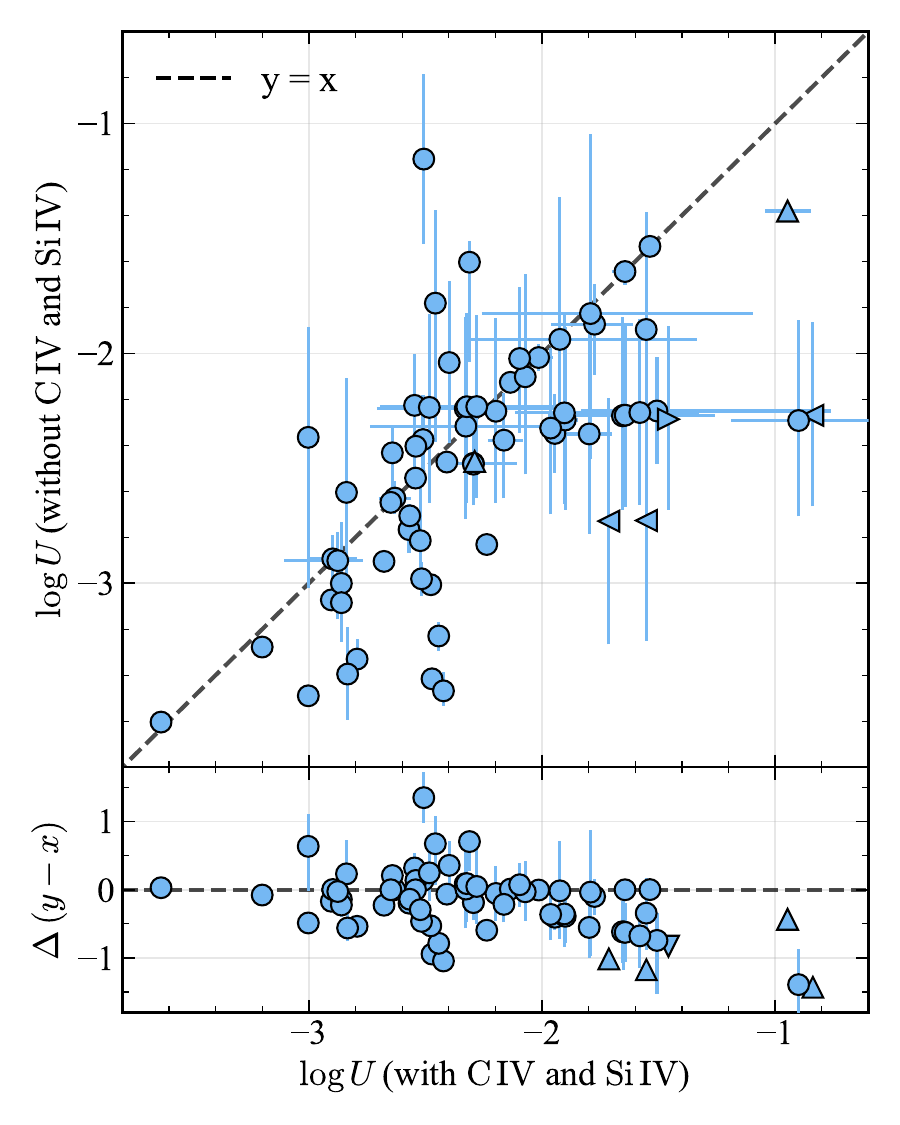}
\par\end{centering}
\end{minipage}
\par\end{centering}
\caption{
\textit{Left panel:} Comparison of median $\left[\rm{X/H}\right]$ (within 68\% CI) for the BRIDGE survey estimated from models with and without high ions \CIV\, and \SiIV. The black dashed line is the 1:1 relationship. The lower panels shows the residual.
\textit{Right panel:} Similar comparison for $\log U$.
}\label{fig:w_wo_h}
\end{figure*} 

\subsection{Ions Included in Photoionization Models}\label{app:ionsincluded}

The ions available to constrain the photoionization models vary with the wavelength coverage and redshift of each absorber. At $z<1$, \citetalias{wotta2019} explored the impact of including higher ions (in their case \OIV) on photoionization modeling, and found that it resulted in systematically lower densities while the metallicities remained largely unaffected. \citetalias{lehner2022} performed a similar analysis for \CIV\ at $2.2\le z\le 3.6$, and did not find a significant impact on metallicity, and ionization parameter (or density). For a subset of our sample, we have access to both low (\CII, \CIII, \SiII, \MgII, \FeII) and high ions (\CIV, \SiIV), while for others (especially for $z_{\rm{abs}}\sim1$) we can only access low ions (like \MgII, \FeII). We perform an analogous test for our sample at Cosmic Noon.

For absorbers with coverage of \CIV\, and/or \SiIV, we carry out photoionization modeling both with and without these high ions to assess their systematic effect. We plot our estimated metallicities and ionization parameters for the two cases in Figure~\ref{fig:w_wo_h}. Among the 70 absorbers with \CIV\ and \SiIV\, coverage, two situations arise: (i) at least one other low ion is detected (55 absorbers), and (ii) only high ions are detected (15 absorbers). In the latter case, excluding high ions yields an upper limit on metallicity that is consistent with the value obtained when the high ions are included. For these 15 absorbers, we inspect the velocity profile of the components. In 8 cases, the component structures for \CIV\, and \SiIV\, closely match that of \HI\, (with Ly$\alpha$ accessible in the optical spectra), and we adopt the models that include high ions. For the remaining 7 absorbers, the component structure was ambiguous, and we adopt upper limits on \CIV\, and \SiIV. In such cases, the observed \CIV\ or \SiIV\ may be associated with an unresolved, blended, lower $N_{\HI}$ system buried in the strong Ly$\alpha$ absorption profile. 

Overall, metallicities derived with and without high ions are consistent. For the 54 absorbers that are not limits, the mean absolute difference in metallicity is $0.13\pm0.04$ with only 7\% (5/70) of absorbers showing a metallicity difference $>0.4$ dex. In contrast, the ionization parameter $\log U$ ($n_{\rm{H}}$) is systematically higher (lower) in models that include \CIV\, and/or \SiIV. The mean absolute difference between estimated $\log U$ is $0.33\pm0.04$ with 38\% absorbers exceeding a difference $>0.4$ dex. Because overdensities are derived directly from $n_{\rm H}$, they inherit the same systematics. Thus, while including high ions introduces a systematic in the estimated densities, it does not significantly bias the metallicity estimates, consistent with results at both lower \citep{lehner2019} and higher \citepalias{lehner2022} redshifts.

The systematic offset in densities is also propagated to the inferred physical length scales, for which the mean offset is $\Delta(\log l[\rm{kpc}])=0.70\pm0.10$. We therefore caution against over-interpreting the length scale estimates, which are very sensitive to the modeling assumptions. 

\section{Metallicities for the Literature Sample}\label{app:lit_sample}

We compile a sample of 17 SLLSs from the literature \citep{dessauges2003,som2013,prochaska2015}, adopting their reported \HI\ and metal-ion column densities. \cite{fumagalli2016} derived metallicities for these absorbers with the same Bayesian framework we use here, but adopted the HM12 EUVB.  For the 11 absorbers with $\log N_{\HI}<20.0$, we re-estimate metallicities using the HM05 EUVB for consistency with our sample. The remaining 6 absorbers have $\log N_{\HI}>=20.0$. Since our Cloudy grid only goes upto $\log N_{\HI}$ we build smaller grids to estimate the metallicity and ionization parameter by maximizing the likelihood function following \cite{lehner2013}. We do this for both HM05 and HM12 EUVB. Table~\ref{tab:lit_sample} compares the \cite{fumagalli2016} values with ours. 

Our HM12 metallicities agree with those of \cite{fumagalli2016} with a median offset of $0.15$ dex, likely due to modeling choices (inclusion of dust in their models). For these absorbers the HM05 and HM12 estimates are generally consistent within the mean offset of $\langle[\rm{X/H}]_{\rm HM12}-[\rm{X/H}]_{\rm HM05}\rangle=0.16\pm0.13$ dex derived in Appendix~\ref{app:euvb}.

\input{table5_lit_met}

\newpage
\section{Data tables}\label{appendix:tables}
\input{table6_sample}
\input{table7_metal_cols}
\input{table8_cloudyres}
\input{table9_physprop}
\section{Figures}\label{appendix:figs}
\begin{center}
\includegraphics[width=\textwidth]{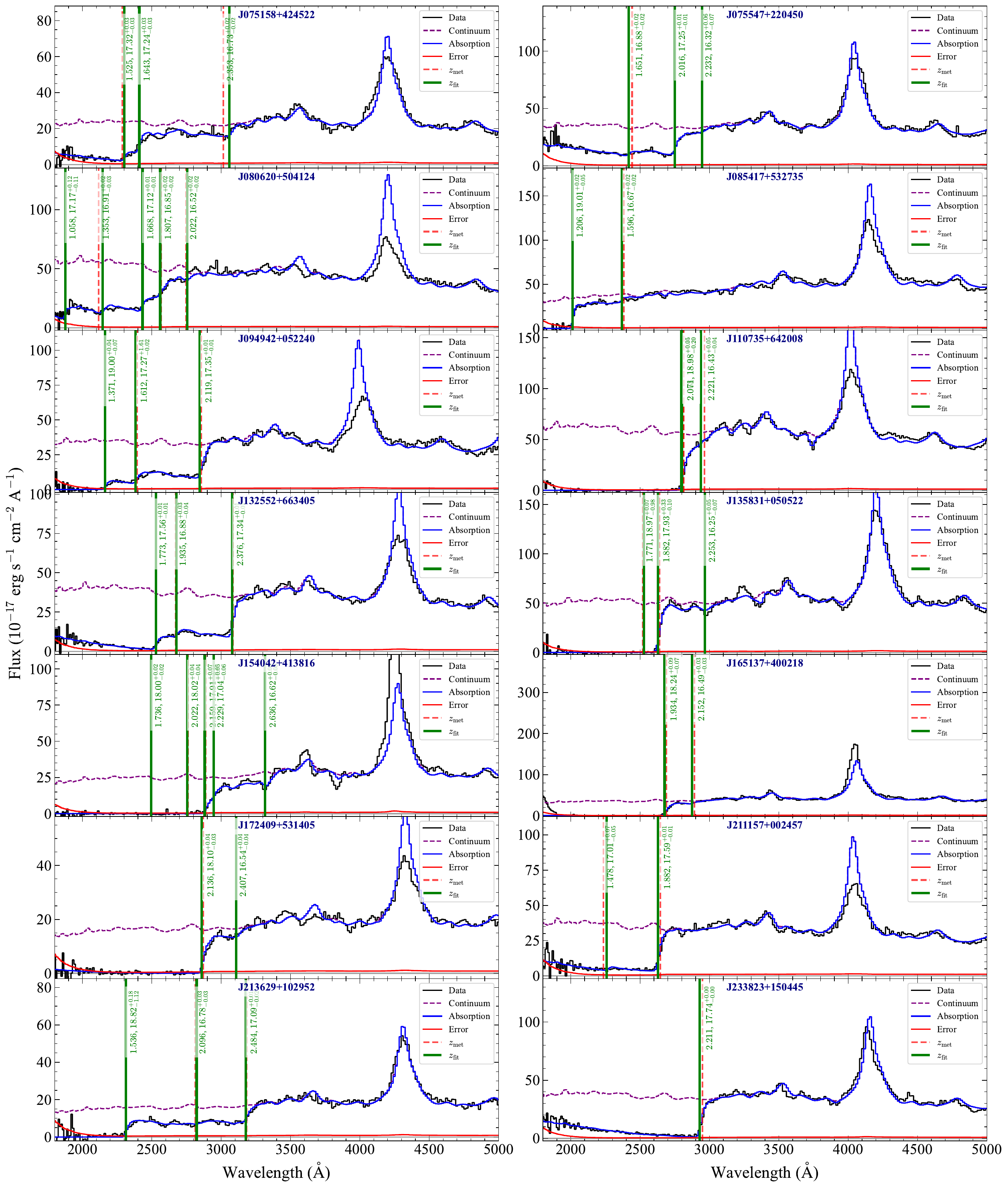}

\figcaption{Continuum and absorption-line fits for the WFC3 spectra in our
sample. The black, red, purple dashed, and blue lines represent the observed flux, 1$\sigma$ error, fitted quasar continuum, and the best-fit absorption model, respectively. Green vertical
lines mark the fitted absorber redshifts, with best-fit $z$ and $\log N_{\rm HI}$ values labeled for each system. The vertical red dotted lines are redshifts determined using metal lines in the optical spectra. \label{fig:wfc3_fits_mosaic}}
\end{center}

\begin{center}
\includegraphics[width=\textwidth]{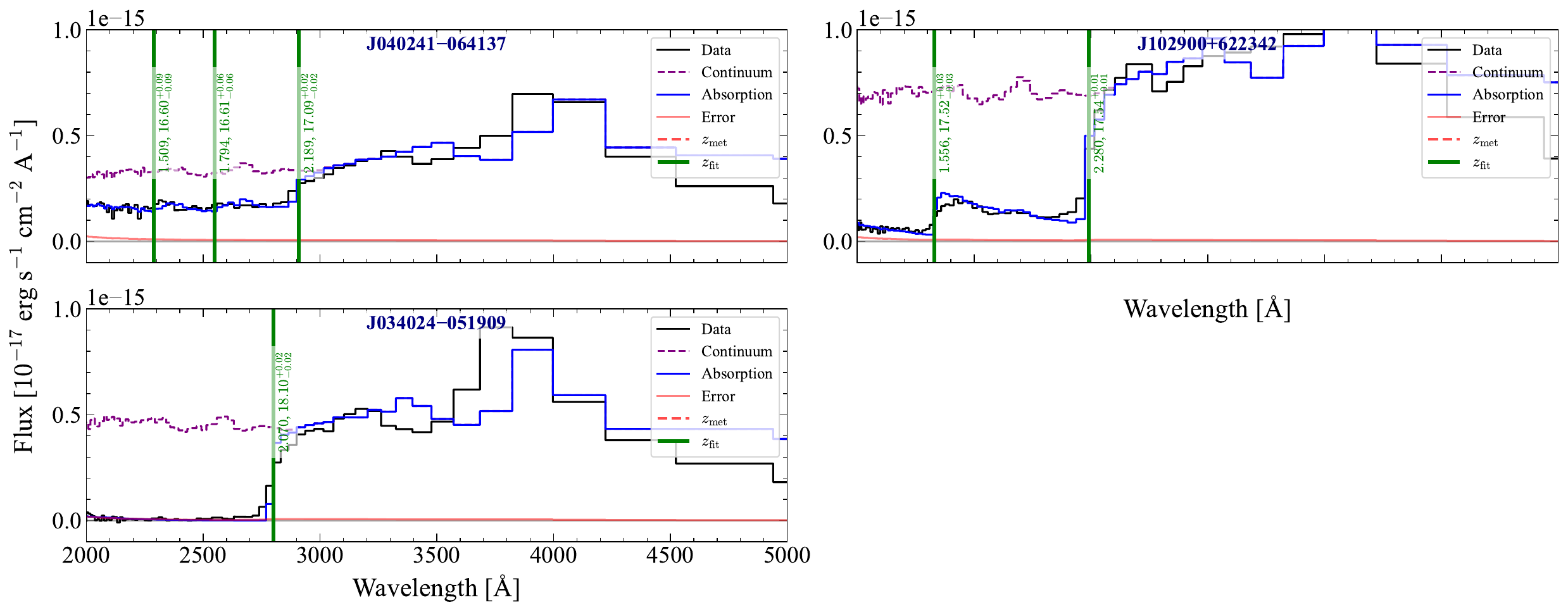}

\figcaption{Same as Figure~\ref{fig:wfc3_fits_mosaic}, but for ACS spectra.
\label{fig:acs_fits}}
\end{center}
\input{table10_lit_met_cols}
\end{document}

%% file: table1_hiresqsos.tex
\begin{deluxetable}{lcccc}
\tablecaption{Quasars observed by Keck/HIRES} \label{tab:hires}

\tablehead{
    \colhead{QSO} &
    \colhead{$z_{\rm em}$} &
    \colhead{Coverage (\AA)} &
    \colhead{S/N} 
}
\startdata
J034024$-$051909 & 2.320 & 3036, 6175 & 12, 32 \\
J040241$-$064137 & 2.432 & 2995, 5882 &  7, 16 \\
J075158$+$424522 & 2.453 & 3021, 5883 &  7, 23 \\
J075547$+$220450 & 2.319 & 2996, 5946 &  8, 19 \\
J080620$+$504124 & 2.457 & 3021, 5885 &  4, 10 \\
J085417$+$532735 & 2.418 & 3210, 6288 &  5, 16 \\
J094942$+$052240 & 2.282 & 2996, 5882 &  9, 18 \\
J102900$+$622342 & 2.470 & 2995, 5881 & 12, 20 \\
J105315$+$400756 & 2.482 & 2995, 5881 &  8, 20 \\
J110411$+$024655 & 2.532 & 2995, 5881 & 10, 22 \\
J110735$+$642008 & 2.316 & 2995, 5882 &  8, 23 \\
J114358$+$052445 & 2.557 & 2995, 5951 & 18, 37 \\
J130240$+$025457 & 2.916 & 2995, 5881 & 13, 20 \\
J132552$+$663405 & 2.511 & 3035, 5881 & 12, 19 \\
J134100$+$412314 & 1.204 & 2995, 5882 & 12, 30 \\
J135831$+$050522 & 2.455 & 2995, 5880 & 12, 19 \\
J141528$+$370621 & 2.374 & 2995, 5881 & 13, 37 \\
J153514$+$483659 & 2.542 & 3031, 5882 & 12, 23 \\
J154042$+$413816 & 2.516 & 2995, 5882 & 12, 32 \\
J161003$+$442353 & 2.595 & 2995, 5881 & 10, 22 \\
J163201$+$373749 & 1.478 & 2995, 5879 & 25, 50 \\
J165137$+$400218 & 2.341 & 2995, 5956 & 20, 40 \\
J172409$+$531405 & 2.547 & 3030, 7987 & 10, 30 \\
J211157$+$002457 & 2.325 & 2996, 6777 & 11, 26 \\
J213629$+$102952 & 2.555 & 2996, 7302 & 15, 44 \\
J233823$+$150445 & 2.419 & 2996, 6297 & 12, 38 \\
\enddata
\tablecomments{These spectra were observed between November 2010 and August 2011. We report S/N per resolution element for Keck/HIRES for the blue and red portions of the spectra separately, estimated from the unabsorbed quasar continuum near the transitions used in our analysis. Blue-side S/N is measured near low-ionization lines such as \HI\,(1215 \AA), \SiII\,(1190 \AA), and \OI\,(1302 \AA); red-side S/N is measured near \SiII\,(1526 \AA), \CIV\,(1548 \AA), \AlIII\,(1854 \AA), and similar transitions. The specific lines used for each estimate depend on the wavelength coverage and absorber redshifts along each sightline.}
\end{deluxetable}

%% file: table2_metpdf.tex
\begin{deluxetable*}{ccccccccc}
\tablecaption{Metallicity statistics for CCC ($z<1$), BRIDGE ($1\le z\le2.2$), and KODIAQ-Z ($2.2<z\le3.6$).} \label{tab:meanZ}
\tablehead{
    \colhead{Abs. Type} &
    \colhead{$\log N_{\rm HI}$} &
    \colhead{$N_{\rm abs}$} &
    \colhead{$\langle[{\rm X/H}]\rangle$} &
    \colhead{$\widetilde{[{\rm X/H}]}$} &
    \colhead{IQR} &
    \colhead{\parbox{2cm}{\centering Fraction with\\ $[{\rm X/H}]<-1.7$}} &
    \colhead{\parbox{2cm}{\centering Fraction with\\ $[{\rm X/H}]<-2.4$}} &
    \colhead{\parbox{2cm}{\centering Fraction with\\ $[{\rm X/H}]<-3.0$}}
}
\startdata
\multicolumn{9}{c}{CCC ($z<1$)} \\
\hline
pLLSs+LLSs+SLLSs & [16.2, 20.3) & $133$ & $-1.20\pm1.31$ & $-1.18$ & [$-1.95$, $-0.39$] & $27\%$--$40\%$ & $5\%$--$13\%$ & $0.0\%$--$2.0\%$\\
pLLSs & [16.2, 17.2) & $95$ & $-1.42\pm1.33$ & $-1.35$ & $[-2.14$, $-0.55]$ & $33\%$--$40\%$ & $7\%$--$18\%$ & $0.0\%$--$2.8\%$\\
LLSs & [17.2, 19.0) & $30$ & $-0.63\pm1.19$ & $-0.69$ & $[-1.44$, $-0.14]$ & $8\%$--$31\%$ & $1\%$--$14\%$ & $0\%$--$8\%$\\
SLLSs & [19.0, 20.3) & $8$ & $-0.72\pm0.56$ & $-0.55$ & $[-1.17$, $-0.29]$ & $0\%$--$25\%$ & $0\%$--$25\%$ & $0\%$--$25\%$\\
DLAs & $\ge20.3$ & $17$ & $-0.30\pm0.31$ & $-0.19$ & $[-0.44$, $-0.14]$ & $0\%$--$14\%$ & $0\%$--$14\%$ & $0\%$--$14\%$ \\
\hline
\multicolumn{9}{c}{BRIDGE ($1\le z\le2.2$)} \\
\hline
pLLSs+LLSs+SLLSs & [16.2, 20.3) & $80$ & $-1.27\pm1.10$ & $-1.19$ & $[-1.70$, $-0.49]$ & $18\%$--$34\%$ & $9\%$--$21\%$ & $1\%$--$7\%$\\
pLLSs & [16.2, 17.2) & $20$ & $-1.57\pm1.13$ & $-1.36$ & $[-1.84$, $-0.78]$ & $16\%$--$48\%$ & $13\%$--$43\%$ & $1\%$--$20\%$\\
LLSs & [17.2, 19.0) & $41$ & $-1.34\pm1.15$ & $-1.20$ & $[-1.88$, $-0.53]$ & $17\%$--$39\%$ & $8\%$--$26\%$ & $1\%$--$10\%$\\
SLLSs & [19.0, 20.3) & $19$ & $-0.80\pm0.78$ & $-0.88$ & $[-1.56$, $-0.22]$ & $6\%$--$34\%$ & $0\%$--$12\%$ & $0\%$--$12\%$\\
DLAs & $\ge20.3$ & $63$ &  $-0.63\pm0.54$  & $-0.64$ & $[-1.05$, $-0.24]$ & $0.4\%$--$6.8\%$ & $0.0\%$--$4.1\%$ & $0.0\%$--$4.1\%$\\
\hline
\multicolumn{9}{c}{KODIAQ-Z ($2.2< z\le3.6$)} \\
\hline
pLLSs+LLSs+SLLSs & [16.2, 20.3) & $188$ & $-2.11\pm0.93$ & $-2.02$ & $[-2.67$, $-1.54]$ & $62\%$--$73\%$ & $29\%$--$40\%$ & $9\%$--$17\%$\\
pLLSs & [16.2, 17.2) & $40$ & $-2.34\pm1.05$ & $-2.25$ & $[-2.98$, $-1.51]$ & $57\%$--$80\%$ & $31\%$--$55\%$ & $14\%$--$35\%$\\
LLSs & [17.2, 19.0) & $68$ & $-2.20\pm0.84$ & $-2.11$ & $[-2.75$, $-1.66]$ & $61\%$--$79\%$ & $28\%$--$47\%$ & $9\%$--$23\%$\\
SLLSs & [19.0, 20.3) & $80$ & $-1.92\pm0.89$ & $-1.88$ & $[-2.47$, $-1.33]$ & $55\%$--$72\%$ & $20\%$--$36\%$ & $4\%$--$14\%$\\
DLAs & $\ge20.3$ & $168$ & $-1.14\pm0.56$ & $-1.22$ & $[-1.56$, $-0.75]$ & $12\%$--$21\%$ & $0.0\%$--$1.6\%$ & $0.0\%$--$1.6\%$\\
\enddata
\tablecomments{$\widetilde{\left[\rm{X/H}\right]}$ is the median metallicity, and IQR is the inter-quartile range that we derive directly using walkers from the MCMC optimization of ionization models. Fraction ranges quote the 90\% Wilson confidence interval. CCC and BRIDGE (6 absorbers) SLLS statistics include point-estimate absorbers without MCMC chains, incorporated via pseudo-walker draws from their reported metallicity and error. The mean, median, and IQR for DLAs are calculated using DLA metallicity estimates from \cite{decia2018}.}
\end{deluxetable*}

%% file: table3_physicspdf.tex
\begin{deluxetable*}{cccccccccc}
\tablecaption{Physical properties of BRIDGE-I absorbers}
\label{tab:physprop}
\tablehead{
    \colhead{Type} &
    \colhead{N} &
    \colhead{$\overline{\log N_{\rm HI}}$} &
    \colhead{$\widetilde{\log N_{\rm HI}}$} &
    \colhead{$\overline{\log N_{\rm H}}$} &
    \colhead{$\widetilde{\log N_{\rm H}}$} &
    \colhead{$\overline{\log n_{\rm H}}$} &
    \colhead{$\widetilde{\log n_{\rm H}}$} &
    \colhead{$\overline{\log l}$} &
    \colhead{$\widetilde{\log l}$} \\
    \colhead{ } &
    \colhead{ } &
    \colhead{$[\rm{cm}^{-2}]$} &
    \colhead{$[\rm{cm}^{-2}]$} &
    \colhead{$[\rm{cm}^{-2}]$} &
    \colhead{$[\rm{cm}^{-2}]$} &
    \colhead{$[\rm{cm}^{-3}]$} &
    \colhead{$[\rm{cm}^{-3}]$} &
    \colhead{$[\rm{kpc}]$} &
    \colhead{$[\rm{kpc}]$} 
}
\startdata
\\ [-2.5ex] 
\hline
\multicolumn{10}{c}{Entire Sample} \\
\hline
all & 75 & $17.87 \pm 1.00$ & $17.46$ & $20.32 \pm 0.74$ & $20.33$ & $-2.24 \pm 0.70$ & $-2.19$ & $1.07 \pm 1.38$ & $1.02$ \\ 
pLLSs & 20 & $16.88 \pm 0.18$ & $16.88$ & $20.42 \pm 0.86$ & $20.22$ & $-2.75 \pm 0.66$ & $-2.56$ & $1.68 \pm 1.50$ & $1.28$ 
\\ 
LLSs  & 41 & $17.81 \pm 0.64$ & $17.51$ & $20.17 \pm 0.68$ & $20.20$ & $-2.04 \pm 0.62$ & $-2.02$ & $0.72 \pm 1.27$ & $0.76$ \\ 
SLLSs & 14 & $19.49 \pm 0.36$ & $19.50$ & $20.61 \pm 0.61$ & $20.60$ & $-2.12 \pm 0.57$ & $-2.08$ & $1.24 \pm 1.16$ & $1.20$ \\ 
\\ [-2.5ex] 
\hline
\multicolumn{10}{c}{Restricted Sample without the $\log U $ Constraint} \\
\hline
all & 63 & $17.99 \pm 1.06$ & $17.51$ & $20.37 \pm 0.70$ & $20.35$ & $-2.23 \pm 0.76$ & $-2.19$ & $1.12 \pm 1.40$ & $1.04$ \\
pLLSs & 16 & $16.87 \pm 0.17$ & $16.87$ & $20.57 \pm 0.79$ & $20.25$ & $-2.89 \pm 0.61$ & $-2.59$ & $1.97 \pm 1.39$ & $1.34$ \\
LLSs  & 33 & $17.84 \pm 0.64$ & $17.51$ & $20.19 \pm 0.64$ & $20.21$ & $-2.02 \pm 0.65$ & $-2.03$ & $0.72 \pm 1.27$ & $0.79$ \\
SLLSs & 14 & $19.49 \pm 0.36$ & $19.50$ & $20.61 \pm 0.61$ & $20.60$ & $-2.12 \pm 0.57$ & $-2.08$ & $1.24 \pm 1.16$ & $1.20$ \\
\enddata
\tablecomments{The bar and tilde above each parameter represent the mean ($\pm$ standard deviation) and the median, respectively, of the logarithmic values of the parameter. We list the number of each absorber type, $\rm{N}$. Our entire statistical sample consists of 99 absorbers with $\log N_{\HI} \le20.0$, here we only include models for which $\log U$ is well-constrained. We also report statistics of the more restricted sample which only includes absorbers for which we did not assume a Gaussian prior on $\log U$.}
\end{deluxetable*}

%% file: table4_meanlogU.tex
\begin{deluxetable}{lccc}
\tablecaption{Mean ionization parameter $\overline{\log U}$ by absorber type and sample. \label{tab:meanlogU}}
\tablehead{\colhead{} & \multicolumn{3}{c}{Mean ionization parameter, $\overline{\log U}$}\\
\colhead{Absorber type} & \colhead{CCC} & \colhead{BRIDGE} & \colhead{KODIAQ-Z}
}
\startdata
SLFS & $-2.56 \pm 0.43$ & \nodata & $-1.52 \pm 0.45$ \\
pLLS & $-2.89 \pm 0.38$ & $-2.09 \pm 0.52$ & $-2.03 \pm 0.51$ \\
LLS & $-3.07 \pm 0.19$ & $-2.67 \pm 0.52$ & $-2.19 \pm 0.46$ \\
SLLS & \nodata & $-2.59 \pm 0.38$ & $-2.50 \pm 0.66$ \\
\enddata
\tablecomments{The sample mean, $\overline{\log U}$, and its standard deviation for each absorber type are derived from the individual measurements shown in Figure~\ref{fig:logUzstat}. BRIDGE does not include SLFS absorbers and CCC does not include SLLSs.}
\end{deluxetable}

%% file: table5_lit_met.tex
\begin{deluxetable}{lcccccc}
\tablecaption{Comparison of Literature and Re-estimated Metallicities \label{tab:lit_sample}}
\tablewidth{0pt}
\tablehead{
\colhead{Target} & \colhead{$z_{\rm abs}$} & \colhead{$\log N_{\rm HI}$} & \colhead{[X/H]$_{\rm lit, HM12}$} & \colhead{[X/H]$_{\rm HM12}$} & \colhead{[X/H]$_{\rm HM05}$} & \colhead{Reference}
}
\startdata
J000345$-$232346 & $2.187$ & $19.65\pm0.15$ & $-1.57^{+0.07}_{-0.09}$ & $-1.43^{+0.03}_{-0.03}$ & $-1.60^{+0.04}_{-0.05}$ & P15 \\
J034024$-$051909 & $2.174$ & $19.35\pm0.20$ & $-1.87^{+0.08}_{-0.12}$ & $-1.81^{+0.07}_{-0.07}$ & $-1.90^{+0.02}_{-0.04}$ & P15 \\
J082849$+$085854 & $2.044$ & $19.90\pm0.10$ & $-0.59^{+0.12}_{-1.21}$ & $-0.44^{+0.07}_{-0.07}$ & $-0.43^{+0.07}_{-0.07}$ & P15 \\
J092705$+$562114 & $1.775$ & $19.00\pm0.10$ & $+0.54^{+0.06}_{-0.06}$ & $+1.00^{+0.08}_{-0.08}$ & $+0.94^{+0.08}_{-0.09}$ & P15 \\
J101939$+$524627 & $1.834$ & $19.10\pm0.30$ & $-0.68^{+0.21}_{-1.38}$ & $-0.30^{+0.04}_{-0.03}$ & $-0.43^{+0.04}_{-0.05}$ & P15 \\
J103921$-$271916 & $2.139$ & $19.55\pm0.10$ & $-0.41^{+0.22}_{-2.86}$ & $-0.25^{+0.03}_{-0.01}$ & $-0.25^{+0.01}_{-0.01}$ & S13 \\
J145649$-$193852 & $2.170$ & $19.75\pm0.20$ & $-0.65^{+0.07}_{-0.07}$ & $-0.55^{+0.05}_{-0.05}$ & $-0.90^{+0.04}_{-0.04}$ & P15 \\
J150932$+$111311 & $1.821$ & $18.50\pm0.50$ & $-1.13^{+0.11}_{-0.13}$ & $-1.22_{-0.18}^{+0.15}$ & $-1.30^{+0.10}_{-0.15}$ & P15 \\
\enddata
\tablecomments{17 absorbers: 2 from \citealt{som2013} (S13), 3 from \citealt{dessauges2003} (D03), 12 from \citealt{prochaska2015} (P15). Literature metallicities are drawn from \cite{fumagalli2016} who adopted the HM12 EUVB; we re-estimate metallicities using the HM05 EUVB for consistency with the rest of the sample (see \S\ref{sec:metallicity}).}
\end{deluxetable}

%% file: table6_sample.tex
\startlongtable
\begin{deluxetable*}{lcrcccccrc}
\tablecaption{The BRIDGE sample.} \label{tab:full_sample}

\tablehead{
    \colhead{Target} &
    \colhead{RA} &
    \colhead{DEC} &
    \colhead{$z_{\rm qso}$} &
    \colhead{$z_{\rm abs}$} &
    \colhead{Sample} & \colhead{UV data} &
    \colhead{$\log N_{\HI}$} &
    \colhead{\HI\ flag} &
    \colhead{Optical data}\\
    \colhead{} &
    \colhead{(deg)} &
    \colhead{(deg)} &
    \colhead{} &
    \colhead{} &
    \colhead{} & \colhead{} &
    \colhead{[cm$^{-2}$]} &
    \colhead{} &
    \colhead{}    
}
\startdata
J000345$-$232346 & $0.938$ & $-23.396$ & $2.262$ & $2.187$ & P15 & \nodata & $19.65\pm0.15$ & $0$ & MIKE \\
J000448$-$415728 & $1.201$ & $-41.958$ & $2.760$ & $2.301$ & R11 & FOS$-$H & $18.08\pm0.27$ & $-2$ & UVES \\
J001210$-$012207 & $3.045$ & $-1.369$ & $1.998$ & $1.726$ & R11 & STIS & $17.44_{-0.02}^{+0.01}$ & $0$ & UVES \\
J002151$-$012834 & $5.466$ & $-1.476$ & $1.588$ & $1.409$ & R11 & STIS & $17.27_{-0.06}^{+0.04}$ & $0$ & UVES \\
J002151$-$012834 & $5.466$ & $-1.476$ & $1.588$ & $1.503$ & R11 & STIS & $17.19_{-0.04}^{+0.03}$ & $0$ & UVES \\
J004342$-$262209 & $10.928$ & $-26.369$ & $3.053$ & $1.892$ & R11 & STIS & $17.34\pm0.05$ & $0$ & UVES \\
J010054+021136 & $15.225$ & $+2.194$ & $1.954$ & $1.464$ & R11 & FOS$-$H & $17.09\pm0.03$ & $0$ & UVES \\
J012017+213346 & $20.072$ & $+21.563$ & $1.500$ & $1.048$ & R11 & FOS$-$L & $17.62\pm0.05$ & $0$ & UVES \\
\enddata
\tablecomments{We list the quasar sightlines (Target), their coordinates (RA and DEC in degrees), and redshifts ($z_{\rm qso}$) for each system. The absorber redshifts $z_{\rm abs}$ listed here are estimated using metal lines in the optical spectra. We provide a ``Sample name" for each system, specifying the reference for the UV spectra: \citealt{ribaudo2011} (R11), \citealt{omeara2013} (O13), \citealt{prochaska2015} (P15), \citealt{som2013} (S13), and \citealt{dessauges2003} (D03). We also list the UV instrument used to obtain the spectra. The $N_{\rm HI}$ values and their uncertainties (either adopted from published work or reassessed in this work) are listed next. We assign an \HI\, flag of $0$ for unsaturated measurements, and $-2$ for saturated absorbers; in the latter case, $N_{\rm HI}$ is constrained to a range, with a lower limit from the Lyman limit and an upper limit from the Lyman series lines accessible in the optical spectra. For saturated absorbers we list the central value and symmetric uncertainties that bound these limits. The last column lists the instrument used to obtain the optical spectra: HIRES on the Keck telescope, UVES on the VLT, or MIKE on the Magellan telescope.}
\end{deluxetable*}

%% file: table7_metal_cols.tex
\startlongtable
\begin{deluxetable}{cccrrrccrcc}
\tablecaption{Summary of the Metal Ion Column Densities of the Absorbers\label{tab:summary}}
\tablewidth{0pt}
\tablehead{
    \colhead{Target} & \colhead{$z_{\rm abs}$} & \colhead{Ion} & \colhead{$v_1$} & \colhead{$v_2$} & \colhead{$v$} & \colhead{$\sigma_v$} & \colhead{$\log N$} & \colhead{Detection} & \colhead{Reliability} & \colhead{Unique} \\
    \colhead{} & \colhead{} & \colhead{} & \colhead{(km~s$^{-1}$)} & \colhead{(km~s$^{-1}$)} & \colhead{(km~s$^{-1}$)} & \colhead{(km~s$^{-1}$)} & \colhead{[cm$^{-2}$]} & \colhead{Flag} & \colhead{Flag} & \colhead{ID}
}
\startdata
J000448$-$415728 & $2.301$ & Al~\textsc{ii} & $-50.0$ & $200.0$ & $81.1$ & $0.1$ & $>13.43$ & $-2$ & $3$ & $1$ \\
J000448$-$415728 & $2.301$ & Al~\textsc{iii} & $-50.0$ & $200.0$ & $66.3$ & $0.2$ & $13.48 \pm 0.01$ & $0$ & $1$ & $1$ \\
J000448$-$415728 & $2.301$ & C~\textsc{ii} & $-151.6$ & $203.3$ & $60.7$ & $6.3$ & $>15.35$ & $-2$ & $3$ & $1$ \\
J000448$-$415728 & $2.301$ & C~\textsc{iii} & $-150.0$ & $300.0$ & $57.5$ & $31.5$ & $>15.01$ & $-2$ & $3$ & $1$ \\
J000448$-$415728 & $2.301$ & C~\textsc{iv} & $-127.4$ & $300.2$ & $70.5$ & $0.5$ & $>15.17$ & $-2$ & $1$ & $1$ \\
J000448$-$415728 & $2.301$ & Fe~\textsc{ii} & $-50.0$ & $200.0$ & $78.7$ & $0.4$ & $14.12 \pm 0.01$ & $0$ & $1$ & $1$ \\
J000448$-$415728 & $2.301$ & Mg~\textsc{i} & $-70.0$ & $175.1$ & $62.5$ & $0.5$ & $12.55 \pm 0.01$ & $0$ & $1$ & $1$ \\
J000448$-$415728 & $2.301$ & Mg~\textsc{ii} & $-70.0$ & $200.0$ & $75.2$ & $0.7$ & $>14.36$ & $-2$ & $1$ & $1$ \\
\enddata
\tablecomments{
A value of $999.0$ for $v$ and $\sigma_v$ indicates an upper limit (non-detection), for which the velocity centroid is not measured. The detection flag is defined as follows: $0$ = detection; $-1$ = upper limit (non-detection); $-2$ = lower limit (saturation). The reliability flag is defined as follows: $1$ = non-detection or a measurement based on multiple transitions; $2$ = at least one detection consistent with upper limits; $3$ = a measurement based on only a single transition, or on several saturated transitions. Column densities affected by mild saturation in a doublet have been corrected following the prescription of \cite{savage1991}, assuming a factor of two difference in $f$-values between the doublet members. We report measurements for the 88 absorbers for which we measured the metal ion column densities. For the remaining 17 absorbers from the literature, we refer the reader to the respective papers.
}
\end{deluxetable}

%% file: table8_cloudyres.tex
\begingroup
\setlength{\tabcolsep}{1pt}
\startlongtable
\begin{deluxetable*}{lccccc}
\tablecaption{Metallicities of the Absorbers in the BRIDGE Sample \label{tab:bridge_metallicity}}
\tablehead{
    \colhead{QSO Name} &
    \colhead{$z_{\rm abs}$} &
    \colhead{$\log N_{\rm H\,I}$} &
    \colhead{[X/H]} &
    \colhead{$\log U$} &
    \colhead{[C/$\alpha$]} \\
    \colhead{} &
    \colhead{} &
    \colhead{[cm$^{-2}$]} &
    \colhead{} &
    \colhead{} &
    \colhead{}
}
\startdata
J000345$-$232346 & $2.187$ & $19.65 \pm 0.15$ & $-1.65$, $-1.60$, $-1.56$ & $-2.54$, $-2.51$, $-2.48$ & $-0.01$, $0.07$, $0.16$ \\
J000448$-$415728 & $2.301$ & $18.08 \pm 0.27$ & $-0.50$, $-0.50$, $-0.50$ & $-2.57$, $-2.57$, $-2.56$ & $0.46$, $0.63$, $0.86$ \\
J001210$-$012207 & $1.726$ & $17.44 \pm 0.01$ & $<-4.72$, $<-3.45$, $<-1.98$ & $<-4.37$, $<-3.11$, $<-1.55$ & $<-0.83$, $<-0.10$, $<0.77$ \\
J002151$-$012834 & $1.409$ & $17.27 \pm 0.05$ & $<-4.69$, $<-3.50$, $<-2.36$ & $-3.25$, $-2.70$:, $-2.18$ & $-0.26$, $0.08$:, $0.42$ \\
J002151$-$012834 & $1.503$ & $17.19 \pm 0.04$ & $<-4.73$, $<-3.59$, $<-2.51$ & $-2.65$, $-2.25$:, $-1.85$ & $-0.27$, $0.08$:, $0.43$ \\
J004342$-$262209 & $1.892$ & $17.34 \pm 0.05$ & $-1.38$, $-1.31$, $-1.25$ & $-2.88$, $-2.84$, $-2.80$ & $0.59$, $0.66$, $0.73$ \\
J010054+021136 & $1.464$ & $17.09 \pm 0.03$ & $-0.42$, $-0.37$, $-0.33$ & $-1.57$, $-1.54$, $-1.50$ & $0.23$, $0.38$:, $0.61$ \\
J012017+213346 & $1.048$ & $17.62 \pm 0.05$ & $-1.14$, $-1.09$, $-1.05$ & $-3.66$, $-3.63$, $-3.61$ & $-0.27$, $0.08$:, $0.42$ \\
\enddata
\tablecomments{The lower to upper bounds for each quantity represent the 68\% CI for detections and the 80\% CI for the upper or lower limits; the middle number is the median value. A colon (:) following the median value in the $\log U$ or [C/$\alpha$] column indicates that a Gaussian prior was adopted for that quantity in the absorber's photoionization model. Upper and lower limits are denoted with $<$ and $>$, respectively.}
\tablenotetext{a}{The literature absorbers with $\log N_{\HI}>20$ that we analysed using a smaller grid of $n_{\rm H}$ and $[\rm{X/H}]$ spanning $-4.5$--$0.0$ and $-5.0$--$2.5$ (see \S\ref{subsec:sampleg20}). These are not a part of the statistical sample. }
\end{deluxetable*}
\endgroup

%% file: table9_physprop.tex
\startlongtable
\begin{deluxetable*}{lccccccc}
\tablecaption{Physical Properties of the BRIDGE Sample \label{tab:bridge_physical}}
\tablehead{
    \colhead{QSO Name} &
    \colhead{$z_{\rm abs}$} &
    \colhead{$\log N_{\rm H\,I}$} &
    \colhead{$\log N_{\rm H}$} &
    \colhead{$\log n_{\rm H}$} &
    \colhead{$\log T$} &
    \colhead{$l$} \\
    \colhead{} &
    \colhead{} &
    \colhead{[cm$^{-2}$]} &
    \colhead{[cm$^{-2}$]} &
    \colhead{[cm$^{-3}$]} &
    \colhead{[K]} &
    \colhead{(kpc)}
}
\startdata
J000345-232346 & 2.187 & $19.65 \pm 0.15$ & $20.64$, $20.67$, $20.71$ & $-2.19$, $-2.16$, $-2.13$ & $4.22$, $4.23$, $4.23$ & $19.07$, $22.30$, $25.93$ \\
J000448-415728 & 2.301 & $18.08 \pm 0.27$ & $20.46$, $20.46$, $20.47$ & $-2.12$, $-2.11$, $-2.11$ & $4.10$, $4.10$, $4.10$ & $12.01$, $12.24$, $12.48$ \\
J001210-012207 & 1.726 & $17.44 \pm 0.01$ & $18.55$, $19.66$, $21.07$ & $-3.13$, $-1.57$, $-0.31$ & $4.07$, $4.19$, $4.42$ & $0.00$, $0.56$, $233.95$ \\
J002151-012834 & 1.409 & $17.27 \pm 0.05$ & $19.38$, $20.00$, $20.61$ & $-2.57$, $-2.04$, $-1.50$ & $4.18$, $4.26$, $4.37$ & $0.24$, $3.58$, $49.39$ \\
J002151-012834 & 1.503 & $17.19 \pm 0.04$ & $19.99$, $20.46$, $20.94$ & $-2.87$, $-2.47$, $-2.07$ & $4.28$, $4.35$, $4.45$ & $3.74$, $27.12$, $211.69$ \\
J004342-262209 & 1.892 & $17.34 \pm 0.05$ & $19.80$, $19.87$, $19.94$ & $-1.87$, $-1.83$, $-1.79$ & $4.20$, $4.21$, $4.22$ & $1.27$, $1.61$, $2.04$ \\
J010054+021136 & 1.464 & $17.09 \pm 0.03$ & $20.84$, $20.91$, $20.98$ & $-3.23$, $-3.19$, $-3.15$ & $4.19$, $4.21$, $4.22$ & $317.54$, $407.52$, $522.36$ \\
J012017+213346 & 1.048 & $17.62 \pm 0.05$ & $19.15$, $19.20$, $19.24$ & $-1.28$, $-1.26$, $-1.24$ & $4.09$, $4.09$, $4.10$ & $0.08$, $0.09$, $0.11$ \\
\enddata
\tablecomments{Columns list the 16th, 50th, and 84th percentile values from
the MCMC posterior for $\log N_{\rm H}$, $\log n_{\rm H}$, $\log T$, and $l$.}
\end{deluxetable*}

%% file: table10_lit_met_cols.tex
\startlongtable
\begin{deluxetable}{lccRcccr}
\tablecaption{Column Densities adopted for the Literature Absorbers\label{tab:lit_met_cols}}
\tablewidth{0pt}
\tablehead{
    \colhead{Target} & \colhead{Sample} & \colhead{$z_{\rm abs}$} & \colhead{[$v_{\rm min}$}, $v_{\rm max}]$ & \colhead{Data} & \colhead{Ion} & \colhead{$\log N$} & \colhead{Detection}\\
    \colhead{} & \colhead{} & \colhead{} & \colhead{(km~s$^{-1}$)} & \colhead{} & \colhead{} & \colhead{[cm$^{-2}$]} & \colhead{Flag}
}
\startdata
J103921$-$271916 & S13 & $2.139$ & $-100$, $325$ & MIKE & \HI\  &  $19.55\pm0.10$ & $0$\\
$\,$ & $\,$ & $\,$ & $\,$ & $\,$ & \MgII\ &  $12.98\pm0.01$ & $0$\\
$\,$ & $\,$ & $\,$ & $\,$ & $\,$ & \AlII\ &   $>13.66$ & $-2$\\
$\,$ & $\,$ & $\,$ & $\,$ & $\,$ & \CII\ &  $>15.08$ & $-2$\\
$\,$ & $\,$ & $\,$ & $\,$ & $\,$ & \SII\  &  $14.75\pm0.04$ & $0$\\
J131119$-$012030 & S13 & $1.762$ & $0$, $550$ & MIKE & \HI\ &  $20.00\pm0.08$ & $0$\\
$\,$ & $\,$ & $\,$ & $\,$ & $\,$ & \MgI\  & $12.25\pm0.14$ & $0$\\
$\,$ & $\,$ & $\,$ & $\,$ & $\,$ & \MgII\  & $>14.15$ & $-2$\\
$\,$ & $\,$ & $\,$ & $\,$ & $\,$ & \AlII\  &  $>12.90$ & $-2$\\
$\,$ & $\,$ & $\,$ & $\,$ & $\,$ & \AlIII\ &  $<11.85$ & $-1$\\
J144653+011355 & D03 & $2.087$ & $-180$, $200$ & UVES & \HI\ & $20.18\pm0.10$ & $0$\\
$\,$ & $\,$ & $\,$ & $\,$ & $\,$ & \OI\  & $>16.08$ & $-2$\\
$\,$ & $\,$ & $\,$ & $\,$ & $\,$ & \CII\  & $>16.30$ & $-2$\\
$\,$ & $\,$ & $\,$ & $\,$ & $\,$ & \SiII\  & $14.83\pm0.02$ & $0$\\
$\,$ & $\,$ & $\,$ & $\,$ & $\,$ & \AlII\  & $13.42\pm0.03$ & $0$\\
$\,$ & $\,$ & $\,$ & $\,$ & $\,$ & \FeII\  & $14.41\pm0.03$ & $0$\\
J151352+085555 & D03 & $2.088$ & $-160$, $160$ & UVES & \HI\ & $19.47\pm0.10$ & $0$\\
$\,$ & $\,$ & $\,$ & $\,$ & $\,$ & \OI\ & $<15.30$ & $-1$\\
$\,$ & $\,$ & $\,$ & $\,$ & $\,$ & \SiII\ & $14.05\pm0.10$ & $0$\\
$\,$ & $\,$ & $\,$ & $\,$ & $\,$ & \AlII\ & $12.93\pm0.14$ & $0$\\
$\,$ & $\,$ & $\,$ & $\,$ & $\,$ & \AlIII\ & $<11.84$ & $-1$\\
$\,$ & $\,$ & $\,$ & $\,$ & $\,$ & \FeII\ & $13.26\pm0.03$ & $0$\\
$\,$ & $\,$ & $\,$ & $\,$ & $\,$ & \FeIII\ & $<13.75$ & $-1$\\
J211927$-$353740 & D03 & $1.996$ & $-100$, $200$ & UVES & \HI & $20.06\pm0.10$ & $0$\\
$\,$ & $\,$ & $\,$ & $\,$ & $\,$ & \OI\ & $>16.68$ & $-2$\\
$\,$ & $\,$ & $\,$ & $\,$ & $\,$ & \SiII\ & $15.33\pm0.12$ & $0$\\
$\,$ & $\,$ & $\,$ & $\,$ & $\,$ & \SII\ & $<14.95$ & $-1$\\
$\,$ & $\,$ & $\,$ & $\,$ & $\,$ & \AlII\ & $>13.94$ & $-2$\\
$\,$ & $\,$ & $\,$ & $\,$ & $\,$ & \AlIII\ & $<13.25$ & $-1$\\
$\,$ & $\,$ & $\,$ & $\,$ & $\,$ & \FeII\ & $14.77\pm0.09$ & $0$\\
$\,$ & $\,$ & $\,$ & $\,$ & $\,$ & \FeIII\ & $<14.60$ & $-1$\\
$\,$ & $\,$ & $\,$ & $\,$ & $\,$ & \ZnII\ & $12.29\pm0.09$ & $0$\\
$\,$ & $\,$ & $\,$ & $\,$ & $\,$ & \CIV\ & $>15.28$ & $-2$\\
$\,$ & $\,$ & $\,$ & $\,$ & $\,$ & \SiIV\ & $14.38\pm0.02$ & $0$\\
J000345$-$232346 & P15 & $2.187$ & $-213$, $64$ & MIKE & \HI\ & $19.65\pm{0.15}$ & $0$ \\
$\,$ & $\,$ & $\,$ & $\,$ & $\,$ &\CII\ &    $>14.45$ &  $-2$ \\
$\,$ & $\,$ & $\,$ & $\,$ & $\,$ &\CIV\ &    $14.26\pm0.05$ &  $0$ \\
$\,$ & $\,$ & $\,$ & $\,$ & $\,$ &\OI\  &    $14.16\pm0.05$ &  $0$ \\
$\,$ & $\,$ & $\,$ & $\,$ & $\,$ &\MgI\ &  $<11.72$  &  $-1$  \\
$\,$ & $\,$ & $\,$ & $\,$ & $\,$ &\MgII\ & $13.64\pm0.05$  &  $0$ \\
$\,$ & $\,$ & $\,$ & $\,$ & $\,$ &\AlII\ & $>13.00$ &  $-2$ \\
$\,$ & $\,$ & $\,$ & $\,$ & $\,$ &\AlIII\ & $<12.40$ &  $-1$  \\
$\,$ & $\,$ & $\,$ & $\,$ & $\,$ &\SiII\ & $13.75\pm0.05$  &  $0$ \\
$\,$ & $\,$ & $\,$ & $\,$ & $\,$ &\SiIV\ & $13.74\pm0.05$  &  $0$ \\
$\,$ & $\,$ & $\,$ & $\,$ & $\,$ &\SII\ &  $14.19\pm0.13$  &  $0$ \\
$\,$ & $\,$ & $\,$ & $\,$ & $\,$ &\FeII\ & $13.11\pm0.05$  &  $0$ \\
$\,$ & $\,$ & $\,$ & $\,$ & $\,$ &\ZnII\ & $<12.38$  & $-1$  \\
J020455+364918 & P15 &  $1.955$ & $-102$, $218$ & HIRES & \HI\ & $20.10\pm{0.20}$ & $0$ \\
$\,$ & $\,$ & $\,$ & $\,$ & $\,$ &\AlII\ & $>13.77$ &  $-2$ \\
$\,$ & $\,$ & $\,$ & $\,$ & $\,$ &\AlIII\ & $13.61\pm0.05$ &  $0$  \\
$\,$ & $\,$ & $\,$ & $\,$ & $\,$ &\SiII\ & $15.12\pm0.09$  &  $0$ \\
$\,$ & $\,$ & $\,$ & $\,$ & $\,$ &\ZnII\ & $<12.39$  & $-1$  \\
J034024$-$051909 & P15 & $2.174$ & $-72$, $57$ & HIRES & \HI\ & $19.35\pm{0.20}$ & $0$\\
$\,$ & $\,$ & $\,$ & $\,$ & $\,$ &\CII\ & $>14.40$ &  $-2$ \\
$\,$ & $\,$ & $\,$ & $\,$ & $\,$ &\CIV\ & $13.86\pm0.05$ &  $0$  \\
$\,$ & $\,$ & $\,$ & $\,$ & $\,$ &\OI\ & $>14.56$  &  $-2$ \\
$\,$ & $\,$ & $\,$ & $\,$ & $\,$ &\AlII\ & $12.64\pm0.05$ &  $0$  \\
$\,$ & $\,$ & $\,$ & $\,$ & $\,$ &\AlIII\ & $12.49\pm0.14$ &  $0$  \\
$\,$ & $\,$ & $\,$ & $\,$ & $\,$ &\SiII\ & $13.84\pm0.05$  &  $0$ \\
$\,$ & $\,$ & $\,$ & $\,$ & $\,$ &\SiIV\ & $13.39\pm0.05$ &  $0$  \\
J082849+085854 & P15 & $2.044$ & $-249$, $90$ & HIRES & \HI\ & $19.90\pm{0.10}$ & $0$ \\
$\,$ & $\,$ & $\,$ & $\,$ & $\,$ &\CII\ & $>15.14$ &  $-2$ \\
$\,$ & $\,$ & $\,$ & $\,$ & $\,$ &\OI\ & $>15.49$  &  $-2$ \\
$\,$ & $\,$ & $\,$ & $\,$ & $\,$ &\AlIII\ & $13.59\pm0.05$ &  $0$  \\
$\,$ & $\,$ & $\,$ & $\,$ & $\,$ &\SiII\ & $15.25\pm0.10$  &  $0$ \\
$\,$ & $\,$ & $\,$ & $\,$ & $\,$ &\FeII\ & $14.88\pm0.05$  &  $0$ \\
J092705+562114 & P15 & $1.775$ & $-261$, $198$ & HIRES & \HI\ & $19.00\pm{0.10}$ & $0$ \\
$\,$ & $\,$ & $\,$ & $\,$ & $\,$ &\CII\ & $>15.40$ &  $-2$ \\
$\,$ & $\,$ & $\,$ & $\,$ & $\,$ &\OI\ & $>15.63$  &  $-2$ \\
$\,$ & $\,$ & $\,$ & $\,$ & $\,$ &\AlII\ & $>13.92$ &  $-2$ \\
$\,$ & $\,$ & $\,$ & $\,$ & $\,$ &\AlIII\ & $14.05\pm0.05$ &  $0$  \\
$\,$ & $\,$ & $\,$ & $\,$ & $\,$ &\SiII\ & $15.58\pm0.05$  &  $0$ \\
$\,$ & $\,$ & $\,$ & $\,$ & $\,$ &\SII\ & $15.29\pm0.05$ &  $0$  \\
$\,$ & $\,$ & $\,$ & $\,$ & $\,$ &\ZnII\ & $<12.62$  & $-1$  \\
J095309+523029 & P15 & $1.768$ & $-200$, $300 $& HIRES & \HI\ & $20.10\pm{0.10}$ & $0$ \\
$\,$ & $\,$ & $\,$ & $\,$ & $\,$ &\CII\ & $>15.44$ &  $-2$ \\
$\,$ & $\,$ & $\,$ & $\,$ & $\,$ &\CIV\ & $>15.21$ &  $-2$  \\
$\,$ & $\,$ & $\,$ & $\,$ & $\,$ &\OI\ & $>15.68$  &  $-2$ \\
$\,$ & $\,$ & $\,$ & $\,$ & $\,$ &\MgII\ & $<15.43$  & $-1$  \\
$\,$ & $\,$ & $\,$ & $\,$ & $\,$ &\AlII\ & $>13.96$ &  $-2$ \\
$\,$ & $\,$ & $\,$ & $\,$ & $\,$ &\AlIII\ & $13.86\pm0.05$ &  $0$  \\
$\,$ & $\,$ & $\,$ & $\,$ & $\,$ &\SiII\ & $15.67\pm0.05$  &  $0$ \\
$\,$ & $\,$ & $\,$ & $\,$ & $\,$ &\SiIV\ & $>14.57$ &  $-2$ \\
$\,$ & $\,$ & $\,$ & $\,$ & $\,$ &\SII\ & $15.35\pm0.05$ &  $0$  \\
$\,$ & $\,$ & $\,$ & $\,$ & $\,$ &\FeII\ & $14.99\pm0.10$  &  $0$ \\
$\,$ & $\,$ & $\,$ & $\,$ & $\,$ &\ZnII\ & $12.89\pm0.05$  & $0$  \\
J101939+524627 & P15 & $1.834$ & $-198$, $78$ & HIRES & \HI\ & $19.10\pm{0.30}$ & $0$ \\
$\,$ & $\,$ & $\,$ & $\,$ & $\,$ &\CIV\ & $>14.93$ &  $-2$ \\
$\,$ & $\,$ & $\,$ & $\,$ & $\,$ &\AlII\ & $>13.34$ &  $-2$ \\
$\,$ & $\,$ & $\,$ & $\,$ & $\,$ &\AlIII\ & $13.62\pm0.05$ &  $0$  \\
$\,$ & $\,$ & $\,$ & $\,$ & $\,$ &\SiII\ & $15.32\pm0.05$  &  $0$ \\
$\,$ & $\,$ & $\,$ & $\,$ & $\,$ &\SiIV\ & $>14.14$ &  $-2$ \\
$\,$ & $\,$ & $\,$ & $\,$ & $\,$ &\FeII\ & $14.19\pm0.05$  &  $0$ \\
$\,$ & $\,$ & $\,$ & $\,$ & $\,$ &\ZnII\ & $12.43\pm0.09$  & $0$  \\
J115940$-$003203 & P15 & $1.904$ & $-380$, $90$ & HIRES & \HI\ & $20.05\pm{0.15}$ & $0$\\
$\,$ & $\,$ & $\,$ & $\,$ & $\,$ & \CII\ & $>15.38$ &  $-2$ \\
$\,$ & $\,$ & $\,$ & $\,$ & $\,$ & \CIV\ & $>15.22$ &  $-2$  \\
$\,$ & $\,$ & $\,$ & $\,$ & $\,$ & \OI\ & $>15.66$  &  $-2$ \\
$\,$ & $\,$ & $\,$ & $\,$ & $\,$ & \AlII\ & $>13.98$ &  $-2$ \\
$\,$ & $\,$ & $\,$ & $\,$ & $\,$ & \AlIII\ & $13.82\pm0.05$ &  $0$  \\
$\,$ & $\,$ & $\,$ & $\,$ & $\,$ & \SiII\ & $15.14\pm0.10$  &  $0$ \\
$\,$ & $\,$ & $\,$ & $\,$ & $\,$ & \SiIV\ & $>14.54$ &  $-2$ \\
$\,$ & $\,$ & $\,$ & $\,$ & $\,$ & \SII\ & $15.19\pm0.05$ &  $0$  \\
J145649$-$193852 & P15 & $2.170$ & $-173$, $145$ & MIKE & \HI\ & $19.75\pm{0.20}$ & $0$\\
$\,$ & $\,$ & $\,$ & $\,$ & $\,$ &\MgI\ & $12.35\pm0.05$ &  $0$  \\
$\,$ & $\,$ & $\,$ & $\,$ & $\,$ &\MgII\ & $>14.23$ &  $-2$ \\
$\,$ & $\,$ & $\,$ & $\,$ & $\,$ &\AlII\ & $>13.38$ &  $-2$ \\
$\,$ & $\,$ & $\,$ & $\,$ & $\,$ &\AlIII\ & $12.99\pm0.05$ &  $0$  \\
$\,$ & $\,$ & $\,$ & $\,$ & $\,$ &\SiII\ & $<14.84$  & $-1$  \\
$\,$ & $\,$ & $\,$ & $\,$ & $\,$ &\FeII\ & $14.26\pm0.05$  &  $0$ \\
$\,$ & $\,$ & $\,$ & $\,$ & $\,$ &\ZnII\ & $<12.28$  & $-1$  \\
J150932+111311 & P15 & $1.821$ & $-85$, $105$ & HIRES & \HI\ & $18.50\pm{0.50}$ & $0$\\
$\,$ & $\,$ & $\,$ & $\,$ & $\,$ &\CIV\ & $>14.83$ &  $-2$ \\
$\,$ & $\,$ & $\,$ & $\,$ & $\,$ &\AlII\ & $>13.12$ &  $-2$ \\
$\,$ & $\,$ & $\,$ & $\,$ & $\,$ &\AlIII\ & $13.04\pm0.05$ &  $0$  \\
$\,$ & $\,$ & $\,$ & $\,$ & $\,$ &\SiII\ & $14.21\pm0.05$  &  $0$ \\
$\,$ & $\,$ & $\,$ & $\,$ & $\,$ &\SiIV\ & $>14.17$ &  $-2$ \\
$\,$ & $\,$ & $\,$ & $\,$ & $\,$ &\FeII\ & $13.76\pm0.11$  &  $0$ \\
$\,$ & $\,$ & $\,$ & $\,$ & $\,$ &\ZnII\ & $<12.59$  & $-1$  \\
J160843+071508 & P15 & $1.763$ & $-144$, $156$ & HIRES & \HI\ & $19.40\pm{0.30}$ & $0$\\
$\,$ & $\,$ & $\,$ & $\,$ & $\,$ &\AlIII\ & $13.53\pm0.05$ &  $0$  \\
$\,$ & $\,$ & $\,$ & $\,$ & $\,$ &\SiII\ & $<15.80$  & $-1$  \\
J212329$-$005052 & P15 & $2.059$ & $-286$, $137$ & HIRES & \HI\ & $19.25\pm{0.15}$ & $0$\\
$\,$ & $\,$ & $\,$ & $\,$ & $\,$ &\CII\ & $>15.11$ &  $-2$ \\
$\,$ & $\,$ & $\,$ & $\,$ & $\,$ &\CIV\ & $>14.61$ &  $-2$  \\
$\,$ & $\,$ & $\,$ & $\,$ & $\,$ &\AlII\ & $>13.45$ &  $-2$ \\
$\,$ & $\,$ & $\,$ & $\,$ & $\,$ &\AlIII\ & $13.15\pm0.05$ &  $0$  \\
$\,$ & $\,$ & $\,$ & $\,$ & $\,$ &\SiII\ & $14.60\pm0.05$  &  $0$ \\
$\,$ & $\,$ & $\,$ & $\,$ & $\,$ &\SiIV\ & $13.96\pm0.05$ &  $0$  \\
$\,$ & $\,$ & $\,$ & $\,$ & $\,$ &\FeII\ & $14.39\pm0.05$  &  $0$ \\
\enddata
\tablecomments{17 absorbers: 2 from S13, 3 from D03, 12 from P15.
}
\end{deluxetable}